\pdfoutput=1

\documentclass[a4paper,12pt]{article}

\usepackage{amsmath}
\usepackage{graphicx, psfrag, amsfonts, bm}
\usepackage{enumerate}
\usepackage{algorithm}
\usepackage{algpseudocode}
\usepackage{multirow}
\usepackage[makeroom]{cancel}
\usepackage{hhline}
\usepackage{fullpage}
\usepackage{url}
\usepackage{subcaption}
\usepackage[labelformat=simple]{subcaption}

\usepackage{natbib}
\usepackage{authblk}
\usepackage[titletoc,title]{appendix}

\DeclareMathOperator*{\argmin}{arg\,min}

\newcommand{\Mn}{\mbox{Mn}}
\newcommand{\Be}{\mbox{Be}}
\newcommand{\Dir}{\mbox{Dir}}
\newcommand{\unif}{\mbox{Unif}}

\newcommand{\ti}[1]{\tilde{#1}}
\newcommand{\tpi}{\tilde\pi}
\newcommand{\tC}{\tilde{C}}
\newcommand{\tN}{\tilde{N}}
\newcommand{\tn}{\tilde{n}}

\newcommand{\tp}{\tilde{p}}
\newcommand{\tbx}{\tilde{\bm{x}}}

\newcommand{\pbar}{\bar{p}}

\def\bh {\bm h}
\def\bL {\bm L}

\def\bs {\bm s}

\def\bz {\bm z}
\def\bn {\bm n}
\def\brho {\bm \rho}
\def\bx {\bm x}
\def\bZ {\bm Z}
\def\bw {\bm w}
\def\bu {\bm u}
\def\tbu {\tilde{\bm{u}}}
\newcommand{\Cmin}{C_{\mbox{min}}}
\newcommand{\Cmax}{C_{\mbox{max}}}

\newcommand{\Chat}{\hat{C}}
\newcommand{\Zhat}{\hat{\bZ}}
\newcommand{\what}{\hat{\bw}}
\newcommand{\ptkghat}{\hat{p}_{tkg}}

\newcommand{\umle}{\bm u^*}
\newcommand{\wts}{w_{t\star}}

\newcommand{\DD}{\mathcal{D}}
\newcommand{\HH}{\mathcal{H}}
\newcommand{\true}{^{\text{TRUE}}}

\newcommand{\BC}{\text{BC}}

\makeatletter
\renewcommand\paragraph{%
  \@startsection{paragraph}
    {4}
    {\z@}
    {3.25ex \@plus1ex \@minus.2ex}
    {-1em}
    {\normalfont\normalsize\bfseries\maybe@addperiod}%
}
\newcommand{\maybe@addperiod}[1]{%
  #1\@addpunct{.}%
}
\makeatother

\providecommand{\keywords}[1]{\textit{Keywords:  } #1}

\begin{document}

\title{PairClone: A Bayesian Subclone Caller Based on Mutation Pairs}
\author[1]{Tianjian Zhou}
\author[2]{Peter M\"{u}ller\thanks{Email: pmueller@math.utexas.edu}}
\author[3]{Subhajit Sengupta}
\author[3, 4]{Yuan Ji\thanks{Email: yji@health.bsd.uchicago.edu }}
\affil[1]{Department of Statistics and Data Sciences, The University of Texas at Austin}
\affil[2]{Department of Mathematics, The University of Texas at Austin}
\affil[3]{Program for Computational Genomics and Medicine, NorthShore University HealthSystem}
\affil[4]{Department of Public Health Sciences, The University of Chicago}

\maketitle

\begin{abstract}
Tumor cell populations can be thought of as being composed of
homogeneous cell subpopulations, with each subpopulation being
characterized
by overlapping sets of single nucleotide variants (SNVs).  
Such
subpopulations are known as subclones and are an important target for
precision medicine.  Reconstructing such subclones from
next-generation sequencing (NGS) data is one of the major challenges
in precision medicine.  We present PairClone as a new tool to
implement this reconstruction.
The main idea of PairClone is to model short reads mapped to
pairs of proximal
SNVs. In contrast, most existing methods use only marginal 
reads for
unpaired SNVs.  Using Bayesian nonparametric models, we estimate
posterior probabilities of the number, genotypes and population
frequencies of subclones in one or more tumor sample.  We use the
categorical Indian buffet process (cIBP)
as a prior probability model for subclones that are represented
as vectors of categorical matrices that record the corresponding
sets of mutation  pairs.  Performance of PairClone is assessed
using simulated and real 
datasets. 
An open source software package can be obtained at \url{http://www.compgenome.org/pairclone}.
\end{abstract}

\noindent\keywords{
Categorical Indian buffet process; Latent feature model; Local haplotype; Next-generation sequencing; Random categorical matrices; Subclone; Tumor heterogeneity.
}

\section{Introduction}
We explain intra-tumor heterogeneity by representing tumor cell
populations as a mixture of 
subclones. We reconstruct unobserved  subclones by
utilizing information from 
pairs of
proximal mutations
that are obtained from next-generation sequencing (NGS) data. 
We exploit the fact that some short reads
in NGS data cover pairs of phased mutations that reside on
two sufficiently proximal loci. Therefore haplotypes of the
mutation pairs can be observed and used for subclonal inference. 

We develop a suitable sampling model that represents
the paired nature of the data, and construct a nonparametric Bayesian
feature allocation model as a prior for the hypothetical
subclones. Both models together allow us to develop a fully probabilistic
description of the composition of the tumor as a mixture of
homogeneous underlying subclones, including the genotypes  and number of
such subclones.

\subsection{Background}
\begin{figure}[h!]
\begin{center}
\begin{subfigure}[t]{.28\textwidth}
\centering
\includegraphics[width=\textwidth]{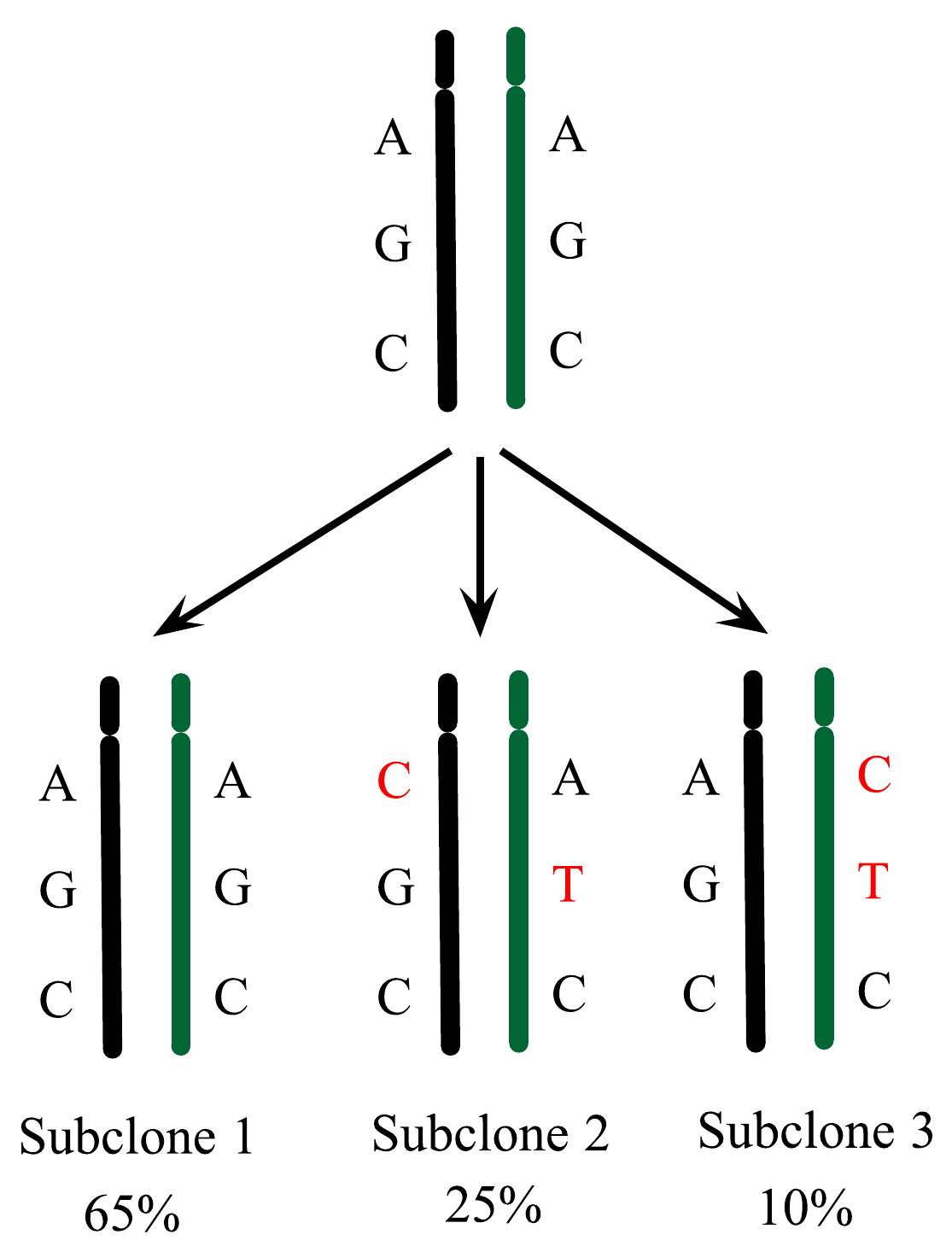}
\caption{Tumor evolution}
\label{fig:tm_evo}	
\end{subfigure}
\hspace{5mm}
\begin{subfigure}[t]{.67\textwidth}
\centering
\includegraphics[width=\textwidth]{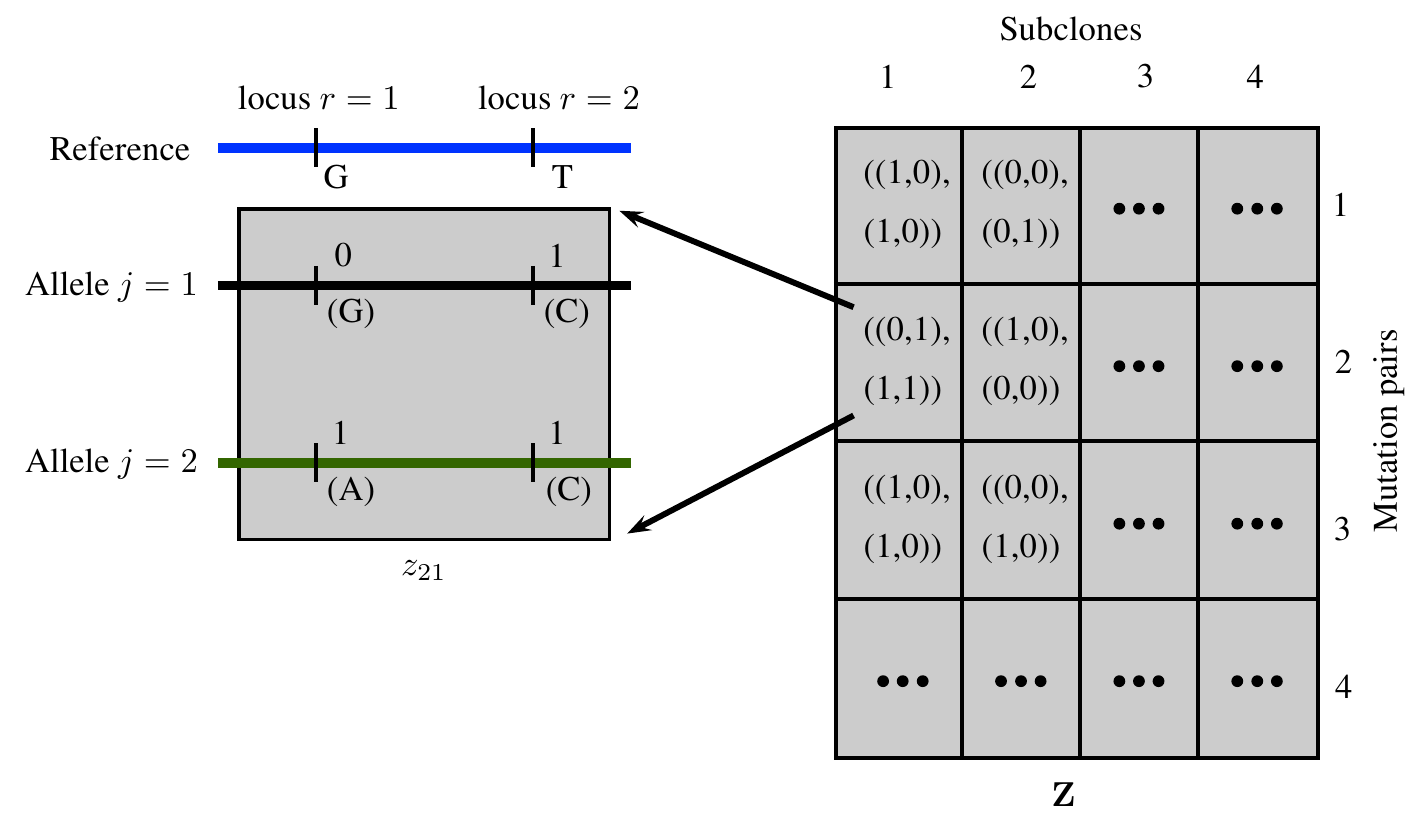}
\caption{Subclone structure matrix $\bZ$}
\label{fig:Z_example}	
\end{subfigure}
\begin{subfigure}[t]{.85\textwidth}
\centering
\vspace{7mm}
\includegraphics[width=\textwidth]{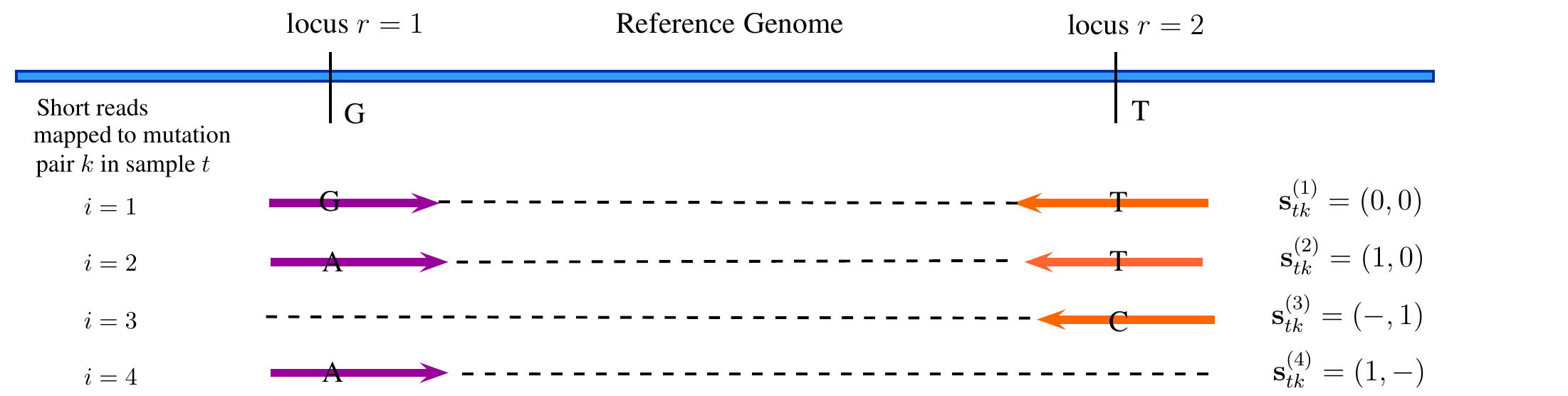}
\caption{Paired-end reads data}
\label{fig:mut_example}
\end{subfigure}
\end{center}
\caption{(a) Illustration of tumor evolution, emergence of subclones and
  their population frequencies. 
  (b) Illustration of the subclone structure matrix $\bZ$.  
  {\bf Right panel:} A subclone is represented by one column of $\bZ$. Each
  element of a column represents the subclonal genotypes for a
  mutation pair. For example, the genotypes for mutation  pair
  2 in subclone 1 is ((0, 1), (1, 1)), which is shown in detail on the
  left panel. 
  {\bf Left panel:} The reference genome for mutation pair $2$ is (G, T) and
  the corresponding genotype of subclone $1$ is ((G, C), (A, C)),
  which gives rise to $\bz_{21} = ((0, 1), (1, 1))$.  
  (c) Illustration of  paired-end reads   data for a mutation pair.  Shown are four short reads mapped to mutation pair $k$ in sample $t$. Some reads are mapped to both loci of the mutation pair, and others are mapped to only one of the two loci. The two ends of the same read are marked with opposing arrows in purple and orange.  }
\end{figure}

NGS technology~\citep{mardis2008next}
has enabled researchers to develop bioinformatics tools that are
being used to understand the landscape of tumors within and across
different samples.  An important related task is to reconstruct cellular
subpopulations in one or more tumor samples, known as subclones.
Mixtures of such subclones with varying population frequencies 
across spatial
locations in the same tumor, across tumors from different
time points,  or across tumors from the primary and metastatic
sites can provide information about the
mechanisms of tumor evolution and metastasis. 
 Heterogeneity of cell populations is seen, for example, in varying
frequencies of distinct somatic mutations.
The hypothetical tumor subclones are homogeneous. That is, a subclone
is characterized by unique genomic variants in its genome~
\citep{marjanovic2013cell,almendro2013cellular,polyak2011heterogeneity,stingl2007molecular,shackleton2009heterogeneity,dexter1978heterogeneity}.
Such subclones arise as the result of cellular evolution,
which can be described by a phylogenetic tree that records how a
sequence of somatic mutations gives rise to different cell subpopulations.
Figure~\ref{fig:tm_evo} provides a stylized and
simple illustration in which a homogeneous sample with one original normal 
clone evolves into a heterogeneous sample with three
subclones. Subclone 1 is the original parent cell population, and
subclones 2 and 3 are descendant subclones of subclone 1, each
possessing somatic mutations marked by the red letters. Each
subclone  possesses  two homologous chromosomes (in black and green), and
each chromosome in Figure~\ref{fig:tm_evo} is marked by a triplet of letters
representing the nucleotide on  the three genomic loci. Together, the
three subclones include four different haplotypes,
(A, G, C), (A, G, T), (C, G, C), and (A, A, T), at these three genomic
loci.  
In addition, each subclone has a different population frequency shown
as the percentage values in
Figure~\ref{fig:tm_evo}.  

We use NGS data to infer such tumor heterogeneity.
In an NGS experiment, DNA fragments are first produced by extracting
the DNA molecules from the cells in a tumor sample. The fragments are
then sequenced using short reads.  For the three subclones in
Figure~\ref{fig:tm_evo}, there are four aforementioned haplotypes at
the three loci. Consequently, short reads that cover some of these
three loci may manifest different alleles. 
For example, if a large
number of 
reads cover the first two loci, we might observe (A, G), (C, G), (A, T) and
(C, T), four alleles 
for the mutation pair. Observing four alleles 
is direct evidence supporting the presence of subclones~\citep{sengupta2015NAR}.
This is because, in the absence of copy number variations there can be
only two haploid genomes at any loci for a homogeneous human sample.
Therefore, one can use mutation pairs in copy
neutral regions to develop statistical inference on the presence and
frequency of subclones. This is the goal of our paper.

Almost all mutation-based subclone-calling methods in the literature
use only single nucleotide variants (SNVs)~\citep{oesper2013theta, strino2013trap, jiao2014inferring, miller2014sciclone, roth2014pyclone, zare2014inferring, deshwar2015phylowgs, Bayclone2015,lee2015bayesian,lee2016bayesianjrssc}.
Instead of examining mutation pairs, SNV-based methods
use marginal counts for each recorded locus only.
Consider, for example,  the first locus in
Figure~\ref{fig:tm_evo}. At this locus, 
the reference genome 
has an ``A''
nucleotide while subclones 2 and 3 have a ``C'' nucleotide. In the entire
sample, the ``C'' nucleotide is roughly present in 17.5\% of the DNA
molecules based on the population frequencies illustrated in
Figure~\ref{fig:tm_evo}. 
The percentage of a mutated allele is called variant allele fraction (VAF).
If a sample is 
homogeneous and assuming no copy number variations at the locus, 
the population frequency for the ``C''
nucleotide should be close to  0, 50\%, or 100\%, depending on the
heterozygosity of the locus.  
Therefore,
if the
population frequency of ``C'' deviates from 0\%, 50\%, or 100\%,
the sample is likely to be heterogeneous.
Based on this argument,
SNV-based subclone callers search for SNVs with
VAFs
that are different from these frequencies
(0, 50\%, 100\%), which are evidence for the presence of different (homogeneous)
subpopulations.
In the event of copy
number variations, a similar but slightly more sophisticated reasoning 
can be applied, see for example, Lee et al. (2016).

\subsection{Using mutation pairs} 
NGS data usually contain 
substantially fewer mutation pairs than marginal SNVs.
However, this does not weaken the power of using mutation pairs as
mutation pairs naturally carry important 
phasing information that improves the
accuracy of subclone reconstruction. For example, imagine a tumor
sample that is a mixture of subclones 2 and 3 in Figure
\ref{fig:tm_evo}. Suppose a sufficient amount of short reads cover the
first two loci, we should observe relatively large reads counts for
four alleles (C, G), (A, T), (C, T) and (A, G). One can then reliably
infer that there are heterogeneous cell subpopulations in the tumor
sample.  
In contrast, if we ignore the phasing information and only consider
the (marginal) VAFs for each SNV, then the observed VAFs for both SNVs
are 50\%, which could be heterogeneous mutations from a single cell
population. See Simulation 1 for an illustration. In summary, we
leverage the power of using mutation pairs over marginal SNVs by
incorporating partial phasing information in our model. 
Besides the simulation study we will later also empirically
confirm these considerations in actual data analysis.

The relative advantage of using mutation pairs over marginal
SNV's can be also be understood as a special case of a more general
theme. In biomedical data it is often important to avoid
overinterpretation of noisy data and to distill a relatively weak signal. A
typical example is the probability of expression (POE) model of
\cite{POE:2002}.
Similarly, the 
modeling of mutation pairs is a way to extract the pertinent information
from the massive noisy data.  Due to noise and artifact in NGS data,
such as base-calling or mapping error, many called SNVs might record
unusual population frequencies, for reasons unrelated to the presence
of subclones~\citep{li2014toward}.  Direct modeling of all marginal
read counts one ends up with noise swamping the desired signal
\citep{nik2012mutational, jiao2014inferring}. See our analysis of a
real data set in Section \ref{sec:realdata} for an example. To
mitigate this challenge, most methods use clustering of the VAFs,
including, for
example, \cite{roth2014pyclone}. One would then use the resulting
cluster centers to infer subclones, which is one way of extracting
more concise information. In addition, the vast majority of the methods in the literature show that even though a tumor sample could possess thousands to millions of SNVs, the number of inferred subclones usually is in the low single digit, no more than 10. To this end, we propose instead an alternative approach
to extract useful information by modeling (fewer) mutation pairs, as
mutation pairs contain more information and are of higher quality. We show in our numerical examples later that with a few dozens of these mutation pairs, the inference on the subclones is strikingly similar to cluster-based subclone callers using much more SNVs.


Finally,
using mutation pairs does not exclude the possibility of making use of marginal SNVs. 
In Section \ref{sec:mrc}, we show it is straightforward to jointly model mutation pairs and SNVs. Other biological complexities, such as tumor purity and copy number variations, can also be incorporated in our model.  See Sections \ref{sec:purity} and \ref{sec:cnv} for more details.

\subsection{Representation of subclones}
\label{sec:repsubclone}
We construct a $K \times C$ categorical valued matrix $\bZ$ (Figure
\ref{fig:Z_example}) to represent the subclone structure. Rows of
$\bZ$ are indexed by $k$ and  represent mutation pairs, and a column of $\bZ$, denoted by $\bm
z_c = (\bz_{1c}, \cdots, \bz_{Kc})$, records the  phased mutation
pairs on the two homologous chromosomes    of subclone $c$, $c = 1,
\ldots, C$.
As in Figure \ref{fig:Z_example},
let  $j = 1, 2$ index the two homologous
chromosomes,
 $r=1, 2$ index the two mutation loci,
$\bz_{kc} = (\bz_{kcj}, j=1,2)$ be  the genotype
consisting of   two alleles for mutation pair $k$ in subclone $c$,
and  $\bz_{kcj}=(z_{kcjr}, r=1,2)$ denote the allele of the $j$-th
homologous chromosome.
 Therefore, each entry $\bz_{kc}$ of the matrix $\bZ$ is a $2 \times
2$ binary submatrix itself. 
For example,  in Figure \ref{fig:Z_example} the
entry $z_{21}$ is a pair of 2-dimension binary row vectors, $(0, 1)$
and $(1, 1)$, representing the genotypes for both alleles at mutation
pair $k=2$ of subclone $c=1$; each vector indicates the allele for the mutation pair on
a homologous chromosome. The first vector $(0, 1)$ indicates that locus
$r = 1$ harbors no mutation (0) and locus $r=2$ harbors a mutation (1).
Similarly, the second vector
$(1, 1)$ marks two mutations on both loci.  

In summary, each entry of $\bZ$,
$$
\bz_{kc} = \left( \bz_{kc1}, \bz_{kc2} \right) = \left( (z_{kc11}, z_{kc12}), (z_{kc21}, z_{kc22}) \right)
$$
is a $2 \times 2$ matrix (with the two row vectors horizontally
displayed for convenience).
Each $z_{kcjr}$ is a binary indicator and $z_{kcjr} = 1$ (or $0$)
indicates a mutation (or reference).  Thus, $\bz_{kc}$
can take $Q = 16$ possible values. That is,
$\bz_{kc} \in 
\{\bz^{(1)}, \ldots, \bz^{(16)}\} =$
$\{(00, 00),$ $(00, 01),$ $\ldots,$ $(11,11) \}$,
where we write $00$ short for $(0,0)$ etc., and
$z^{(1)} = (00, 00)$ refers to  the genotype on the
reference genome. Formally, $\bz_{kc}$ is a $2 \times 2$ binary matrix,
and $\bZ$ is a matrix of such binary matrices.
Moreover, we can collapse some $\bz^{(q)}$ values as we do not have
phasing across mutation pairs.
For example, $\bz_{kc}=(01,10)$ and $\bz_{kc}=(10,01)$, etc. have
mirrored rows and are indistinguishable in defining a subclone (a
column of $\bZ$). (More details in Section \ref{sec:prior}).
Typically distinct mutation pairs are distant from each
other, and in NGS data they are almost never phased. Therefore, we can
reduce the number of possible outcomes of $z_{kc}$ to $Q = 10$, due to
the mirrored outcomes. We list them below for later reference:
$z^{(1)} = (00, 00), z^{(2)} = (00, 01), z^{(3)} = (00, 10), z^{(4)} =
(00, 11), z^{(5)} = (01, 01), z^{(6)} = (01, 10), z^{(7)} = (01, 11),
z^{(8)} = (10, 10), z^{(9)} = (10, 11)$ and $z^{(10)} = (11, 11)$. In
summary, the entire matrix $\bZ$ fully specifies the genomes of each
subclone at all the mutation pairs.

Suppose $T$ tumor samples are available from the same patient,
obtained either at different time points (such as initial diagnosis
and relapses), at the same time but from different spatial
locations   within the  same   tumor, or from tumors at different
metastatic sites. We assume those $T$ samples
share the same subclones, while the  subclonal population frequencies
may   vary across samples.
For clinical decisions it can be important to know the population frequencies
of the subclones. 
To facilitate such inference, we introduce a $T
\times (C+1)$ matrix $\bm w$ to represent the population frequencies
of subclones. The element $w_{tc}$ refers to the proportion of
subclone $c$ in sample $t$, where $0 < w_{tc} < 1$ for all $t$ and
$c$, 
and $\sum_{c = 0}^C w_{tc} = 1$. A background subclone, which has no
biological meaning and is indexed by $c = 0$, is included to account
for artifacts and experimental noise. We will discuss more about
this later.

The remainder of this article is organized as follows. In
Sections~\ref{sec:probmodel} and  \ref{sec:posterior}, we propose a
Bayesian feature allocation model and the corresponding posterior
inference scheme to  estimate   the latent subclone structure. In
Section~\ref{sec:simulation}, we  evaluate   the model with three
simulation studies. 
Section~\ref{sec:pipeline} extends the models to accommodate other
biological complexities and present additional simulation results.
Section~\ref{sec:realdata} reports the analysis results for a lung
cancer patient with multiple tumor biopsies.  We conclude with a final
discussion in
Section~\ref{sec:conclusion}.

\section{The PairClone Model}
\label{sec:probmodel}

\subsection{Sampling Model}
\label{sec:splmodel}
Suppose paired-end short reads data are obtained by deep DNA
sequencing of multiple tumor samples. In such data, a short read is
obtained by sequencing two ends of the same DNA fragment. Usually a
DNA fragment is much longer than a short read, and the two ends
do not overlap and must be mapped separately. 
However, since the
paired-end reads are from the same DNA fragment, they are naturally
phased and can be used for inference of alleles and
subclones.
We use \texttt{LocHap}~\citep{sengupta2015NAR} 
to find pairs of mutations that are no more than a fixed number, say 500,
base pairs apart. 
 Such mutation pairs can be mapped by 
paired-end reads,
making them eligible for PairClone analysis. 
See Figure \ref{fig:mut_example} for an example.
For each mutation pair, a number of short reads are mapped to at least
one of the two loci.  Denote the two sequences on short read $i$
mapped to mutation pair $k$ in tissue sample $t$ by $\bs_{tk}^{(i)}
= \left(s_{tkr}^{(i)}, r = 1, 2\right) = \left(s_{tk1}^{(i)},
s_{tk2}^{(i)}\right)$, where $r=1,2$ index the two loci,
$s_{tkr}^{(i)} = 0 \mbox{ or } 1$ indicates that the short read
sequence is a reference or mutation.  Theoretically, each
$s_{tkr}^{(i)}$ can take four values, A, C, G, T, the four nucleotide
sequences. However, at a single locus, the probability of observing
more than two sequences across short reads is negligible since it
would require the same locus to be mutated twice throughout the life
span of the person or tumor, which is unlikely.
We therefore code $s_{tkr}^{(i)}$ as a binary value.
Also, sometimes a short read may cover only one of the two loci in a pair, and we use
$s_{tkr}^{(i)} = -$ to represent a missing base when there is no overlap between a short read and the corresponding SNV. 
Therefore,
$s_{tkr}^{(i)} \in \{ 0, 1, -\}$.  For example, in Figure
\ref{fig:mut_example} locus $r = 1$, $s_{tk1}^{(1)} = 0$ for read $i =
1$, $s_{tk1}^{(2)} = 1$ for read $i = 2$, and $s_{tk1}^{(3)} = -$ for
read $i = 3$. Reads that are not mapped to either locus are excluded
from analysis since they do not provide any information for
subclones. Altogether, $\bs_{tk}^{(i)}$ can take $G = 8$ possible
values, and its sample space is denoted by $\HH = \{\bh_1,
\ldots, \bh_G \} = \{ 00, 01, 10, 11, -0, -1, 0-, 1-\}$.
Each value corresponds to an allele of two loci,
with $-$ being a special ``missing'' coverage.  For mutation pair $k$
in sample $t$, the number of short reads bearing allele $\bh_g$ is
denoted by $n_{tkg} = \sum_i I \left(\bs_{tk}^{(i)} = \bh_g
\right)$, where $I(\cdot)$ is the indicator function,
 and the total number of reads mapped to the mutation pair is then $N_{tk} = \sum_g n_{tkg}$.
Finally, depending upon whether a read covers both loci or only one
locus we distinguish three cases:
(i) a read maps to both loci (complete), taking values
$\bs_{tk}^{(i)} \in \{\bh_1,\ldots,\bh_4\}$;
(ii) a read maps to the second locus only (left missing), $\bs_{tk}^{(i)} \in \{\bh_5,\bh_6\}$;
and (iii) a read maps to the first locus only (right missing), $\bs_{tk}^{(i)} \in \{\bh_7,\bh_8\}$.
We assume a multinomial sampling model for the observed read counts
\begin{align}
(n_{tk1}, \ldots, n_{tk8}) \mid N_{tk} \sim
\Mn(N_{tk};\; p_{tk1}, \ldots, p_{tk8}).
\label{eq:multi}
\end{align}
Here $\bm p = \{ p_{tkg}, g = 1, \ldots, 8\}$ are the probabilities for the 8 possible values of $\bs_{tk}^{(i)}$. For the upcoming
discussion, we separate out the probabilities for the three missingness cases.
Let $v_{tk1}, v_{tk2}, v_{tk3}$ denote the probabilities of observing a short read satisfying cases (i), (ii) and (iii),
respectively. We write $p_{tkg} = v_{tk1} \, \tp_{tkg}, g = 1, \ldots,
4$, $p_{tkg} = v_{tk2} \, \tp_{tkg}, g = 5, 6$, and $p_{tkg} = v_{tk3}
\, \tp_{tkg}, g = 7, 8$. 
Here $\tp_{tkg}$ are the probabilities conditional on case (i),
(ii) or (iii).
That is, $\sum_{g=1}^4 \tp_{tkg} = \sum_{g=5,6} \tp_{tkg} =
\sum_{g=7,8} \tp_{tkg}=1$. We still use a single running index, $g=1,\ldots,8$,
to match the notation in $p_{tkg}$. 
Below we link the multinomial sampling model with the underlying subclone structure by expressing $\tp_{tkg}$ in terms of $\bZ$ and $\bm w$.
Regarding $v_{tk1}, v_{tk2}, v_{tk3}$ we assume non-informative missingness and therefore do not proceed with inference on them (and $v$'s remain constant factors in the likelihood).

\subsection{Prior Model}
\label{sec:prior}

\paragraph{Construction of $\tp_{tkg}$}
The construction of a prior model for $\tp_{tkg}$ is based on
the following generative model.
To generate a short read, we first select a subclone $c$ from which the read
arises, using the population frequencies $w_{tc}$ for sample
$t$. Next we select with probability $0.5$ one of the two 
DNA strands, 
$j= 1 , 2$.
Finally, we record the read $\bh_g$, $g=1,2,3$ or $4$, corresponding to the chosen allele
$\bz_{kcj}=(z_{kcj1},z_{kcj2})$.
In the case of left (or right) missing locus we observe $\bh_g$, $g=5$
or $6$ (or $g=7$ or $8$), corresponding to the observed locus of the chosen allele.
Reflecting these three generative steps, we denote the probability of
observing a short read $\bh_g$ that bears sequence $\bz_{kcj}$ by 
\begin{equation}
 A(\bh_g, \bz_{kc}) =
   \sum_{j = 1}^2 0.5\,\times\,I(h_{g1} = z_{kcj1})\, I(h_{g2}=z_{kcj2}),
 \label{eq:A}
\end{equation}
with the understanding that $I(- = z_{kcjr}) \equiv 1$ for missing reads.
Implicit in \eqref{eq:A} is the restriction
$A(\bh_g, \bz_{kc}) \in \{0, 0.5, 1\}$, depending on the arguments.

Finally, using the definition of $A(\cdot)$ we model the probability
of observing a short read $\bh_g$ as 
\begin{equation}
   \tp_{tkg} = \sum_{c = 1}^C w_{tc}\,A(\bh_g, \bz_{kc}) +
   w_{t0} \, \rho_g. \label{pprior1}
\end{equation}
In \eqref{pprior1} we include $w_{t0} \rho_g$ to model a background subclone
denoted by $c = 0$ with  population frequency   $w_{t0}$. The
background subclone does not exist and has no biological
interpretation. It is only used as a mathematical
device to account for noise and artifacts in the NGS data (sequencing
errors, mapping errors, etc.). The weights $\rho_g$ are the
conditional probabilities of observing a short read $\bs_{tk}^{(i)}$ harboring
allele  $\bh_g$ if the recorded read were due to experimental noise.
Note that $\rho_1+\ldots + \rho_4= \rho_5+\rho_6 = \rho_7+\rho_8
= 1$.

\paragraph{Prior for $C$}
We assume a geometric distribution prior on $C$, $C \sim
\text{Geom}(r)$, to describe the random number of subclones (columns
of $\bZ$), $p(C) = (1 - r)^{C} r, C \in \{1, 2, 3,
\ldots\}$. {\it A priori} $E(C) = 1/r$.  

\paragraph{Prior for $\bZ$}
We use the finite version of the categorical Indian buffet process
(cIBP)~\citep{Sengupta2013cIBP} as the prior for the latent
categorical matrix $\bZ$. The cIBP is a categorical extension of the
Indian buffet process~\citep{griffiths2011indian} and defines feature
allocation~\citep{broderick2013feature} for categorical matrices.
In our application, the mutation pairs are the objects, and the
subclones are the latent features chosen by the objects. The number of
subclones $C$ is random, with the geometric prior $p(C)$. 
Conditional on $C$, we now introduce for
each column of $\bZ$ vector $\bm{\pi}_c = (\pi_{c1},
\pi_{c2}, \ldots, \pi_{cQ})$, where $p(\bz_{kc} = \bz^{(q)}) =
\pi_{cq}$, and $\sum_{q=1}^Q \pi_{cq} = 1$.  Recall that $\bz^{(q)}$ are
the possible genotypes for the mutation pairs defined in Section \ref{sec:repsubclone},
$q=1,\ldots,Q$, for $Q=10$ possible genotypes.

As prior model for $\bm \pi_c$, we use a Beta-Dirichlet distribution
\citep{kim2012bayesian}. Let $\tpi_{cq}= \pi_{cq}/(1-\pi_{c1})$, $q=2,\ldots,Q$.
Conditional on $C$, $\pi_{c1} \sim \Be(1,\alpha / C)$ follows a beta distribution, and
$(\tpi_{c2}, \ldots, \tpi_{cQ}) \sim \Dir(\gamma_2, \ldots,
\gamma_Q)$ follows a Dirichlet distribution. Here $\bz_{kc} = \bz^{(1)}$ corresponds to the
situation that subclone $c$ is not chosen by mutation pair $k$,
because $\bz^{(1)}$ refers to the reference genome. We write
\begin{align*}
\bm \pi_c \mid C \sim \text{Beta-Dirichlet}(\alpha/C, 1, \gamma_2, \ldots, \gamma_Q).
\end{align*}
This construction includes a positive probability for all-zero columns $\bz_c = \bm 0$. In our
application, $\bz_c = \bm 0$ refers to normal cells
with no somatic mutations, which could be included in the
cell subpopulations.

In the definition of the cIBP prior, we would have one more step of dropping all zero columns. This
leaves a categorical matrix $\bZ$ with at most $C$ columns. As shown in \cite{Sengupta2013cIBP}, the marginal limiting
distribution of $\bZ$ follows the cIBP as $C \rightarrow \infty$.

\paragraph{Prior for $\bw$}
We assume $\bm w_t$ follows a Dirichlet prior, 
\begin{align*}
\bm w_t \mid C \stackrel{iid}{\sim} \text{Dirichlet}(d_0, d, \cdots, d),
\end{align*}
for $t = 1, \cdots, T$.
We set $d_0 < d$ to reflect the nature of $c=0$ as a background
noise and model mis-specification term.

\paragraph{Prior for $\bm \rho$}
We complete the model with a prior for $\brho = \{\rho_g\}$.
Recall $\rho_g$ is the
conditional probability of observing a short read with  allele
$\bh_g$ due to experimental noise. We consider complete read, left
missing read and right missing read separately, and assume
\begin{align*}
\rho_{g_1} \sim \text{Dirichlet}(d_1, \ldots, d_1); \quad
\rho_{g_2} \sim  \text{Dirichlet}(2d_1, 2d_1); \quad
\rho_{g_3} \sim  \text{Dirichlet}(2d_1, 2d_1),
\end{align*}
where $g_1=\{1,2,3,4\}$, $g_2=\{5,6\}$ and $g_3 = \{7,8\}$.

\section{Posterior Inference}
\label{sec:posterior}
Let $\bx = (\bZ, \bm \pi, \bm w, \bm \rho )$ denote the unknown
parameters except $C$, where $\bZ = \{z_{kc}\}$, $\bm \pi =
\{\pi_{cq}\}$, $\bm w = \{w_{tc}\}$, and $\bm \rho = \{\rho_{g}\}$.
We use Markov chain Monte Carlo (MCMC) simulations
to generate samples from
the posterior $\bx^{(l)} \stackrel{iid}{\sim}  p(\bx \mid \bn,C)$,
$l = 1, \ldots, L$.  
With fixed $C$ such MCMC simulation is straightforward. 
See, for example, \cite{brooks2011} for a review of MCMC.
Gibbs sampling transition probabilities are used to
update $\bZ$ and $\bm \pi$, and Metropolis-Hastings transition
probabilities are used to update $\bm w$ and $\bm \rho$.
Since $ p(\bx \mid \bn, C)$ is 
expected to be highly multi-modal, we use additional
parallel tempering to improve mixing of the Markov chain.
Details of MCMC simulation and parallel
tempering are described in Appendix \ref{app:sec:mcmc}. 

\paragraph{Updating $C$}
Updating the value of $C$ is more difficult as it involves
trans-dimensional MCMC \citep{green1995}.
At each iteration, we propose a new value $\tilde{C}$ by
generating from a proposal distribution $q(\ti C \mid C)$.
In the later examples we assume that $C$ is {\em a priori} restricted to
$\Cmin \leq C \leq \Cmax$, and use
a uniform proposal
$q(\ti C \mid C) \sim \unif\{\Cmin, \ldots, \Cmax \}$.

Next, we split the data into a training set $\bn'$ and a test
set $\bn''$ with $n_{tkg}' = b n_{tkg}$ and $n_{tkg}'' = (1 - b)
n_{tkg}$, respectively,  for $b \in (0,1)$. 
Denote by $p_b(\bx \mid C) = p(\bx \mid
\bn', C)$ the posterior of $\bx$ conditional on $C$ evaluated on
the training set only. We use $p_b$ in two instances. First, we
replace the original prior $p(\bx \mid C)$ by $p_b(\bx \mid C)$,
and second, we use $p_b$ as a proposal distribution for $\tbx$, 
as $q( \tbx \mid \tC ) = p_b(\tbx \mid
\tC)$. Finally, we evaluate the acceptance probability of
$(\tC,\tbx)$ on the test data by
\begin{align}
  p_{\text{acc}}(C,\bx, \tC,\tbx) = 1 \wedge
  \frac{p(\bn'' \mid \tbx, \tC)}
       {p(\bn'' \mid \bx, C)} \cdot
  \frac{p(\tC) \cancel{p_b(\tbx \mid \tC)}}
       {p(C)   \cancel{p_b(\bx  \mid C  )}} \cdot
  \frac{q(C \mid \tC) \cancel{q(\bx \mid C)}}
  {q(\tC \mid C) \cancel{q(\tbx \mid \tC)}}.
  \label{eq:fbf}
\end{align}
The use of the prior $p_b(\tbx \mid \tC)$ is
similar to the construction of the fractional Bayes factor (FBF)
\citep{ohagan1995} which uses a fraction of the data to define an
informative prior that allows the evaluation of Bayes factors.
In contrast, here $p_b$ is used as an informative proposal
distribution for $\tbx$. Without the use of a training sample it would
be difficult to generate proposals $\tbx$ with reasonable acceptance
rate. In other words, we use $p_b$ to achieve a better mixing Markov
chain Monte Carlo simulation.
The use of the same $p_b$ to replace the original prior avoids the otherwise
prohibitive evaluation of $p_b$ in the acceptance probability
\eqref{eq:fbf}. See more details in Appendix
\ref{app:sec:updatec} and \ref{app:sec:calib}.

\paragraph{Point estimates for parameters}
We use the posterior mode $\Chat$ as a point estimate of
$C$. Conditional on $\Chat$, we follow \cite{lee2015bayesian} to find a
point estimate of $\bZ$. For any two $K \times \Chat$ matrices $\bZ$
and $\bZ'$, a distance between the $c$-th column of $\bZ$ and the
$c'$-th column of $\bZ'$ is defined by $\DD_{cc'}(\bZ, \bm
Z') = \sum_{k = 1}^{K} \| \bz_{kc} - \bz'_{kc'}\|_1$, where $1 \leq c,
c' \leq \Chat$, and we take the vectorized form of $\bz_{kc}$ and $
\bz'_{kc'}$ to compute $L^1$ distance between them.  Then, we define
the distance between $\bZ$ and $\bZ'$ as $d(\bZ, \bZ') =
\min_{\bm \sigma} \sum_{c = 1}^{\Chat} \DD_{c, \bm \sigma_c}(\bm
Z, \bZ')$, where $\bm \sigma = (\sigma_1, \ldots, \sigma_{\Chat})$ is
a permutation of $\{1, \ldots, \Chat\}$, and the minimum is taken over
all possible permutations.
This addresses the potential label-switching
issue across the columns of $\bZ$.  Let $\{\bZ^{(l)}, l = 1, \ldots,
L\}$ be a set of posterior Monte Carlo samples of $\bZ$. A posterior
point estimate for $\bZ$, denoted by $\hat{\bZ}$, is reported as
$\hat{\bZ} = \bZ^{(\hat{l})}$, where
\begin{align*}
   \hat{l} = \argmin_{l \in \{1, \ldots, L\}} \sum_{l' = 1}^L d(\bZ^{(l)}, \bZ^{(l')}).
\end{align*} 
Based on $\hat{l}$, we report posterior point estimates of
$\bm w$ and $\bm \rho$, given by $ \hat{\bm w} = \bm w^{(\hat{l})}$ and $
\hat{\bm \rho} = \bm \rho^{(\hat{l})}$, respectively.

\section{Simulation}
\label{sec:simulation}
We evaluate the proposed model with three simulation studies.
In the first simulation we use single sample data ($T = 1$),
since in 
most current applications only a single sample is available for
analysis. Inferring subclonal structure accurately under only one
sample is a major challenge, and not completely resolved in the 
current literature. The single sample does not rule out meaningful
inference, as the relevant sample size is the number of SNVs or
mutation pairs, or the (even larger) number of reads. In the second
and third simulations we consider multi-sample data,  
similar to the lung cancer data that we analyze later. 
In all simulations, we assume the missing probabilities
$v_{tk2}$ and $v_{tk3}$ to be 30\% or 35\%.  Recall that these
probabilities represent the probabilities that a short read will only
cover one of the two loci in the mutation pair.

Details of the three simulation studies are reported in Appendix \ref{app:sim}.
We briefly summarize the results here.
In the first simulation, we illustrate the advantage of using mutation
pair data over marginal SNV counts.
We generate hypothetical short reads data for $T = 1$ sample and $K = 40$
mutation pairs, using a simulation truth with $C\true=2$. Figure
\ref{fig:sim}(a, d) summarizes the simulation results. 
See Appendix Figures \ref{app:figsim1} and \ref{app:figsim1_pyclone}
for more summaries, including a comparison
with results under methods based on marginal read counts only.

In the second simulation, we  consider data
with $K = 100$ mutation pairs  and a more complicated subclonal
structure with $C\true = 4$ latent subclones and $T=4$ samples.
Inference summaries are show in Figure \ref{fig:sim}(b, e).
Again, more details of the simulation study, including inference on
the weights $\bw$ and a comparison with inference  using marginal cell
counts only, are shown in the appendix.

Finally, in a third simulation we use 
$T = 6$ samples with $C^{\text{TRUE}} = 3$ and latent subclones. 
Some results are summarized in Figure \ref{fig:sim}(c, f), and, again, more
details are shown in the appendix.

In all simulations, panels (a, b, c) vs. (d, e, f) in Figure \ref{fig:sim} show that posterior estimated $\Zhat$ is close to the true $\bZ\true$.

\begin{figure}[h!]
\begin{center}
\begin{subfigure}[t]{.3\textwidth}
\centering
\includegraphics[width=\textwidth]{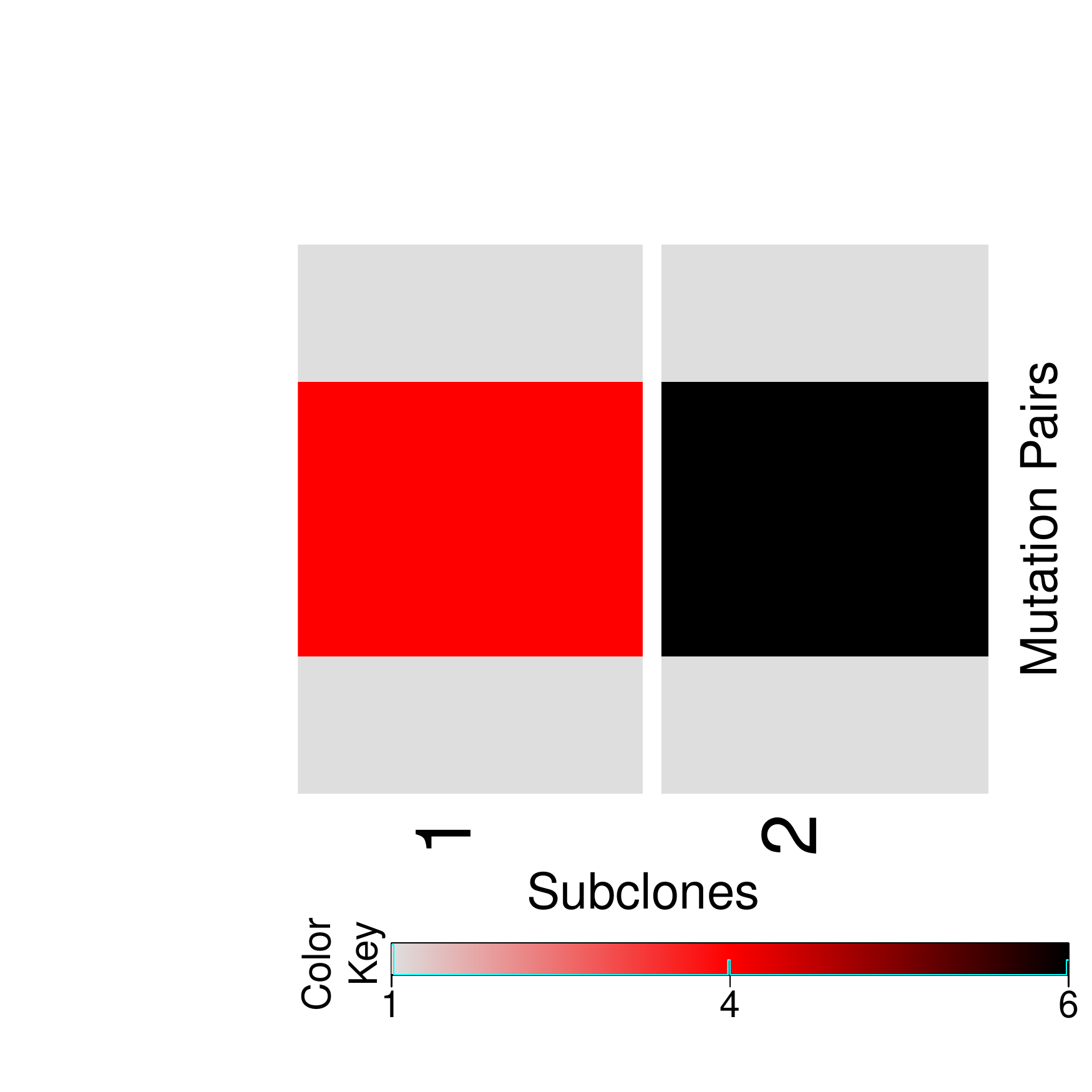}
\caption{Simu. 1, $\bZ\true$}
\end{subfigure}
\begin{subfigure}[t]{.3\textwidth}
\centering
\includegraphics[width=\textwidth]{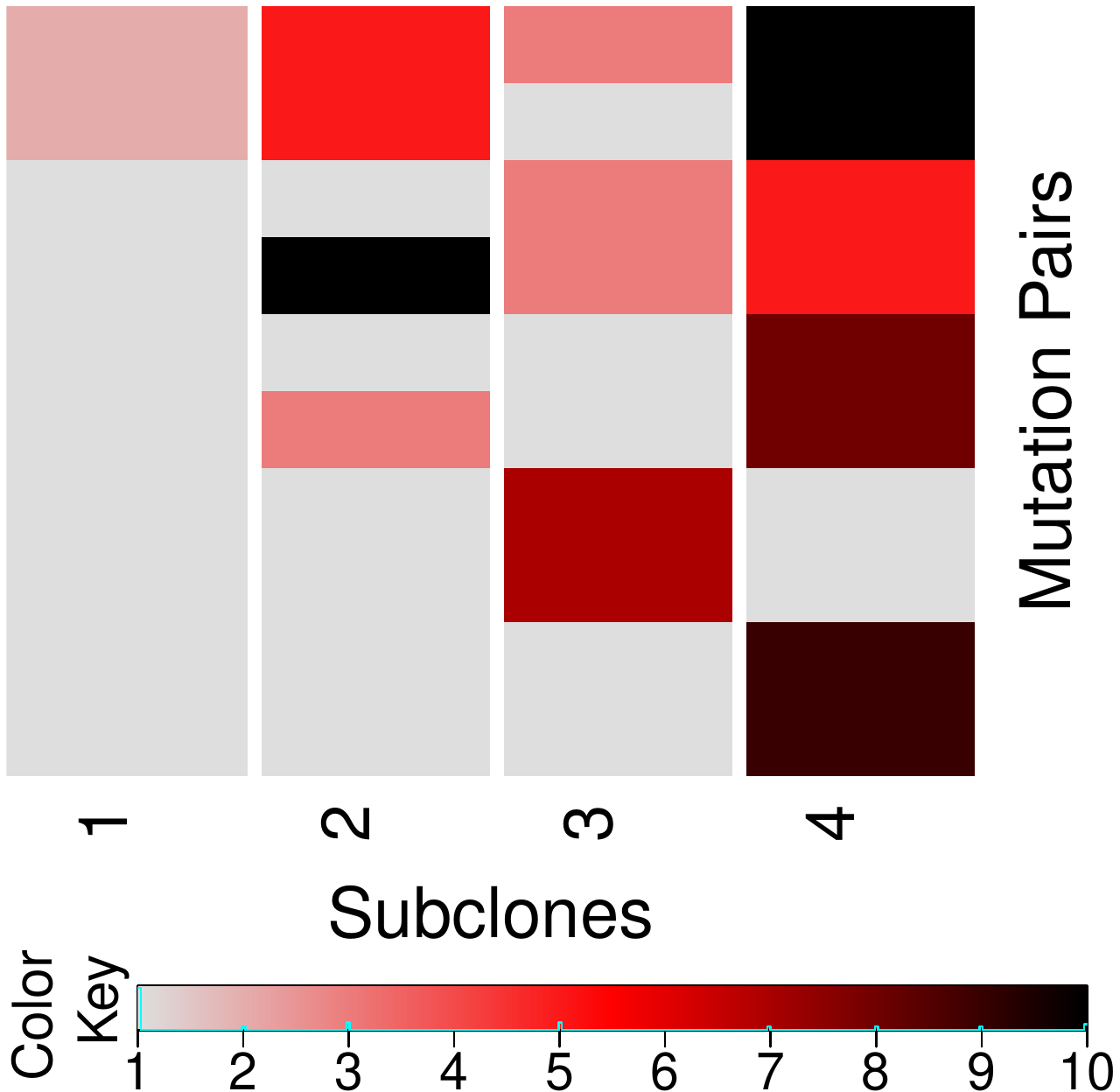}
\caption{Simu. 2, $\bZ\true$}		
\end{subfigure}
\begin{subfigure}[t]{.3\textwidth}
\centering
\includegraphics[width=\textwidth]{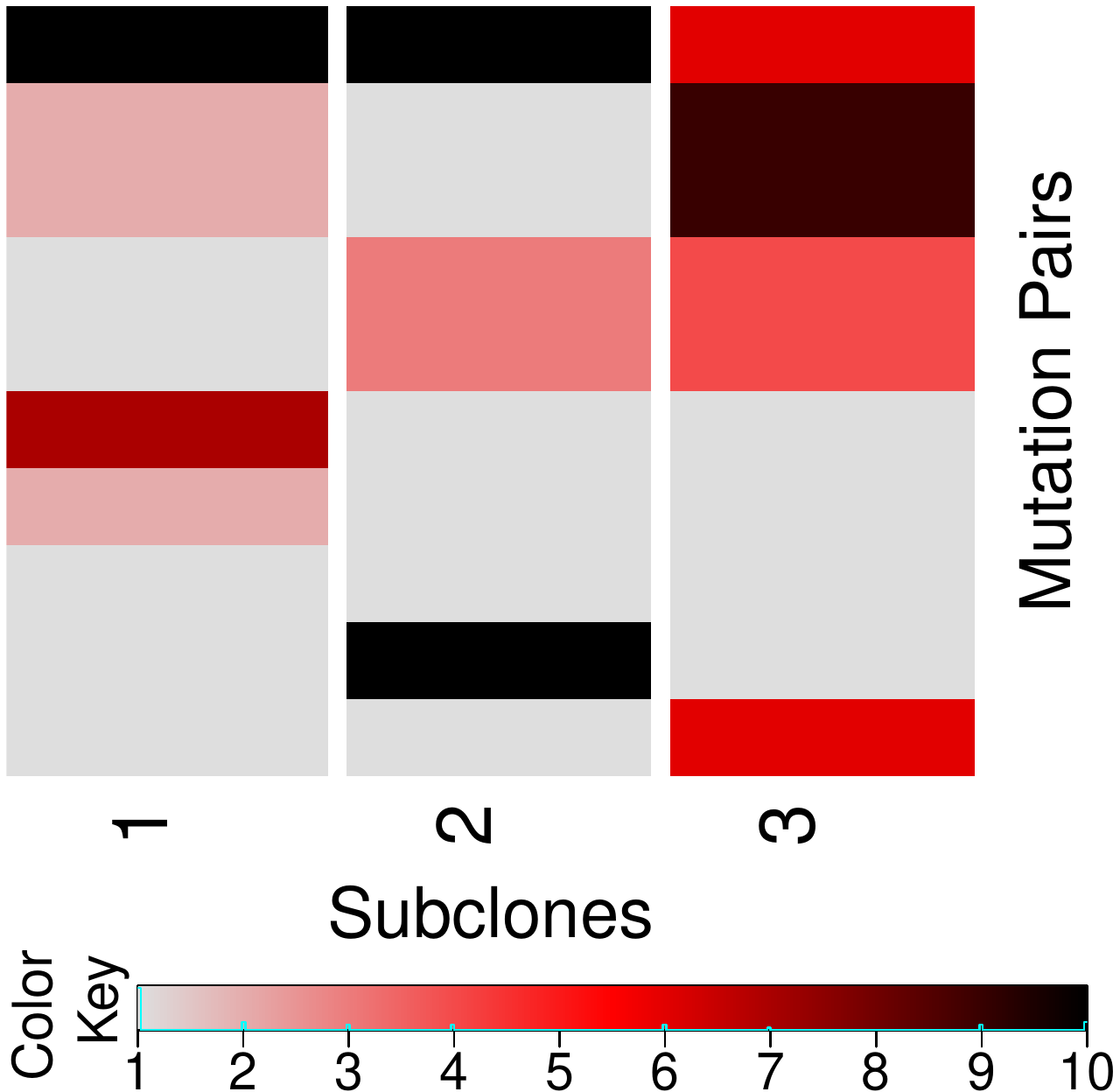}
\caption{Simu. 3, $\bZ\true$}		\vspace{5mm}
\end{subfigure}
\begin{subfigure}[t]{.3\textwidth}
\centering
\includegraphics[width=\textwidth]{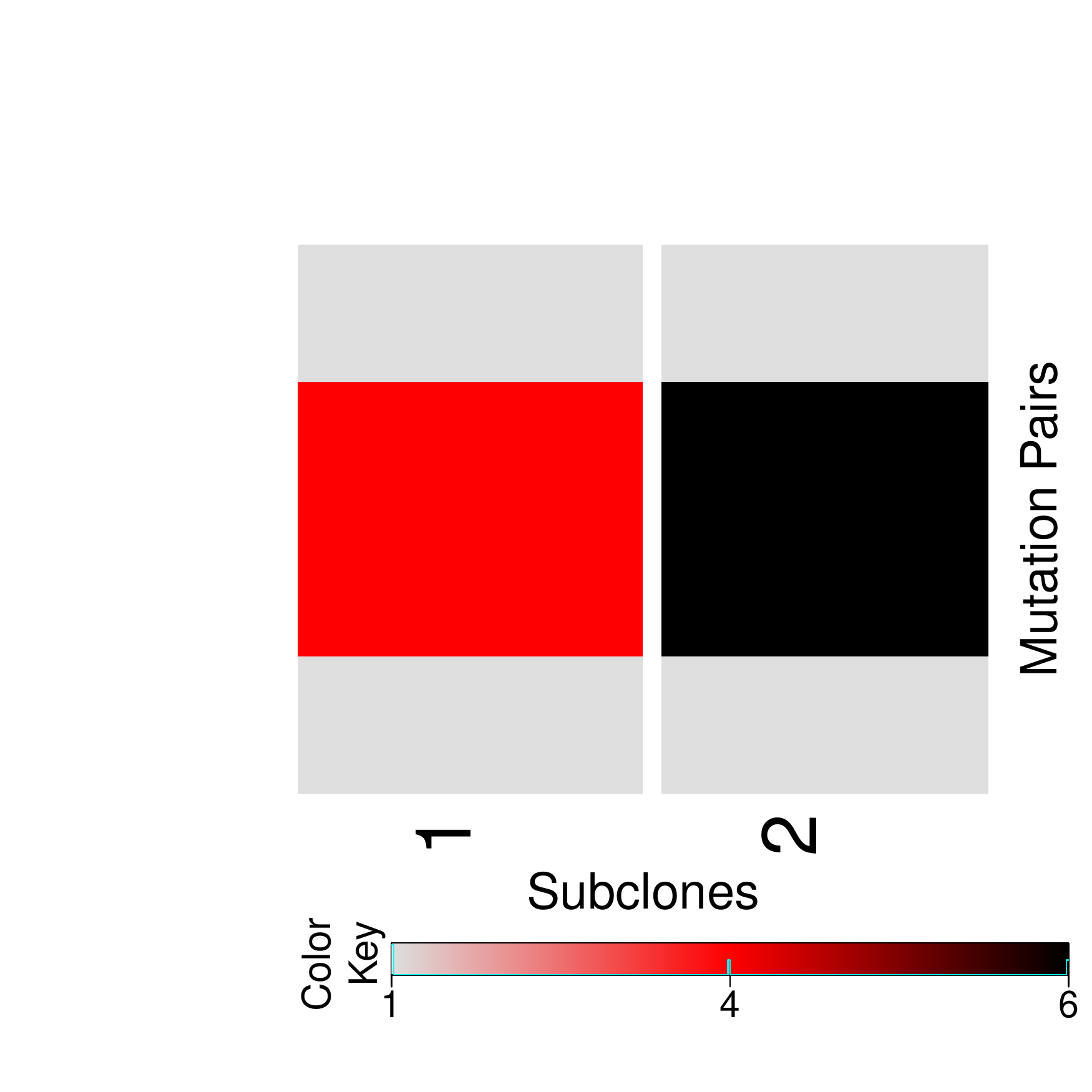}
\caption{Simu. 1, $\Zhat$}
\end{subfigure}
\begin{subfigure}[t]{.3\textwidth}
\centering
\includegraphics[width=\textwidth]{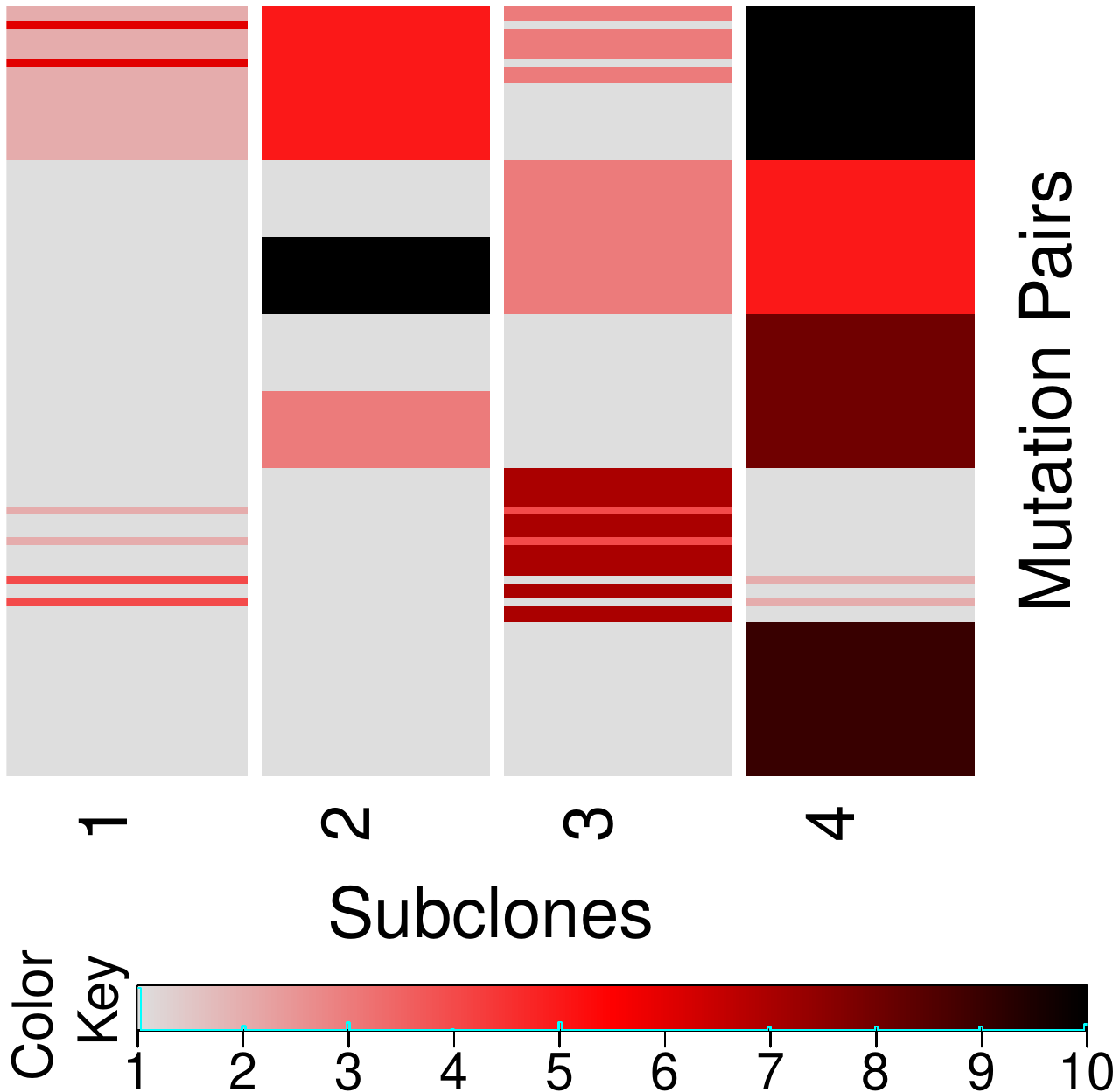}
\caption{Simu. 2, $\Zhat$}		
\end{subfigure}
\begin{subfigure}[t]{.3\textwidth}
\centering
\includegraphics[width=\textwidth]{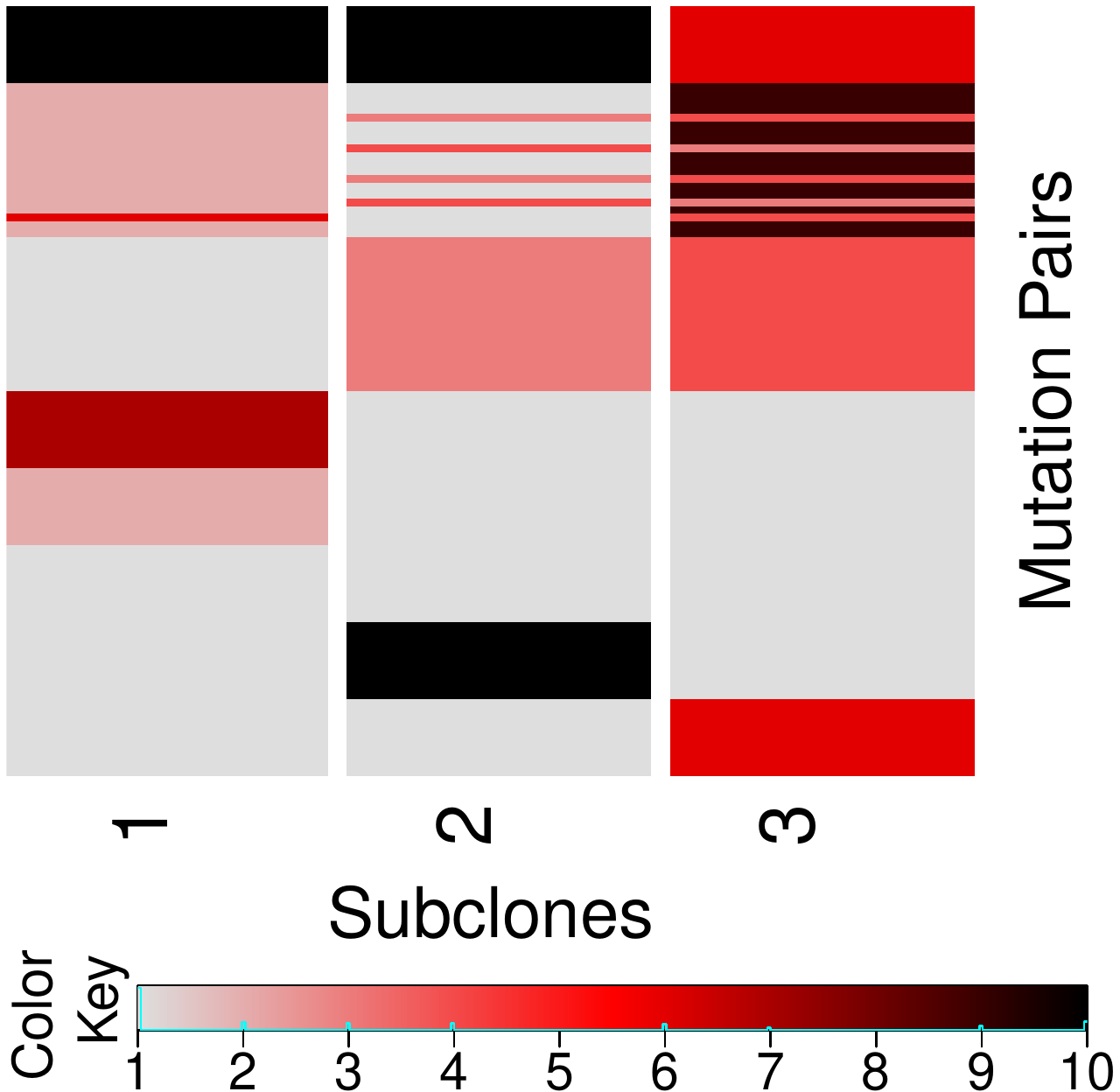}
\caption{Simu. 3, $\Zhat$}		
\end{subfigure}
\end{center}
\caption{Summary of simulation results. 
Simulation truth $\bZ\true$ (a, b, c), and posterior
inference under PairClone (d, e, f) conditional on posterior modes of $C$.}
\label{fig:sim}
\end{figure}

\section{PairClone Extensions}
\label{sec:pipeline}

\subsection{Incorporating Marginal Read Counts}
\label{sec:mrc}
Most somatic mutations are not part of the paired reads that we
use in PairClone. 
We refer to these single mutations as SNVs (single
nucleotide variants) and consider the following simple extension to
incorporate marginal counts for SNVs in PairClone.
We introduce a new $S \times C$ matrix $\bZ^{S}$ to represent the
genotype of the $C$ subclones for these additional SNVs.
To avoid confusion, we denote the earlier $K \times C$ subclone matrix
by $\bZ^{P}$ in this section.
The $(s,c)$ element of $\bZ^S$ reports the  genotype of SNV $s$ in
subclone $c$, with $z^{S}_{sc} \in \{ 0, 0.5, 1\}$ denoting
homozygous wild-type ($0$), heterozygous variant
($0.5$), and homozygous variant ($1$), respectively.
The $c$-th column of $\bZ^{P}$ and $\bZ^{S}$ together define 
subclone $c$. We continue to assume copy number 
neutrality in all SNVs and mutation pairs
(we discuss an extension to incorporating subclonal copy number
variations in the next subsection).
The marginal read counts are easiest
incorporated in the PairClone model by recording them as right (or left)
missing reads (as described in Section \ref{sec:splmodel})
for hypothetical pairs, $k=K+1,\ldots,k+S$. 
Let $\tN_{ts}$ and $\tn_{ts}$ denote
the total count and the number of reads bearing a variant
allele, respectively, for SNV $s$ in sample $t$.
Treating $s$ as a mutation pair $k=K+s$ with missing second read, we
record $n_{tk8}=\tn_{ts}$, $n_{tk1}=\ldots=n_{tk7}=0$ and
$N_{tk}=\tN_{ts}$. 
We then proceed as before, now with $K+S$ mutation pairs. 
Inference reports an augmented $(K+S) \times C$ subclone matrix
$\tilde{\bZ}^P$. We record the first $K$ rows of
$\tilde{\bZ}^P$ as $\bZ^P$, and transform the remaining $S$ rows
to $\bZ^S$ by only recording the genotypes of the observed loci.

\begin{figure}[h!]
\begin{center}
\begin{subfigure}[t]{.6\textwidth}
\centering
Simulation truth
\end{subfigure}
\begin{subfigure}[t]{.3\textwidth}
\centering
~~\vspace{3mm}
\end{subfigure} \\
\begin{subfigure}[t]{.3\textwidth}
\centering
\includegraphics[width=\textwidth]{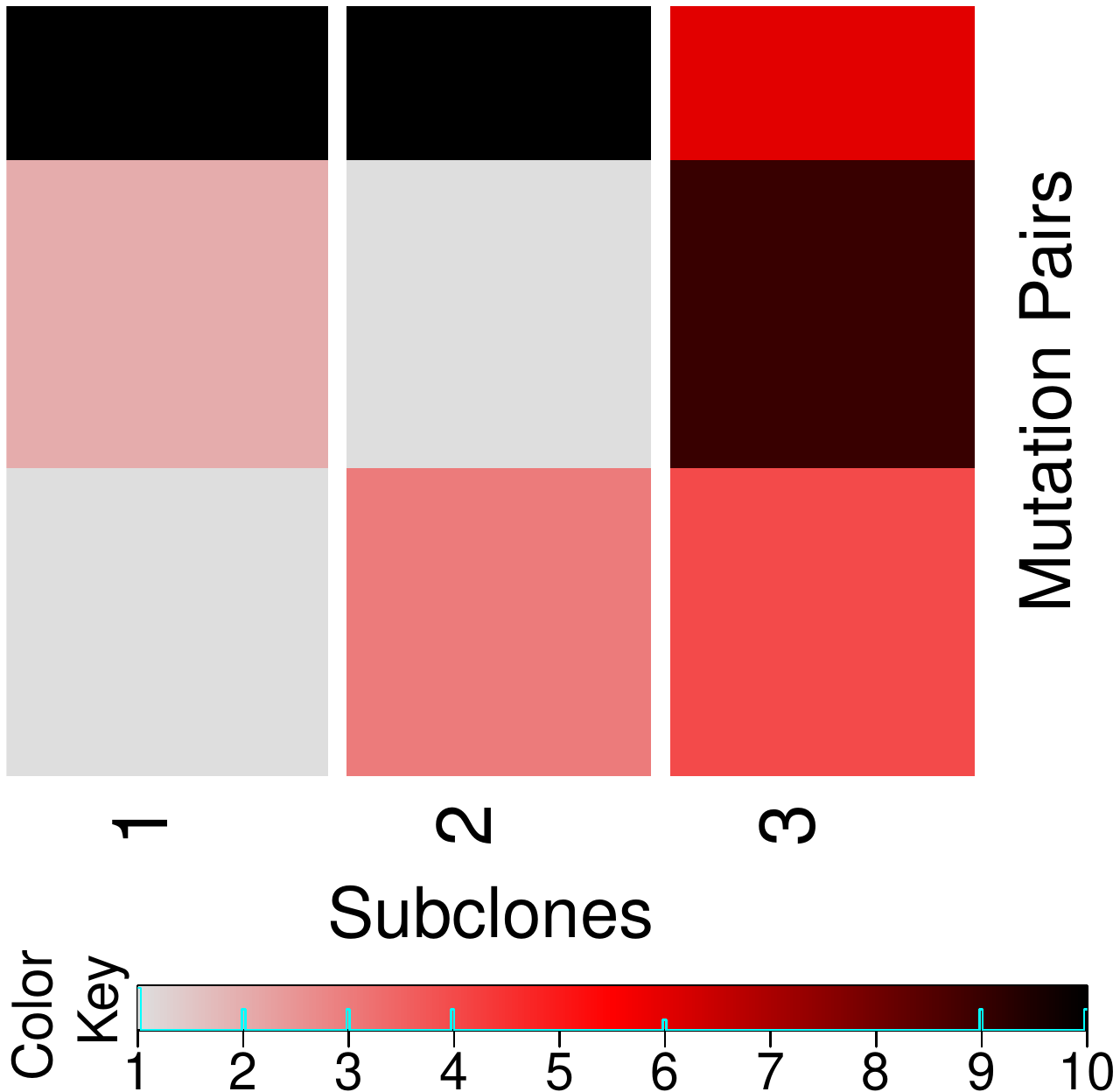}
\caption{$\bZ^{P, \text{TRUE}}$}		
\end{subfigure}
\begin{subfigure}[t]{.3\textwidth}
\centering
\includegraphics[width=\textwidth]{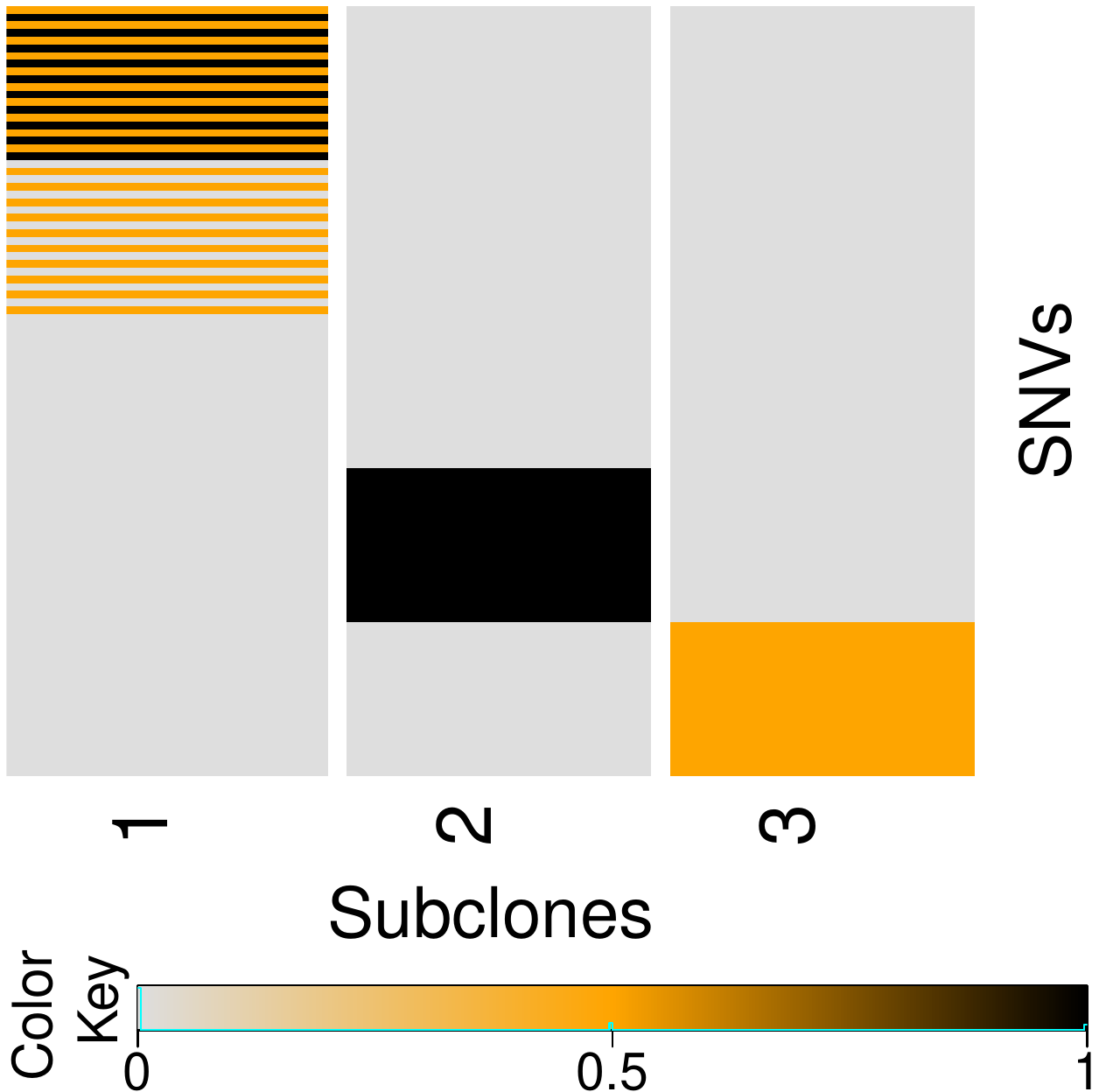}
\caption{$\bZ^{S, \text{TRUE}}$}		
\end{subfigure}
\begin{subfigure}[t]{.3\textwidth}
\centering
\includegraphics[width=\textwidth]{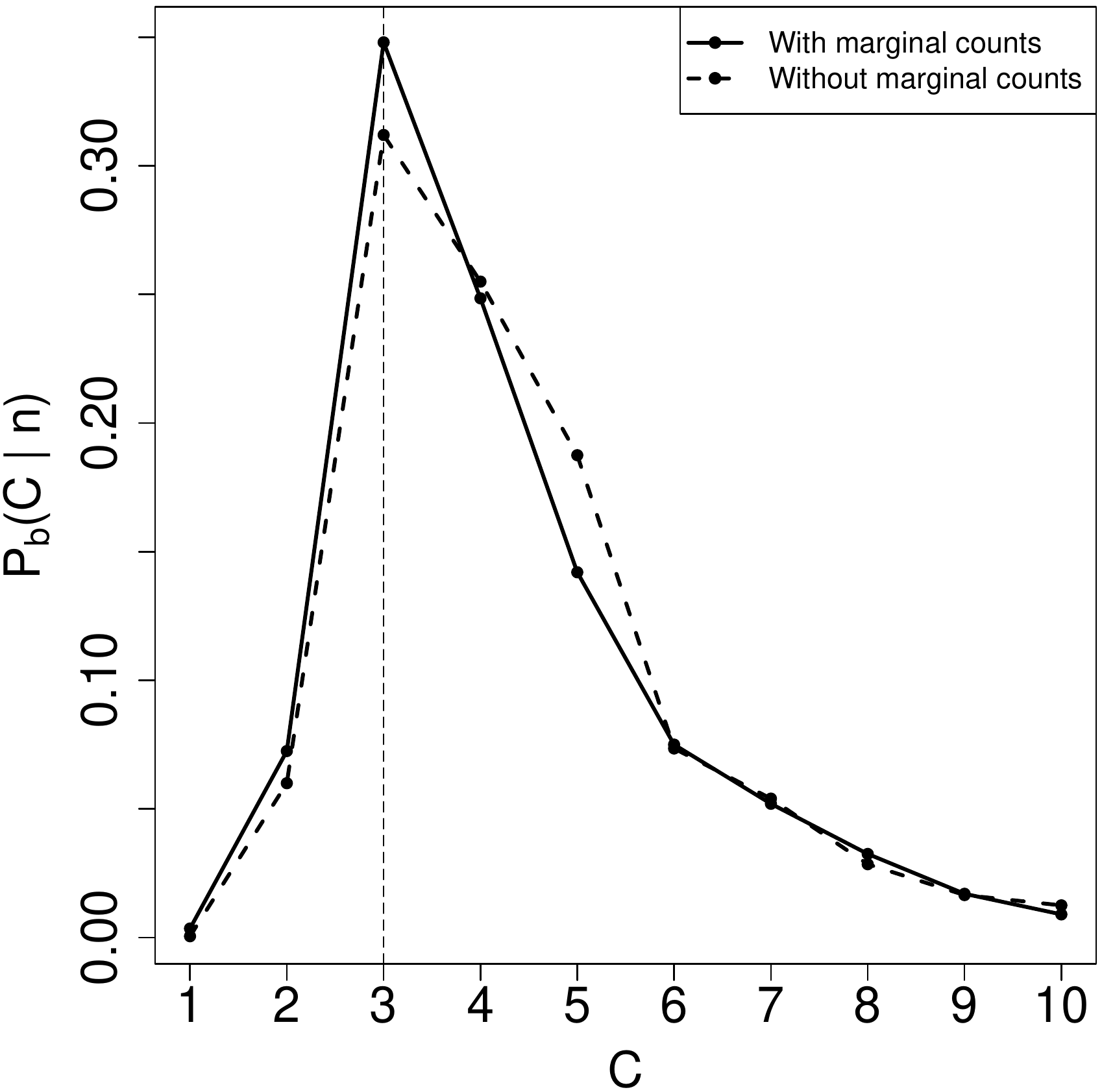}
\caption{$p_b(C \mid \bm n'')$}	
\vspace{8mm}
\end{subfigure}
\begin{subfigure}[t]{.6\textwidth}
\centering
Pairs and SNVs
\end{subfigure}
\begin{subfigure}[t]{.3\textwidth}
\centering
Pairs only \vspace{3mm}
\end{subfigure}  \\
\begin{subfigure}[t]{.3\textwidth}
\centering
\includegraphics[width=\textwidth]{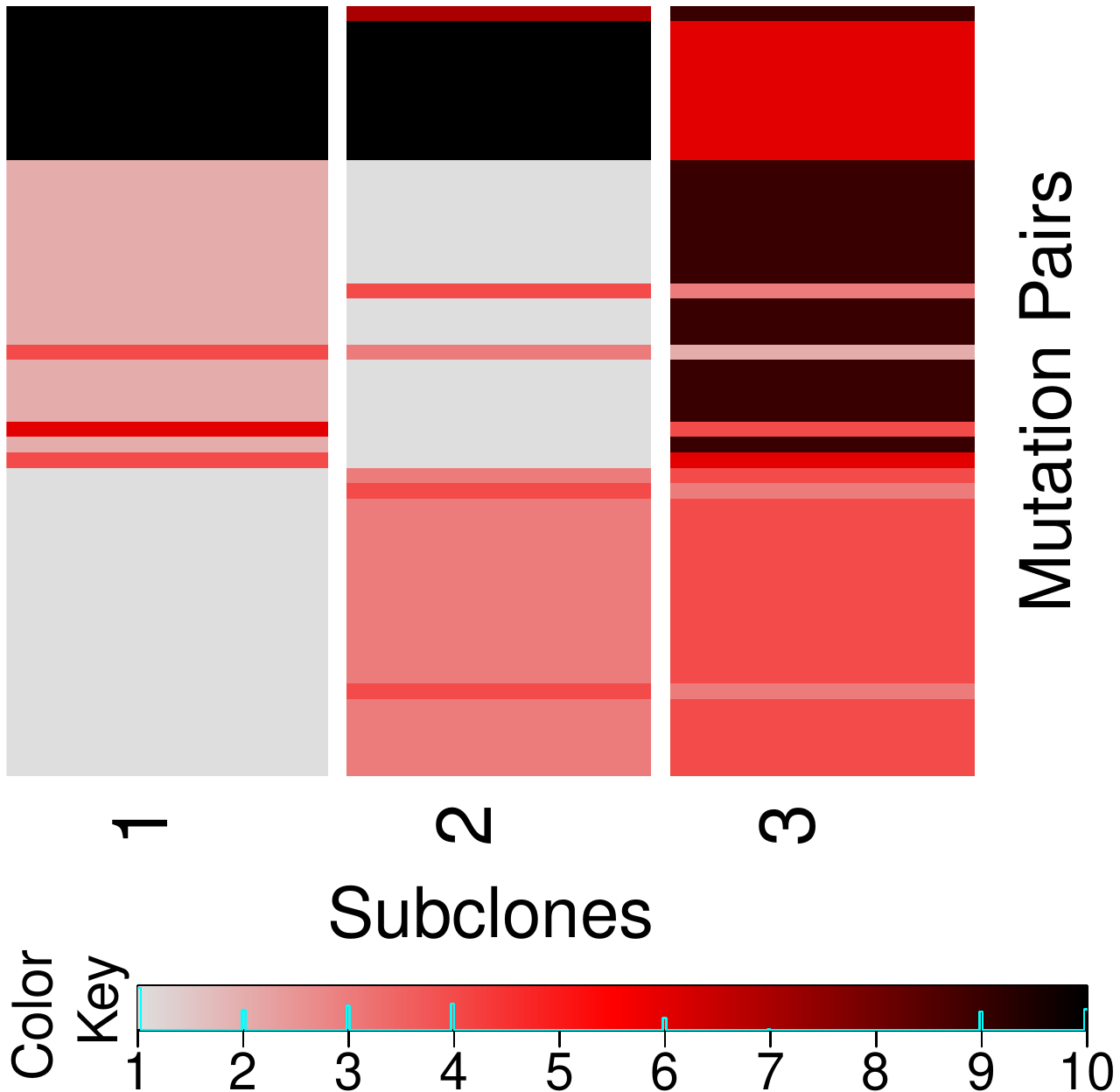}
\caption{$\Zhat^{P}$}		
\end{subfigure}
\begin{subfigure}[t]{.3\textwidth}
\centering
\includegraphics[width=\textwidth]{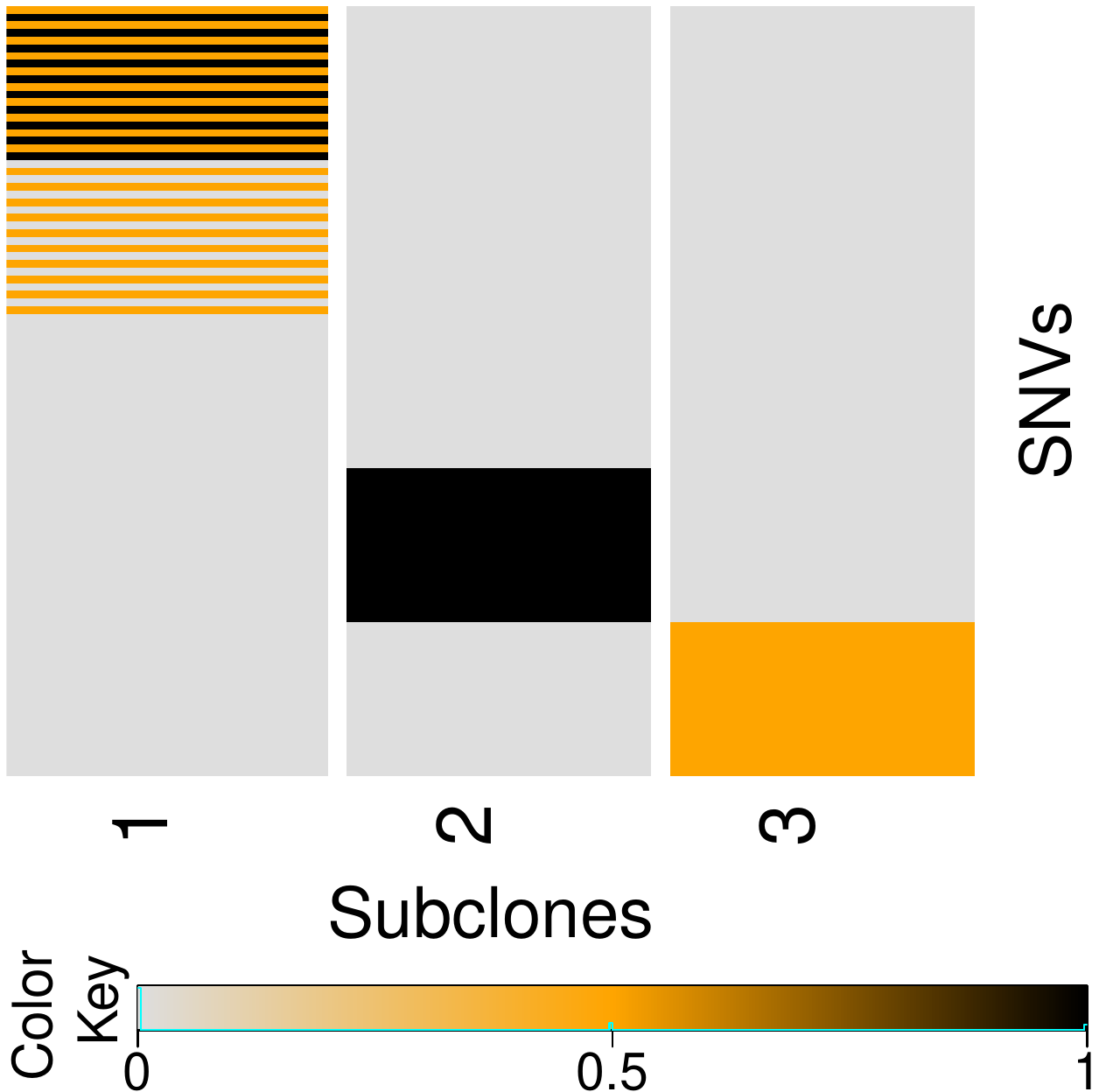}
\caption{$\Zhat^{S}$}		
\end{subfigure}
\begin{subfigure}[t]{.3\textwidth}
\centering
\includegraphics[width=\textwidth]{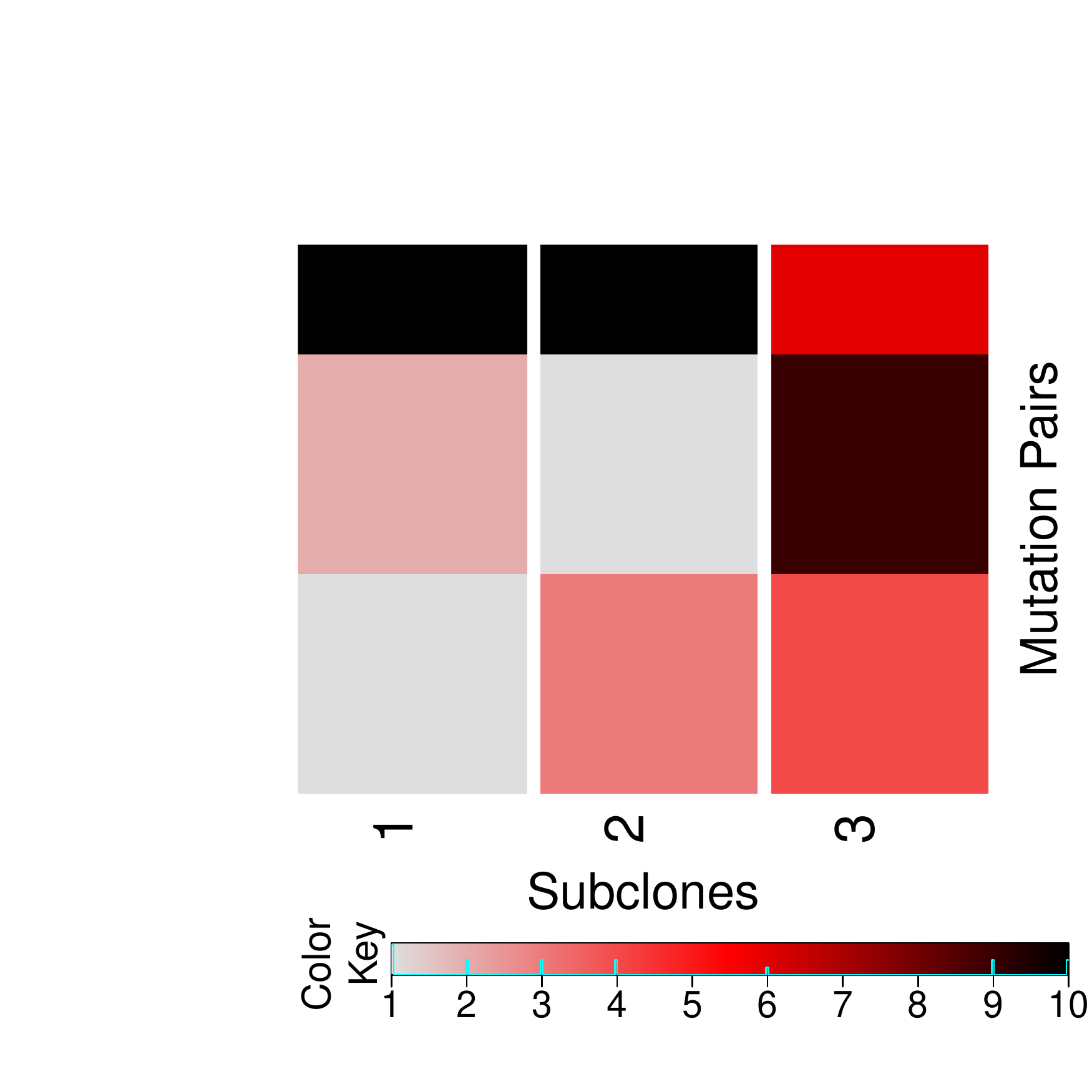}
\caption{$\Zhat_{\text{wom}}^{P}$}		
\end{subfigure}
\end{center}
\caption{Summary of simulation results using additional marginal read counts. Simulation truth $\bZ^{P, \text{TRUE}}$ and $\bZ^{S, \text{TRUE}}$ (a, b), posterior inference with marginal read counts incorporated (c, d, e), and posterior inference without marginal read counts (c, f).}
\label{fig:sim_extension}
\end{figure}

We evaluate the proposed modeling approach
with a simulation study. The simulation setting is the same as
simulation 3 in Section \ref{sec:simulation},   except that we discard
the phasing information of mutation pairs $51-100$ and only record
their marginal read counts. 
Figure \ref{fig:sim_extension}(a)--(f) summarizes the simulation results.
Panels (a, b) show the simulation truth for the mutation pairs and SNVs, respectively. 
Panel (c) shows the posterior $p(C \mid \bn'')$ and panels (d, e) show the estimated 
genotypes $\Zhat^P$ and $\Zhat^S$. Inference for 
the weights $w_{tc}$ recovers the
simulation truth (not shown). 
The result compares favorably to inference under BayClone (Figure
\ref{fig:sim3_BC} in the appendix), due to the additional phasing information for the first 50 mutation pairs.

For a direct evaluation of the information in the additional marginal
counts we also evaluate posterior inference with only the first 50
mutation pairs, shown in Figure \ref{fig:sim_extension} (c, f). 
Comparison with Figure \ref{fig:sim_extension} (c, d) shows that the additional
marginal counts do not noticeably improve inference on tumor
heterogeneity.

\subsection{Incorporating Tumor Purity}
\label{sec:purity}
Usually, tumor samples are not pure in the sense that they contain
certain proportions of normal cells. Tumor purity refers to the
fraction of tumor cells in a tumor sample. To explicitly model tumor
purity, we introduce a normal subclone, the proportion of which in
sample $t$ is denoted by $\wts$, $t = 1, \ldots, T$. The normal
subclone does not possess any mutation  (since we only
consider somatic mutations). 
The tumor purity for sample $t$ is thus $(1 - \wts)$. The normal
subclone is denoted by $\bz_*$, with $\bz_{k*} = \bz^{(1)}$ for all
$k$. 
The remaining subclones are still denoted by
$\bz_c$, $c = 1, \ldots, C$, with proportion $w_{tc}$ 
in sample $t$, and $\sum_{c = 0}^C w_{tc} + \wts = 1$.
 
The probability model needs to be slightly modified to accommodate the
normal subclone. The sampling model remains unchanged as
\eqref{eq:multi}. Same for the prior models for $\bZ$, $\brho$ and $C$.
We only change the construction of $\tp_{tkg}$ and 
$p(\bw)$ as follows. 
With a new normal subclone, the probability of observing a short read
$\bh_g$ becomes 
$ 
   \tp_{tkg} = \sum_{c = 1}^C w_{tc}\,A(\bh_g, \bz_{kc}) +  \wts \,
   A(\bh_g, \bz^{(1)}) +  w_{t0} \, \rho_g,
$
based on the same generative model described in Section \ref{sec:prior}.
Let $\tilde{w}_{tc} = w_{tc} / (1 - \wts)$.
We use a Beta-Dirichlet prior,
$\wts \stackrel{iid}{\sim} \Be(d_1^{*}, d_2^{*})$,
and $\tilde{\bw}_{t} \stackrel{iid}{\sim} \Dir(d_0$, $d$, $\ldots$, $d)$.
An informative prior for $\wts$ could be based on
an estimate from a purity caller, for
example, \cite{van2010allele} or \cite{carter2012absolute}.

We evaluate the modified model with a simulation study. The simulation setting is the same as simulation 3 in Section \ref{sec:simulation},  except that we substitute the first subclone with a normal subclone. 
Posterior inference (not shown) recovers the simulation truth, with posterior mode $\Chat = 2$. Inference on $\bZ$ almost perfectly recovers the simulation truth shown in Figure \ref{fig:sim}(c) (with the first column replaced by an all normal ``subclone''). Similarly for $\bw$. See Appendix \ref{app:sim_purity} for details.


\subsection{Incorporating Copy Number Changes}
\label{sec:cnv}
Tumor cells not only harbor sequence mutations such as SNVs and
mutation pairs, they often undergo copy number changes and produce
copy number variants (CNVs). Genomic regions with CNVs 
have copy number $\ne 2$. 
We briefly outline an extension of PairClone that includes
CNVs in the inference. 
In addition to $\bZ$ which describes sequence variation we 
introduce a $K \times C$ matrix $\bL$ to represent subclonal copy
number variation with $\ell_{kc}$ reporting the copy number for
mutation pair $k$ in subclone $c$. 
We use $\bL$ to augment the sampling model to include the total
read count $N_{tk}$.
Earlier in \eqref{eq:multi}, the multinomial sample size $N_{tk}$ was
considered fixed. We now add a sampling model. Following
\cite{lee2016bayesianjrssc} we assume
$$
  N_{tk} \mid \phi_t, M_{tk} \sim \text{Poisson}(\phi_t M_{tk} / 2)
$$
Here, $\phi_t$ is the expected number of reads in sample $t$ under
copy-neutral conditions, and
$M_{tk}$ is a weighted average copy number across subclones,
$$
   M_{tk} = \sum_{c=1}^C w_{tc} \ell_{kc} + w_{t0} \ell_{k0}.
$$
The last term $w_{t0} \ell_{k0}$ accounts for noise and
artifacts, where $w_{t0}$ and $\ell_{k0}$ are the population frequency
and copy number of the background subclone, respectively. We assume no
CNVs for the background subclone, that is, $\ell_{k0} = 2$ for all $k$. 
We complete the model with a prior $p(\bL)$. 
Assuming $\ell_{kc} \in \{0,\ldots,Q\}$, i.e., a maximum copy number $Q$, 
we use another instance of a finite cIBP. For each
column of $\bL$, we introduce $\bm \pi_c = (\pi_{c0}, \pi_{c1},
\ldots, \pi_{cQ})$ and assume $p(\ell_{kc} = q) = \pi_{cq}$, again
with a Beta-Dirichlet prior for $\bm \pi_c$. 

Recall the construction of $\tp_{tkg}$ in \eqref{pprior1},
including in particular the generative model.
This generative model is now updated to include the varying $\ell_{tc}$.
To generate a short read for mutation pair $k$, we
first select a subclone $c$ from which the read arises, using the
population frequencies $w_{tc} \ell_{kc} / \sum_{c = 0}^C w_{tc} \ell_{kc}$
for sample $t$. Next we select with probability $z_{kcj} / \ell_{kc}$ one
of the four possible alleles, $\bh_g$, $g=1,2,3$ or $4$, 
where we now use $\bz_{kc} = (z_{kcj}, j = 1, \ldots, 4)$ to denote
numbers of alleles having genotypes $00$, $01$, $10$ or $11$, and
$\sum_j z_{kcj} = \ell_{kc}$. 
In the case of left (or right) missing
locus we observe $\bh_g$, $g = 5$ or $6$ (or $g = 7$ or $8$),
corresponding to the observed locus of the chosen allele, similar to before.
In summary, the probability of observing a short read $\bh_g$ can be
written as 
\begin{align*}
\tp_{tkg} =  \sum_{c = 0}^C  \left[ \frac{w_{tc} \ell_{kc}}{\sum_{c =
  1}^C w_{tc} \ell_{kc} + w_{t0} \ell_{k0}} \cdot \frac{A(\bh_g,
  \bz_{kc})}{\ell_{kc}} \right] = \frac{\sum_{c=0}^C w_{tc} A(\bh_g,
  \bz_{kc})}{M_{tk}}, 
\end{align*}
where $A( \cdot )$ corresponds to the described generative model.

\section{Lung Cancer Data}
\label{sec:realdata}
\subsection{Using PairClone}
We apply PairClone to analyze whole-exome in-house data.
Whole-exome sequencing data is generated from four
($T = 4$) surgically dissected tumor samples taken from a single patient
diagnosed with lung adenocarcinoma. The resected tumor is divided into
two portions. One portion is flash frozen and another portion is
formalin fixed and paraffin embedded (FFPE). Four different samples
(two from each portion) are taken. DNA is extracted from all 
four samples. Agilent SureSelect v5+UTR probe kit (targeting coding
regions plus UTRs) is used for exome capture. The exome library is
sequenced in paired-end fashion on an Illumina HiSeq 2000
platform. About 60 million reads are obtained in FASTQ file format,
each of which is 100 bases long. We map paired-end reads to the human
genome (version HG19)~\citep{church2011modernizing} using
BWA~\citep{li2009fast} to generate BAM files for each individual
sample. After mapping the mean coverage of the samples is around $70$
fold. We call variants using UnifiedGenotyper from GATK
toolchain~\citep{mckenna2010genome} and generate a single VCF file for
all of them. A total of nearly $115,000$ SNVs and small indels are
called within the exome coordinates.

Next, using \texttt{LocHap}~\citep{sengupta2015NAR} we find mutation
pair positions, the number of alleles
and number of reads mapped to them.
\texttt{LocHap} searches for multiple SNVs that are scaffolded by
the same pair-end reads,  that is,  they  can be recorded on one
paired end read. We refer to such sets of multiple SNV's as
local haplotypes (LH). When more than two genotypes are exhibited by
an LH, it is called a LH variant (LHV). Using individual BAM files
and the combined VCF file, \texttt{LocHap} generates four individual output
file in HCF format~\citep{sengupta2015NAR}. An HCF file contains LHV
segments with two or three SNV positions. In this analysis, we are
only interested in mutation pair, and therefore filter out all the LHV
segments consisting of more than two SNV locations.
We restrict our analysis to copy number neutral regions. 
To further improve data quality, we drop all LHVs where two
SNVs are very close to each other (within, say, $50$ bps) or close to
any type of structural variants such as indels. We also remove
those LHVs where either of the SNVs is mapped with strand bias by most
reads, or either of the SNVs is mapped towards the
end of the most aligned reads. 
Finally, we only consider mutation pairs that have strong
evidence of heterogeneity.
Since LHVs exhibit $>2$ genotypes in the short reads, by
definition they are somatic mutations.

  At the end of this process, $69$ mutation
pairs are left and we record the read data from HCF files for the
analysis.  In addition, in the hope of utilizing more information
from the data, we randomly choose 69 un-paired SNVs and include them
in the analysis. 
Since in practice, tumor samples often include contamination with 
normal cells, we incorporate inference for tumor purity as described
in Section \ref{sec:purity}. 
 We run MCMC simulation for $30,000$ iterations, discarding the first 
$10,000$ iterations as initial burn-in and 
keeping
every 10th MCMC sample. 
We set the hyperparameter exactly  as in the simulation study 
of Section \ref{sec:purity}. 

\begin{figure}[h!]
\begin{center}
\begin{subfigure}[t]{.3\textwidth}
\centering
\includegraphics[width=\textwidth]{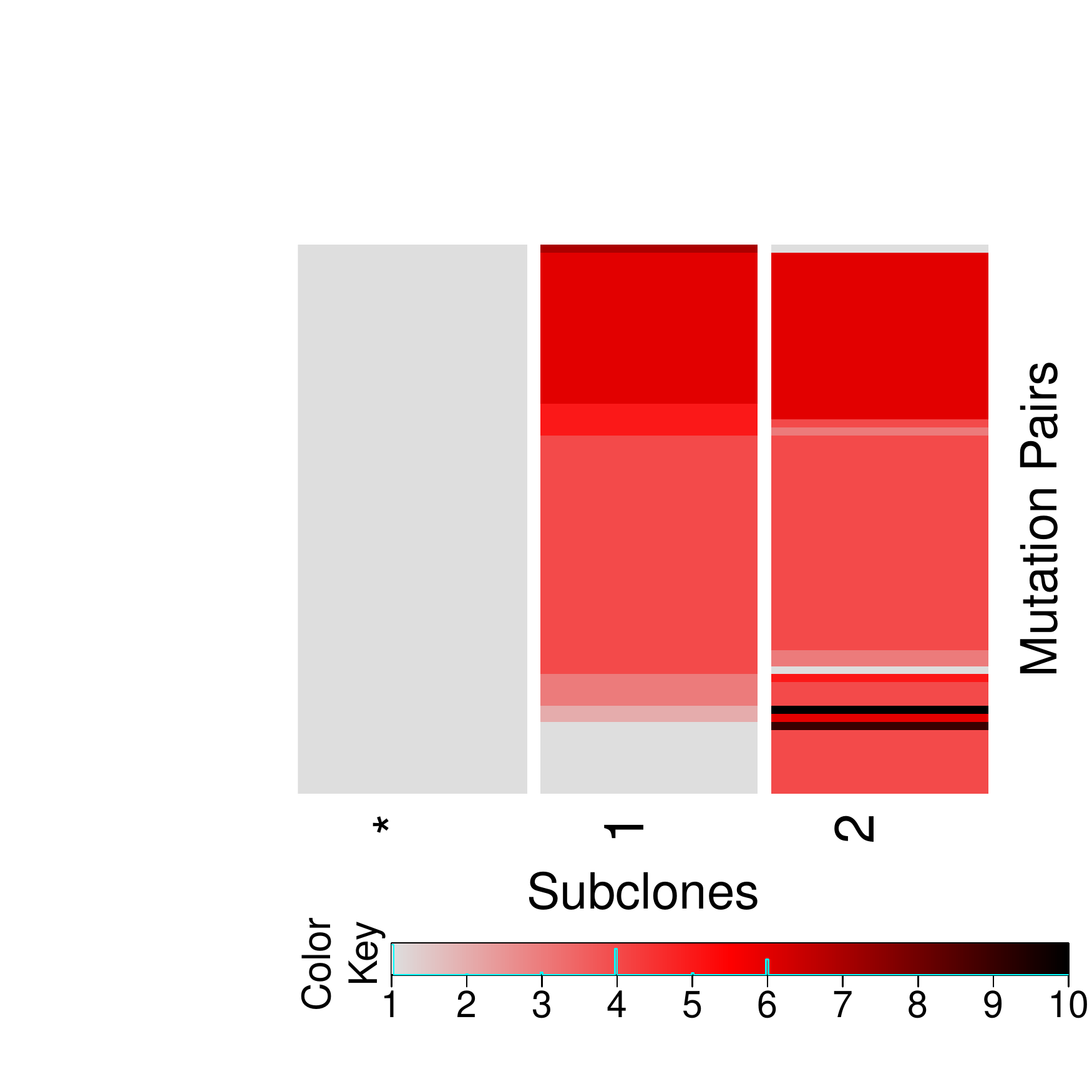}
\caption{$\Zhat^P$}		
\end{subfigure}
\begin{subfigure}[t]{.3\textwidth}
\centering
\includegraphics[width=\textwidth]{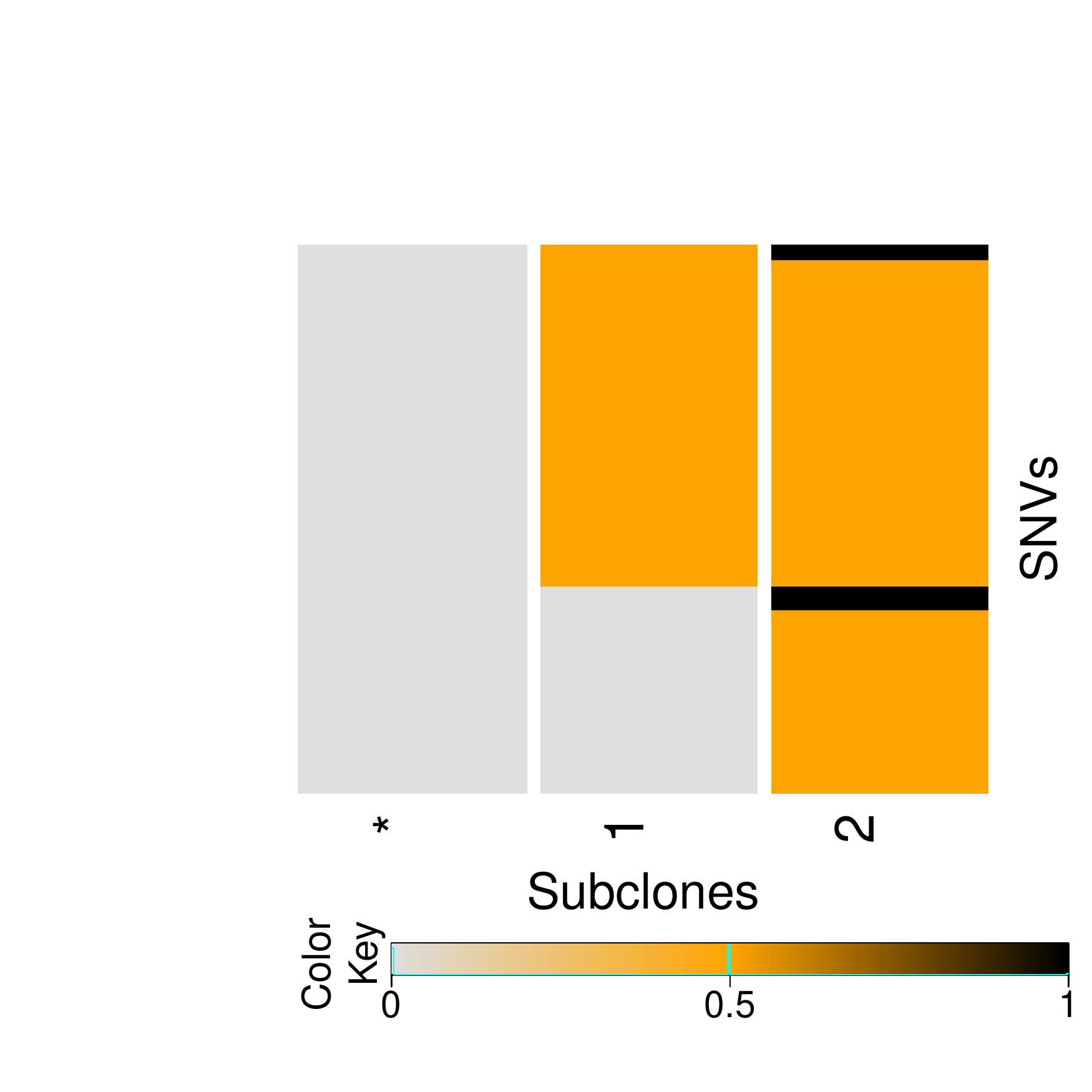}
\caption{$\Zhat^S$}		
\end{subfigure}
\begin{subfigure}[t]{.3\textwidth}
\centering
\includegraphics[width=\textwidth]{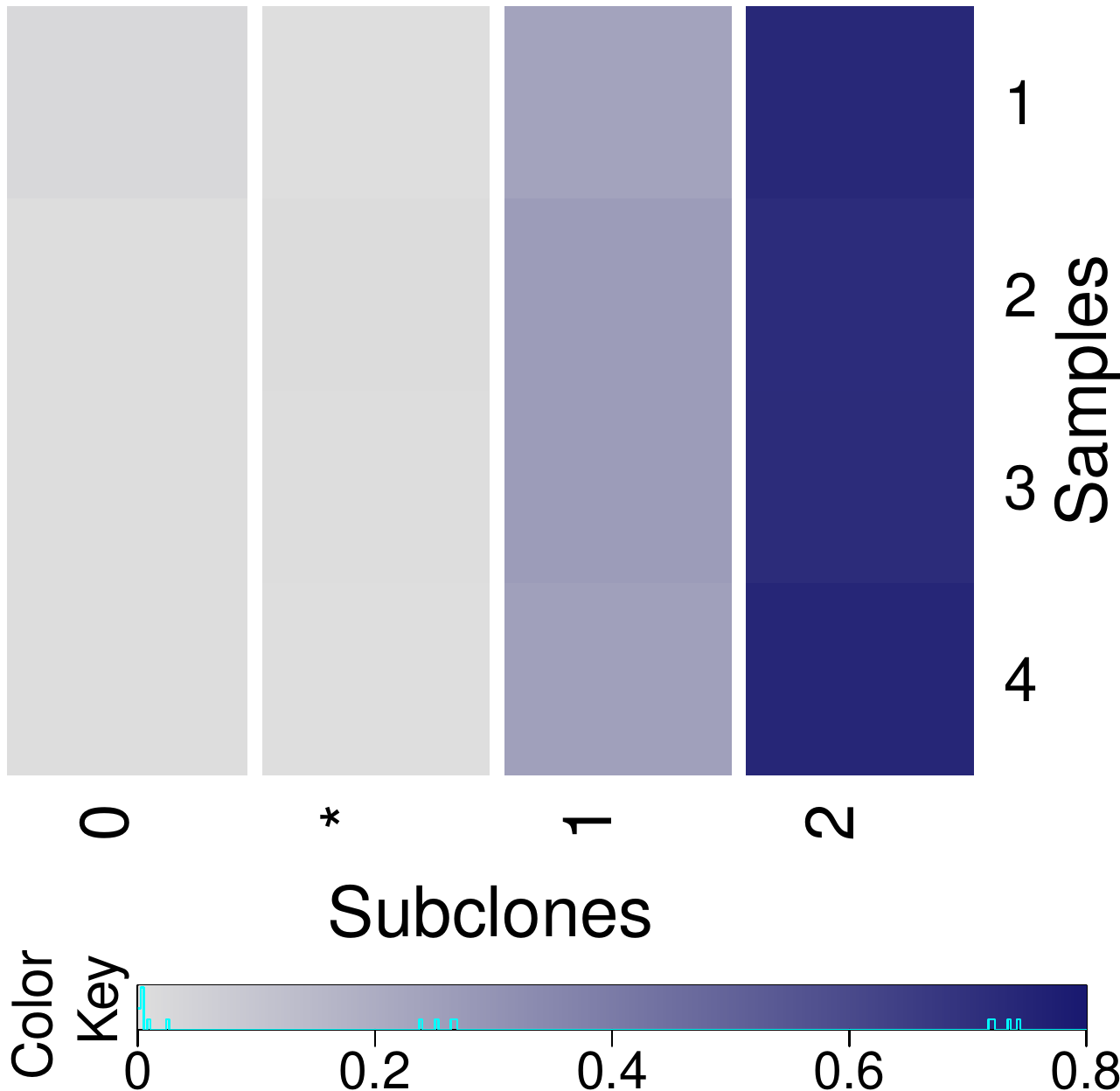}
\caption{$\what$}		
\end{subfigure}
\end{center}
\caption{Lung cancer. Posterior inference under PairClone.}
\label{fig:lung}
\end{figure}

\paragraph{Results}
The posterior distribution $p_b(C \mid \bn'')$ (shown in Appendix Figure \ref{app:figlungC}) reports
$p_b(C \mid \bn'') = 0.24$, $0.31$, $0.17$ and $0.12$ for $C = 1$, $2$, $3$ and $4$, respectively, and then quickly drops below $0.1$, with posterior mode $\Chat = 2$. 
This means, excluding the effect of normal cell contamination, the tumor samples have two subclones. Figure \ref{fig:lung}(a, b) show the estimated subclone matrix $\Zhat^P$ and $\Zhat^S$ corresponding to mutation pairs and SNVs, respectively.  
The first column of $\Zhat^P$ and $\Zhat^S$ represents the normal
subclone.  The rows for both matrices are reordered for a better
display.  Figure \ref{fig:lung}(c)
shows the estimated subclone proportions $\what$ for the four samples.
The second column of $\what$ represents the proportions of normal
subclones in the four samples. The small values indicate high purity
of the tumor samples.
The similar proportions across the four
samples reflect the spatial proximity of the samples. 
Furthermore, excluding a few exceptions that might be due to model
mis-fitting, the subclones form a simple phylogenetic tree: 
$*\rightarrow 1 \rightarrow 2$.  Subclones 1 and 2 share a large portion
of common mutations, while subclone 2 has some private
mutations that are missing in subclone 1.

For informal model checking we inspect a histogram of realized residuals (Appendix Figure \ref{app:figlungresid}). To define residuals, we calculate estimated multinomial probabilities $\{\ptkghat\}$ according to $\Zhat$, $\what$ and empirical values of $\{v_{tk1}, v_{tk2}, v_{tk3}\}$.
Let $\pbar_{tkg} = n_{tkg} / N_{tk}$. The figure plots the residuals $(\ptkghat- \pbar_{tkg})$.
The resulting histogram of residuals is centered around zero with little mass beyond $\pm 0.04$, indicating a good model fit.

\subsection{Using SNVs only}

\begin{figure}[h!]
\begin{center}
\begin{subfigure}[t]{.22\textwidth}
\centering
\includegraphics[width=\textwidth]{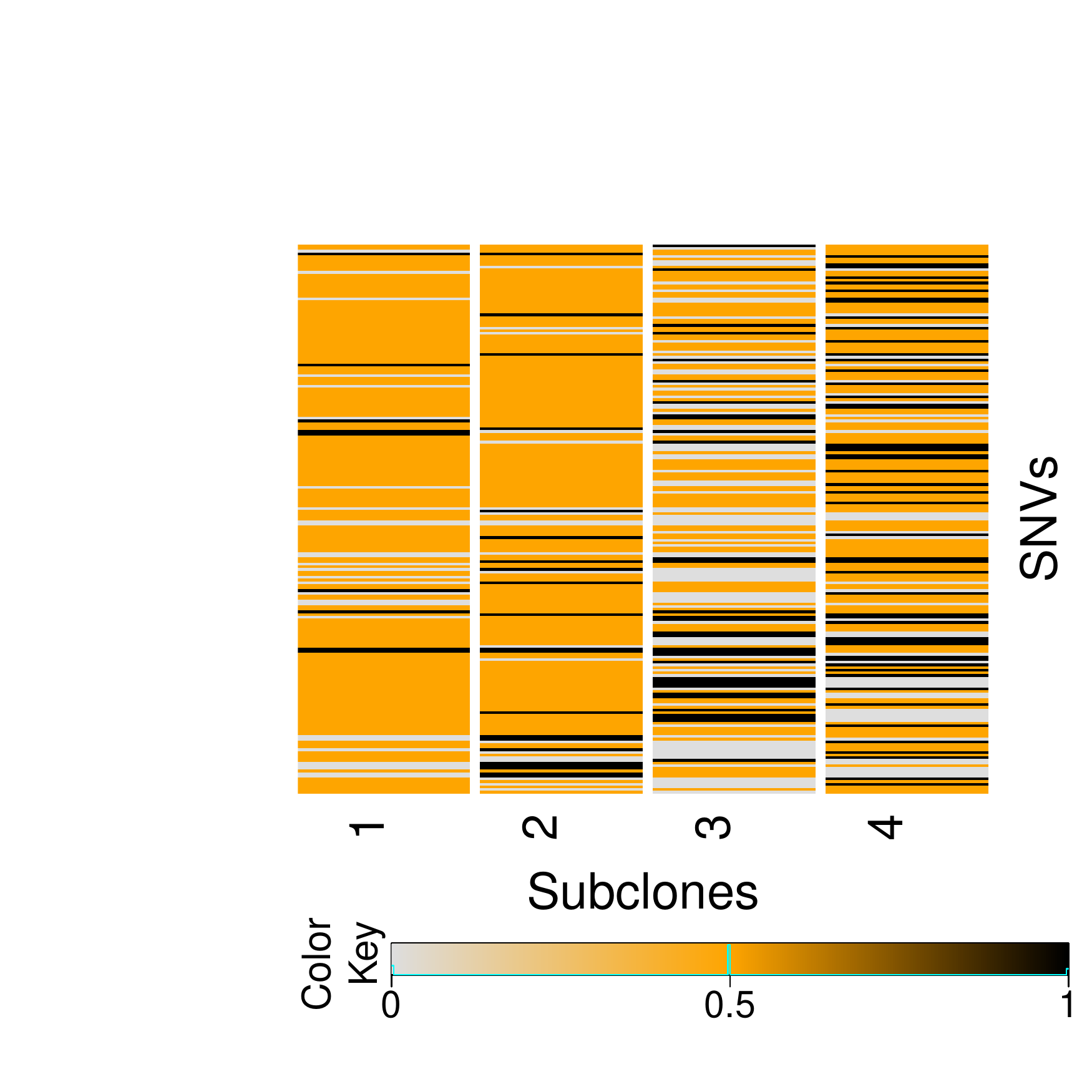}
\caption{$\Zhat_{\text{BC}}$}		
\end{subfigure}
\begin{subfigure}[t]{.22\textwidth}
\centering
\includegraphics[width=\textwidth]{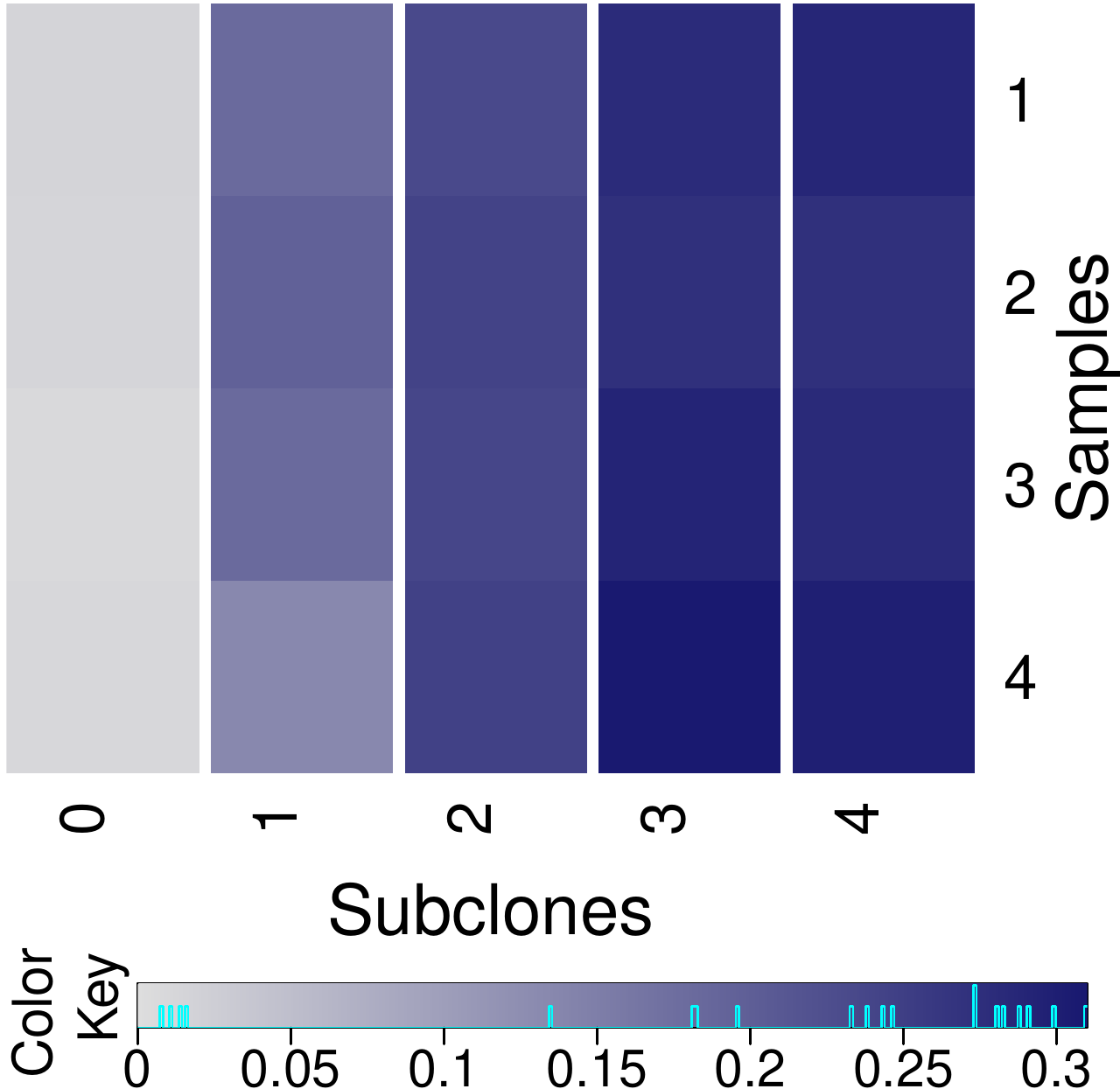}
\caption{$\what_{\text{BC}}$}		
\end{subfigure}
\begin{subfigure}[t]{.25\textwidth}
\centering
\includegraphics[width=\textwidth]{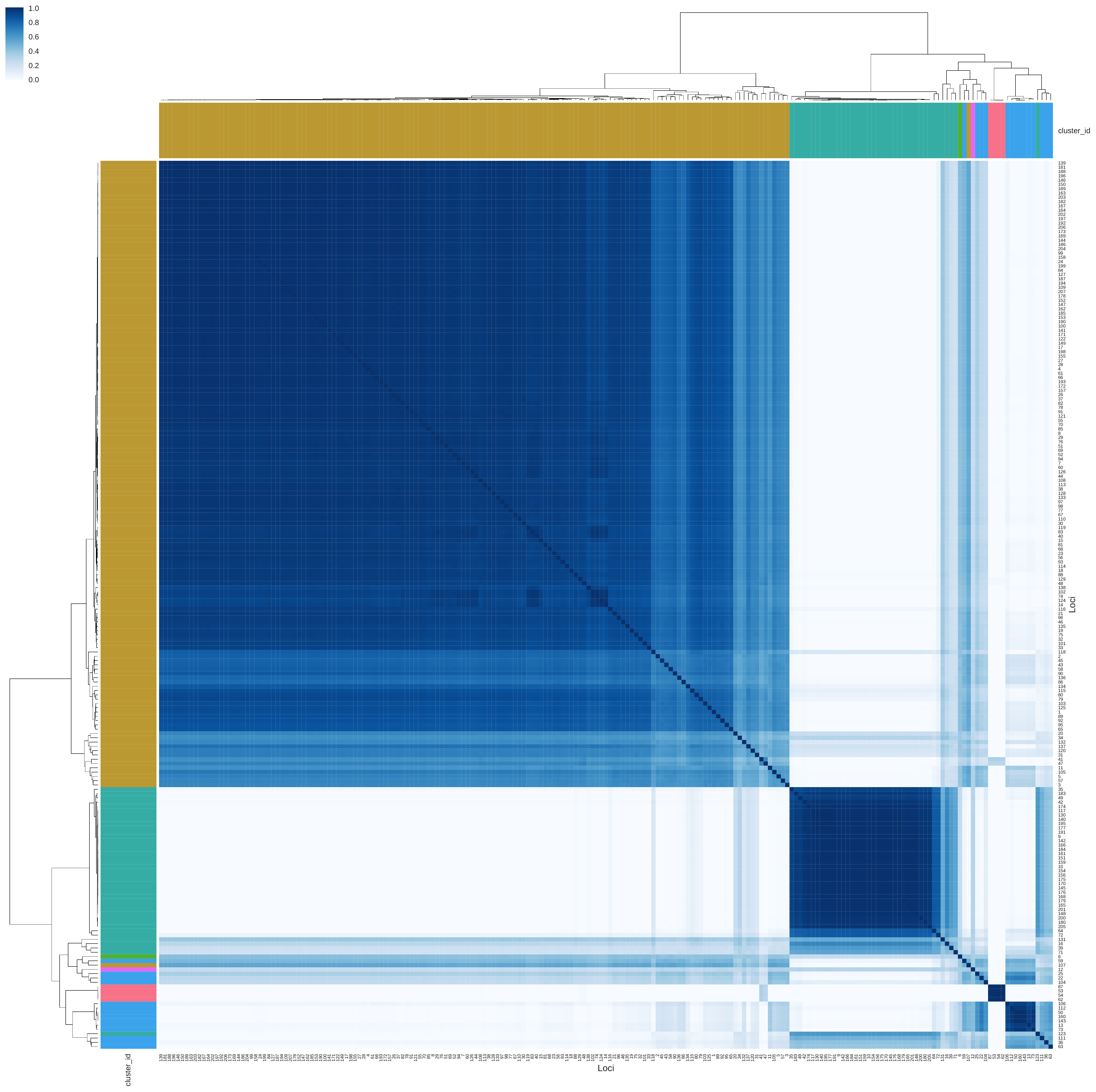}
\caption{Clustering matrix}		
\end{subfigure}
\begin{subfigure}[t]{.27\textwidth}
\centering
\includegraphics[width=\textwidth]{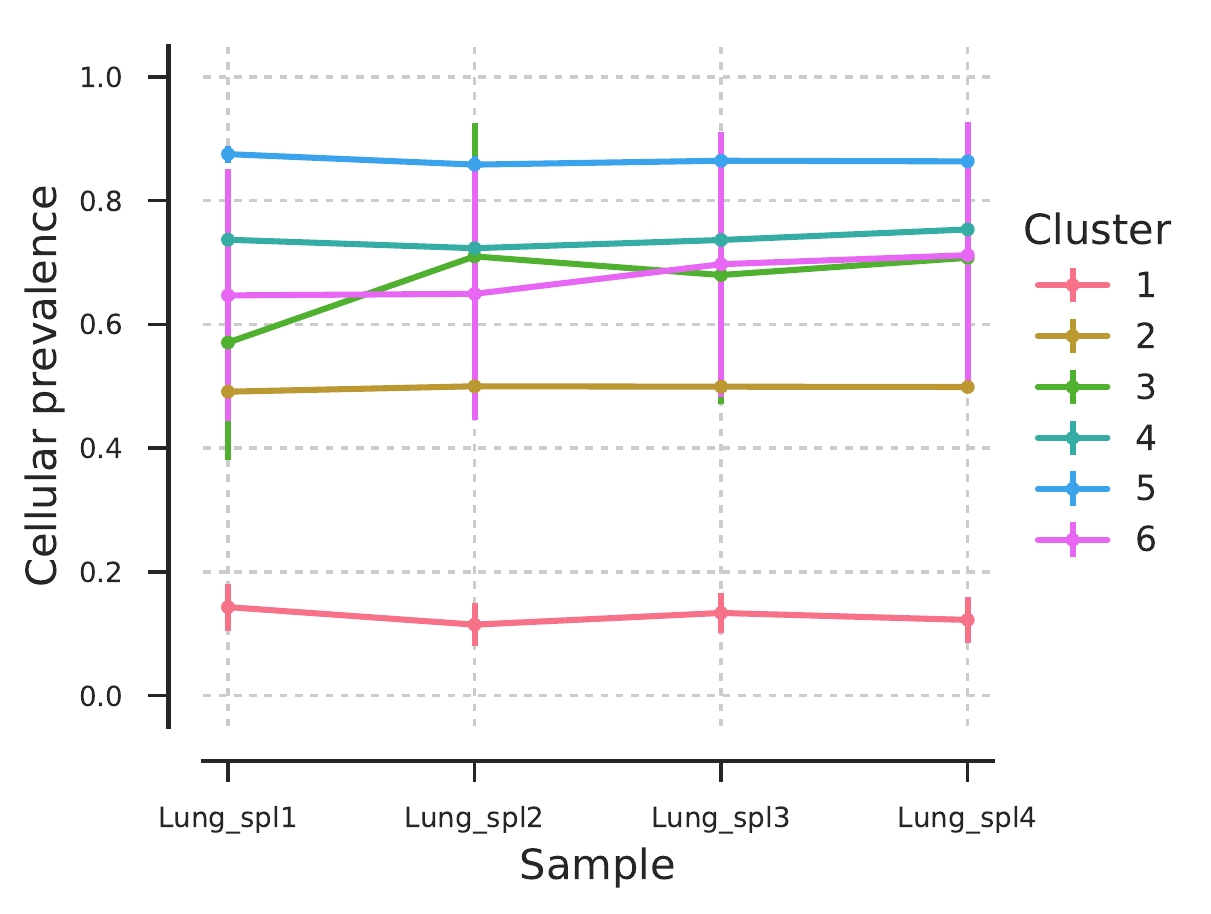}
\caption{Cellular prevalence}		
\end{subfigure}
\end{center}
\caption{Lung cancer. Posterior inference under BayClone (a, b) and PyClone (c, d).}
\label{fig:lungBC}
\end{figure}

For comparison, we also run BayClone and PyClone on the same dataset. Using the log pseudo marginal likelihood (LPML), BayClone reports $\Chat=4$ subclones.
The estimated subclone matrix in
BayClone's format is shown in Figure \ref{fig:lungBC}(a), with the
rows reordered in the same way as in Figure \ref{fig:lung}(a, b).
In light of the earlier simulation results we believe
that the inference under PairClone is more reliable.
Figure \ref{fig:lungBC}(b) shows the estimated subclone proportions
under BayClone. 
Figure \ref{fig:lungBC}(c) shows the estimated clustering of the SNV
loci under PyClone (the color coding along the axes).  
PyClone identifies 6 different clusters. 
The largest cluster (shown in brown) corresponds to loci that have heterozygous variants in
both subclones 1 and 2, the second-largest cluster (shown in blueish green) corresponds to loci
that have homozygous wild types in subclone 1 and homozygous
variants in subclone 2, and the other smaller clusters represent other
less common combinations. The clusters match with clustering of
rows of $\Zhat^P$ and $\Zhat^S$. 
PyClone does not immediately give inference on subclones, but combing clusters with similar cellular prevalence across samples one is able to conjecture subclones. In this sense, PyClone gives similar result compared with PairClone.
Finally, Figure \ref{fig:lungBC}(d) displays PyClone's estimated
cellular prevalences of clusters across different samples. 
The estimated subclone proportions and cellular prevalences across the
four samples remain very similar also under the BayClone and PyClone
output, which strengthens our inference that the four samples possess the same subclonal profile, each with two subclones. 


\begin{figure}[h!]
\begin{center}
\begin{subfigure}[t]{.4\textwidth}
\centering
\includegraphics[width=\textwidth]{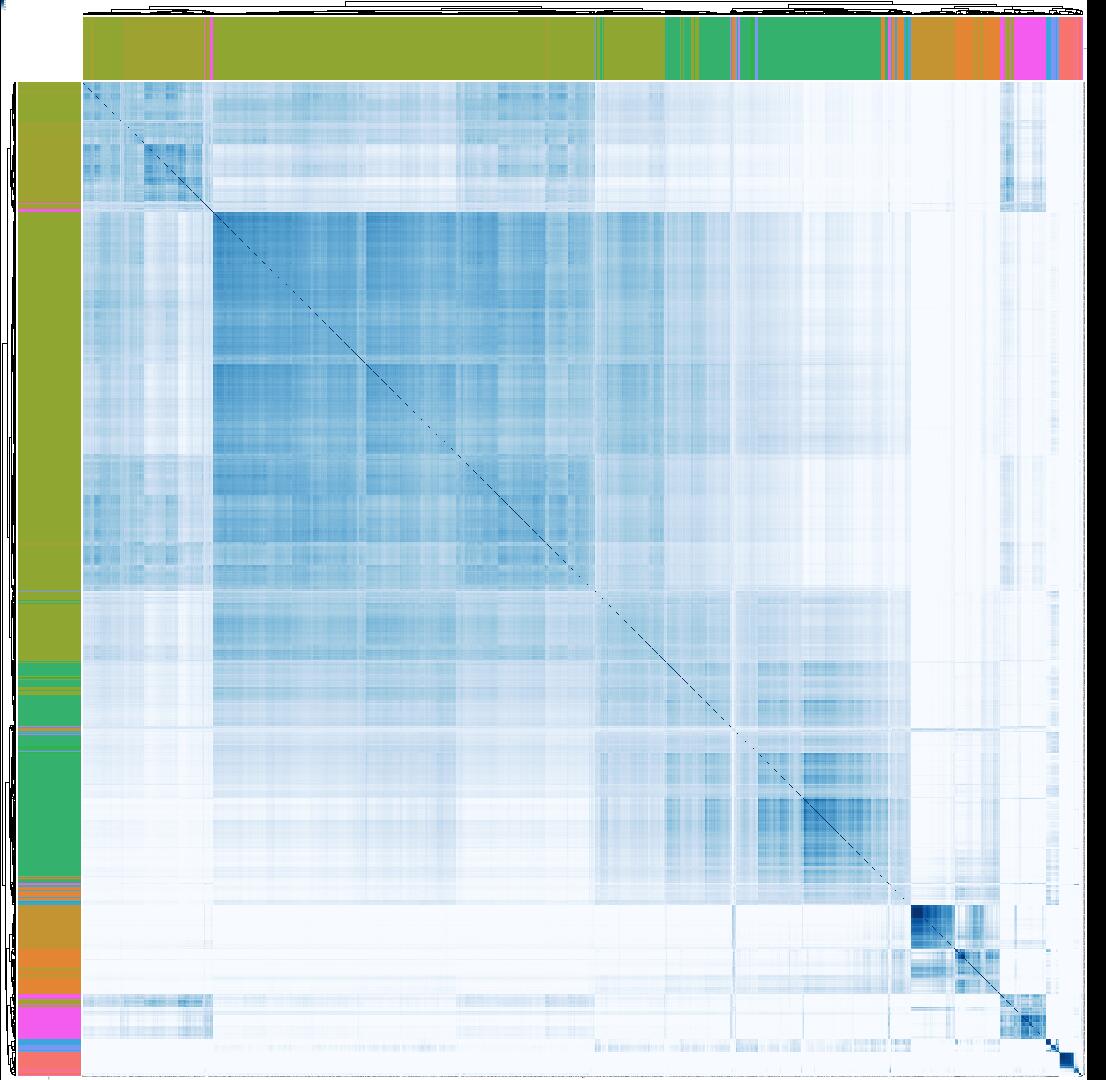}
\caption{Clustering matrix}		
\end{subfigure}
\hspace{7mm}\begin{subfigure}[t]{.45\textwidth}
\centering
\includegraphics[width=\textwidth]{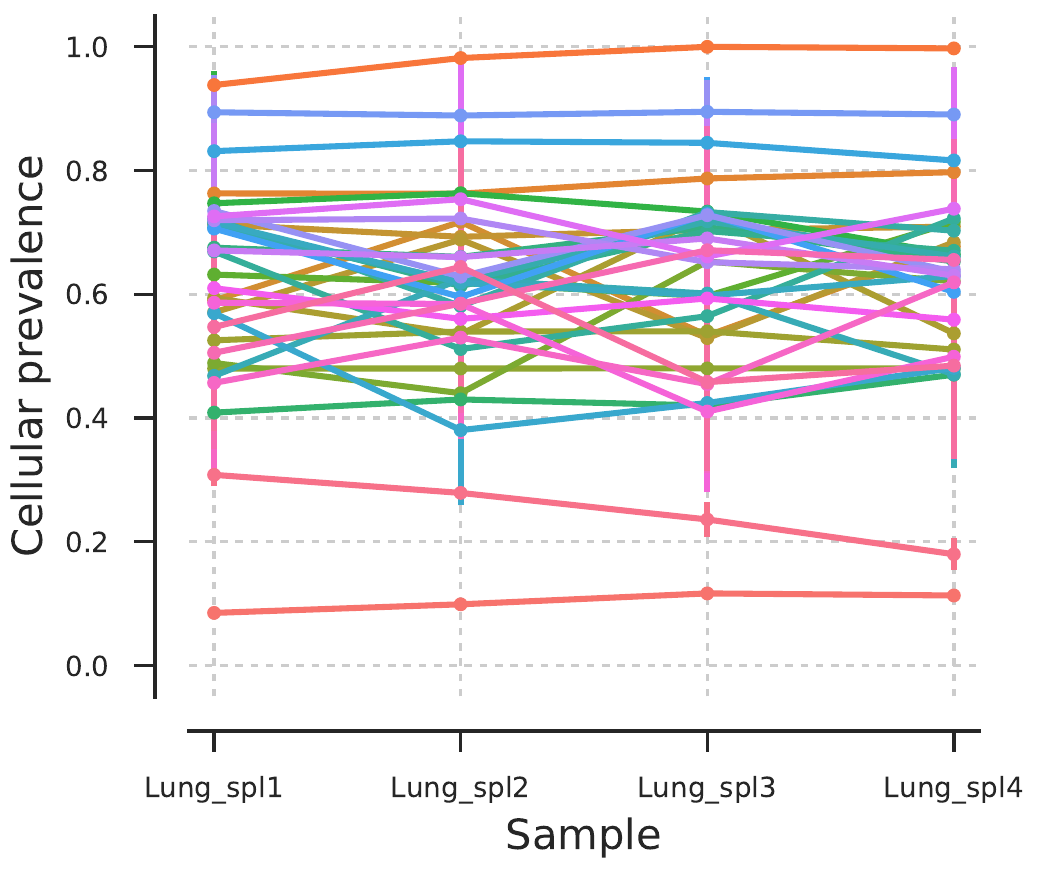}
\caption{Cellular prevalence}		
\end{subfigure}
\end{center}
\caption{Lung cancer. Posterior inference under PyClone using 1800 SNVs. PyClone inferred 34 clusters with two major clusters (olive and green) and many small noisy clusters (other colors).}
\label{fig:lung1800}
\end{figure}

For another comparison, we run PyClone with a much larger number of SNVs ($S = 1800$, which include the 69 pairs and 69 SNVs we ran analysis before) to evaluate the information gain by using additional marginal counts. The results are summarized in Figure \ref{fig:lung1800}, with panel (a) showing the estimated clustering of the 1800 SNVs. PyClone reports 34 clusters. The two largest clusters (olive and green clusters) in Panel (a) match with the two largest clusters (brown and bluish green clusters) in Figure \ref{fig:lungBC}(c) and also corroborate the two subclones inferred by PairClone.
In addition, PyClone infers lots of noisy tiny clusters using 1800 SNVs, which we argue model only noise. 
In summary, this comparison shows the additional marginal counts do not noticeably improve inference on tumor heterogeneity, and modeling mutation pairs is a reasonable way to extract useful information from the data.

\section{Conclusions}
\label{sec:conclusion}

We can significantly enrich our understanding of cancer development by
using high throughput NGS data to infer co-existence of subpopulations
which are genetically different across tumors and within a single
tumor (inter and intra tumor heterogeneity, respectively).
In this paper,
we have presented a novel feature allocation model for reconstructing such subclonal structure 
using mutation pair data.
Proposed inference explicitly models overlapping mutation pairs.
We have shown that more accurate inference can be obtained
using mutation pairs data compared to using only marginal counts for
single SNVs.
Short reads mapped to mutation pairs can provide direct evidence for
heterogeneity in the tumor samples. In this way the proposed approach
is more reliable than methods for subclonal reconstruction that rely
on marginal variant allele fractions only. 

The proposed model is easily extended for data where an
LH segment consists of more than two SNVs. We can easily accommodate
$n$-tuples instead of pairs of SNVs by increasing the number
of categorical values ($Q$) that the entries in the $\bZ$
matrix can take. There are several more interesting
directions of extending the current model.
For example, one could account for the potential
phylogenetic relationship among subclones (i.e the columns in the $\bZ$
matrix). Such extensions would enable one to
infer mutational timing and allow the reconstruction of tumor evolutionary
histories.

Lastly, we focus on statistical inference using bulk sequencing
data on tumor samples. Alternatively, biologists can apply single-cell
sequencing on each tumor cell and study its genome one by one. This is
a gold standard that can examine tumor heterogeneity at the single-cell
level. However, single-cell sequencing is still expensive and cannot
scale up. Also, many bioinformatics and statistical challenges are
unmet in analyzing single-cell sequencing data.

\bibliographystyle{Chicago}
\bibliography{ref_mut_pair_TJ}

\newpage

\begin{appendices}
\makeatletter 
\renewcommand{\thefigure}{A.\@arabic\c@figure}
\makeatother

\makeatletter 
\renewcommand{\thetable}{A.\@arabic\c@table}
\makeatother

\setcounter{figure}{0} 

\section{}
\subsection{ MCMC Implementation Details}
\label{app:sec:mcmc}
We first introduce $\theta_{tc}$ as an unscaled abundance level of
subclone $c$ in sample $t$. Assume $\theta_{t0} \sim \text{Gamma}(d_0,
1)$ and $\theta_{tc} \mid C \sim \text{Gamma}(d, 1)$. Let $w_{tc} =
\theta_{tc} / \sum_{c' = 0}^C \theta_{tc'}$, then $\bw_{t} \sim
\text{Dirichlet}(d_0, d, \ldots, d)$. We make inference on $\bm
\theta$ instead of $\bw$ as the value of $\bm \theta$ is not
restricted in a $C$-simplex. Similarly, we introduce $\rho_g^*$ as an
unscaled version of $\rho_g$. We let $\rho_g^* \sim \text{Gamma}(d_1,
1)$ and $\rho_g = \rho_g^* / \sum_{g' = 1}^4  \rho_{g'}^* $ for $g =
1, \ldots, 4$,  $\rho_g^* \sim \text{Gamma}(2d_1, 1)$ and $\rho_g =
\rho_g^* / \sum_{g' = 5}^6  \rho_{g'}^* $ for $g = 5, 6$, and
$\rho_g^* \sim \text{Gamma}(2d_1, 1)$ and $\rho_g = \rho_g^* /
\sum_{g' = 7}^8  \rho_{g'}^* $ for $g = 7, 8$. 

Conditional on $C$, the posterior distribution for the other parameters is given by
\begin{multline}
p(\bZ, \bm \pi, \bm \theta, \bm \rho^* \mid \bm n, C) \propto
\prod_{t = 1}^T \prod_{k = 1}^K \prod_{g = 1}^G \tp_{tkg}^{n_{tkg}}
\times \prod_{c = 1}^C \prod_{q = 1}^Q \pi_{cq}^{m_{cq}} \times 
\prod_{c = 1}^C \left[ \pi_{c1}^{1 - 1} (1 - \pi_{c1})^{\alpha/C - 1}
  \cdot \prod_{q = 2}^Q \tilde{\pi}_{cq}^{\beta - 1} \right] \times 
\nonumber \\ 
\prod_{t = 1}^T \left[ \theta_{t0}^{d_0 - 1} e^{- \theta_{t0}}
  \prod_{c = 1}^C \left( \theta_{tc}^{d - 1} e^{- \theta_{tc}}
  \right)\right] \times 
\prod_{g = 1}^4 \left( \rho_g^{*d_1 - 1} e^{-\rho_g^{*}} \right)
\cdot \prod_{g = 5}^8 \left( \rho_g^{*2d_1 - 1} e^{-\rho_g^{*}}
\right) .
\nonumber
\end{multline}
where $m_{cq} = \sum_{k = 1}^{K} I(\bz_{kc} = \bz^{(q)})$ counts the number of mutation pairs in subclone $c$ having genotype $\bz^{(q)}$.
\paragraph{Updating $\bZ$.} We update $\bZ$ by
sampling each $\bz_{kc}$ from: 
\begin{align*}
p(\bz_{kc} = \bz^{(q)} \mid \ldots) \propto \prod_{t = 1}^T \prod_{g = 1}^G \left[ \sum_{c' = 1, c' \neq c}^C w_{tc'} \, A(\bh_g, \bz_{kc'}) + w_{tc} \, A(\bh_g, \bz^{(q)}) + w_{t0} \, \rho_g \right]^{n_{tkg}} \cdot \pi_{cq}  
\end{align*}

\paragraph{Updating $\bm \pi$.} The posterior distribution for $\bm \pi$ is
\begin{align*}
p(\bm \pi \mid \ldots) &\propto \prod_{c = 1}^C \left[ \left( \prod_{q = 1}^Q \pi_{cq}^{m_{cq}} \right) \cdot \pi_{c1}^{1 - 1} (1 - \pi_{c1})^{\alpha/C - 1} \cdot \prod_{q = 2}^Q \tilde{\pi}_{cq}^{\beta - 1} \right] \\
&= \prod_{c = 1}^C \left[\pi_{c1}^{m_{c1} + 1 - 1} (1 - \pi_{c1})^{K - m_{c1} + \alpha/C - 1} \cdot \prod_{q = 2}^Q \tilde{\pi}_{cq}^{m_{cq} + \beta - 1} \right].
\end{align*}
For each $c = 1, \ldots, C$, we update $\bm \pi_c$ by sampling from
\begin{align*}
\pi_{c1} \mid \ldots &\sim \text{Beta}(m_{c1} + 1, K - m_{c1} + \alpha / C), \\
(\tilde{\pi}_{c2}, \ldots, \tilde{\pi}_{cQ}) \mid \ldots &\sim \text{Dirichlet}(m_{c2} + \beta, \ldots,  m_{cQ} + \beta),
\end{align*}
and transforming by $(\pi_{c2}, \ldots, \pi_{cQ}) = (1 - \pi_{c1}) \cdot (\tilde{\pi}_{c2}, \ldots, \tilde{\pi}_{cQ})$.

\paragraph{Updating $\bm \theta$.} We update each $\theta_{tc}$ sequentially. For $c = 1, \ldots, C$, 
\begin{align*}
p(\theta_{tc} \mid \ldots) \propto \prod_{k = 1}^K \prod_{g = 1}^G \left[ \sum_{c = 1}^C w_{tc} \, A(\bh_g, \bz_{kc}) + w_{t0} \, \rho_g \right]^{n_{tkg}} \cdot \theta_{tc}^{d - 1} e^{-\theta_{tc}}.
\end{align*}
A Metropolis-Hastings transition probability
is used to update $\theta_{tc}$. At each
iteration, we propose a new $\tilde{\theta}_{tc}$ (on the log scale) by
$\log(\tilde{\theta}_{tc}) \sim N( \log\theta_{tc}, 0.2)$, and evaluate
the acceptance probability by $p_{\text{acc}}(\theta_{tc},
\tilde{\theta}_{tc}) = 1 \wedge \left[ \left( p(\tilde{\theta}_{tc}  \mid
  \ldots) \, p(\theta_{tc} \mid \tilde{\theta}_{tc}) \right) \middle/  \left( p(\theta_{tc} \mid \ldots) \, p(\tilde{\theta}_{tc} \mid \theta_{tc}) \right) \right]$. 
The term  $p(\theta_{tc} \mid \tilde{\theta}_{tc}) / p(\tilde{\theta}_{tc} \mid \theta_{tc}) = \tilde{\theta}_{tc} / \theta_{tc}$ takes into account the Jacobian of the log transformation.
For $c = 0$, the
only difference is to substitute $d$ with $d_0$. 

\paragraph{Updating $\bm \rho^*$.} We update each $\rho_{g}^*$ sequentially. For $g = 1, \ldots, 4$,
\begin{align*}
p(\rho_g^* \mid \ldots) \propto  \prod_{t = 1}^T \prod_{k = 1}^K \prod_{g = 1}^G \left[ \sum_{c = 1}^C w_{tc} \, A(\bh_g, \bz_{kc}) + w_{t0} \, \rho_g \right]^{n_{tkg}} \cdot \rho_g^{*d_1 - 1} e^{-\rho_g^{*}}.
\end{align*}
A Metropolis-Hastings transition probability
is used to update $\rho_g^*$. At each iteration, we propose a new
$\tilde{\rho}_{g}^*$ (on the log scale) by 
$\log(\tilde{\rho}_{g}^*) \sim N(\log{\rho_g^*}, 0.1)$,
and evaluate the acceptance probability by
$p_{\text{acc}}(\rho_g^*, \tilde{\rho}_{g}^*) = 1 \wedge \left[
 \left( p(\tilde{\rho}_{g}^* \mid \ldots) \, p(\rho_g^* \mid \tilde{\rho}_{g}^*) \right) /  \left( p(\rho_g^* \mid \ldots) \, p(\tilde{\rho}_{g}^* \mid \rho_g^* ) \right) \right]$. 
The term  $p(\rho_g^* \mid \tilde{\rho}_{g}^*) / p(\tilde{\rho}_{g}^* \mid \rho_g^* ) = \tilde{\rho}_{g}^* / \rho_g^*$ takes into account the Jacobian of the log transformation.
For $g = 4, \ldots, 8$, the only difference is to substitute
$d_1$ with $2 d_1$. 

\paragraph{Parallel tempering.} Parallel tempering (PT) is a MCMC technique first proposed by \cite{geyer1991markov}. A good review can be found in \cite{liu2008monte}.
PT is suitable for sampling from a multi-modal state space. It helps
the MCMC chain to move freely among local modes which is desired in
our application, and to create a better mixing Markov chain.

\begin{algorithm}
\caption{Parallel Tempering}
\label{PTalgorithm}
\begin{algorithmic}[1]
\State Draw initial state $(\bm x_1^{(0)}, \ldots, \bm x_I^{(0)})$ from appropriate distributions
\For{$l$  in $1, \ldots, L$ }
\State Draw $u \sim \text{Uniform}(0, 1)$
\If{$u \leq u_0$}
\State Conduct the parallel step: update every $\bm x_i^{(l)}$ to $\bm x_i^{(l+1)}$ via respective MCMC scheme
\Else
\State Conduct the swapping step: draw $i \sim \text{Discrete-Uniform}(1, \ldots, I - 1)$, propose a swap between $\bm x_i^{(l)}$ and $\bm x_{i+1}^{(l)}$, accept the swap with probability
\begin{align*}
\min\left\{  1, \frac{\pi_i(\bm x_{i+1}^{(l)}) \pi_{i+1}(\bm x_{i}^{(l)})}{\pi_i(\bm x_{i}^{(l)}) \pi_{i+1}(\bm x_{i+1}^{(l)})} \right\}
\end{align*}
\EndIf
\EndFor
\end{algorithmic}
\end{algorithm}

To sample from the target distribution $\pi(\bm x)$, we consider a
family of distributions $\Pi = \{ \pi_i , i = 1, \ldots, I \}$,  where
$\pi_i(\bm x) \propto \pi(\bm x)^{1 / \Delta_i}$. Without loss of
generality, let $\Delta_I = 1$ and $\pi_I(\bm x) = \pi(\bm x)$.
Denote by $\mathcal{X}_i$ the state space of $\pi_i(\bm x)$. The PT
scheme is illustrated in Algorithm \ref{PTalgorithm}. 

In our application, we find by simulation that PT works well with $I = 10$ temperatures and  $\{ \Delta_1, \ldots, \Delta_{10} \} = \{  4.5, 3.2, 2.5, 2, 1.7, 1.5, 1.35, 1.2, 1.1, 1\}$. We therefore use this parameter setting for all the simulation studies as well as the lung cancer dataset.

\subsection{Updating $C$}
\label{app:sec:updatec}
For updating $C$, we split the data into a training set $\bn'$, and a
test set $\bn''$ with $n_{tkg}' = b n_{tkg}$ and $n_{tkg}'' = (1 -
b) n_{tkg}$. Let $p_b(\bx \mid C) = p(\bx \mid \bn', C)$ denote the
posterior of $\bx$ conditional on $C$ evaluated on the training set
only. We use $p_b$ in two occasions. First, we replace the original
prior $p(\bx \mid C)$ by $p_b(\bx \mid C)$, and second, we use $p_b$
as a proposal distribution of $\tbx$ as $q( \tbx \mid \tilde{C} ) =
p_b(\tbx \mid \tilde{C})$. 
We show that the use of the training sample posterior as proposal
and modified prior in equation (4) (original manuscript) implies an
approximation in the reported marginal posterior for $C$, but leaves
the conditional posterior for all other parameters (given $C$)
unchanged. 

We evaluate the acceptance probability of $\tilde{C}$ on the test data
by
\begin{align*}
p_{\text{acc}} (C, \bx, \tC, \tbx)
   &= 1 \wedge \frac{p(\bn'' \mid \tbx, \tC)}{p(\bn'' \mid \bx, C)} \cdot
                         \frac{p(\tC) p(\tbx \mid \bn', \tC) }{p(C)
                         p(\bx \mid \bn', C)} \cdot \frac{q(C \mid
                         \tC) q(\bx \mid C)}{q(\tC \mid C) q(\tbx \mid
                         \tC)} \\ 
 &= 1 \wedge \frac{p(\bn'' \mid \tbx, \tC)}{p(\bn'' \mid \bx, C)} \cdot
  \frac{p(\tC) }{p(C)} . 
\end{align*}
Under the model $p_b(\cdot)$   with the modified prior,  the implied conditional posterior on $\bx$ satisfies 
\begin{multline*}
p_b(\bx \mid C, \bn) 
= \frac{p_b(\bx \mid C) p(\bn'' \mid \bx, C)}{\int p_b(\bx \mid C) p(\bn'' \mid \bx, C) d \bx} \\
= \frac{p(\bx \mid C) p(\bn' \mid \bx, C) p(\bn'' \mid \bx, C)}{\int p(\bx \mid C) p(\bn' \mid \bx, C) p(\bn'' \mid \bx, C) d \bx} = p(\bx \mid C, \bn), 
\end{multline*}
which indicates the conditional posterior of $\bx$ remains entirely unchanged.
The implied marginal posterior on $C$ is
$p_b(C \mid \bn'') \propto p(C) \, p_b(\bn'' \mid C)$,
with the likelihood on the test data evaluated as $p_b(\bn'' \mid C) =
\int p(\bn'' \mid \bx, C) \, p_b(\bx \mid C) d\bx$.
The use of the prior $p_b( \tbx \mid \tC)$ is similar to the
construction of the fractional Bayes factor (FBF)
\citep{ohagan1995}. Let $\bu = \{ \bm \pi, \bw, \bm \rho\}$ denote the
parameters other than $\bZ$ 
and let $\umle$ denote the maximum likelihood estimate for
$\bu$.
We follow \cite{ohagan1995} to show that inference on $C$ is as if we
were making use of only a fraction $(1-b)$ of the data, with a
dimension penalty.
In short,
$$
  p_b(C \mid \bn'') \propto p(C) p(\bn \mid
    \umle, C)^{1 - b}\, b^{\; p_{C} / 2},
$$
approximately, where $\umle$ is the maximum likelihood estimate of $\bu$, and $p_C$ is the number of unconstrained parameters in $\bu$. To obtain this approximation, consider the marginal
sampling model under $p_b(\cdot)$, after marginalizing with respect to $\bx$:
\begin{multline*}
p_b(\bn'' \mid C) = \int p(\bn'' \mid \bx, C) p_b(\bx \mid C)  d \bx \\ 
= \int p(\bn'' \mid \bx, C)  \,
  \frac{p(\bn' \mid \bx, C) p(\bx \mid C)}
       {\int p(\bn' \mid \bx, C) p(\bx \mid C) d\bx}\,
  d \bx  
= \frac{\int p(\bn \mid \bx, C) p(\bx \mid C) d \bx}{\int p(\bn' \mid
  \bx, C) p(\bx \mid C) d \bx}.  
\end{multline*}
Here we substituted 
the training sample posterior as (new) prior $p_b(\bx \mid C)$. 
The integration
includes a  marginalization with respect to the discrete $\bZ$,
\begin{align*}
  \int p(\bn \mid \bx, C) p(\bx \mid C) d \bx &= \int \sum_{\bZ} p(\bn \mid \bZ, \bu, C) p(\bZ \mid \bu, C) p(\bu \mid C) d\bu \\
  &= \int  p(\bn \mid \bu, C) p(\bu \mid C) d\bu,
\end{align*}
For the remaining real valued parameters $\bu$ we use an appropriate
one-to-one transformation (e.g. logit transformation) $\bu \mapsto
\tbu$, such that $\tbu$ is unconstrained.  To simplify notation we
continue to refer to the transformed parameter as $\bu$ only.
Next, under the binomial sampling model
$p(\bn' \mid \bx,C) \propto p(\bn \mid \bx,C)^b$, leading
to

\begin{align*}
p_b(\bn'' \mid C) &=
\frac{\int p(\bn \mid \bu, C) p(\bu \mid C) d \bu}
     {\int p(\bn' \mid \bu, C) p(\bu \mid C) d \bu} \\
     &=
 \underbrace{\frac{\left[\prod_{t, k} N_{tk}! / (n_{tk1}! \cdots
      n_{tkG}!)\right]^b}{\prod_{t, k} (bN_{tk})! / \left[ (bn_{tk1})!
      \cdots (bn_{tkG})!\right]}}_{m(\bn)} \cdot
\underbrace{\frac{\int p(\bn \mid \bu, C) p(\bu \mid C) d \bu}{\int
    p(\bn \mid \bu, C)^{b} p(\bu \mid C) d \bu}}_{h_b(\bn \mid C)},
\nonumber
\end{align*}

Let $m(\bn)$ and $h_b(\bn \mid C)$ denote the two factors.
The first, $m(\bn)$, is a constant term. And the second factor,
$h_b(\bn \mid C)$, has exactly the same form as equation (12) in
\cite{ohagan1995}, who shows
\begin{align*}
   h_b(\bn \mid C) \approx 
   p(\bn \mid \umle, C)^{1-b} b^{\; p_C / 2}
\end{align*}
Let $N = \sum_{t,k} N_{tk}$. 
The argument of
\cite{gelfand1994bayesian} (case (e)) suggests that the
error in this approximation is of order $O(1/N^2)$
(note that Gelfand and Dey use expansion around the M.A.P. while
O'Hagan uses expansions around the M.L.E.).
This establishes the stated 
approximation of the posterior
$
p_b(C \mid \bn'') \approx k\cdot p(C) p(\bn \mid \umle, C)^{1 - b}\,
b^{\; p_{C} / 2}$, approximately.

\subsection{Simulation Studies and Comparison with Marginal Counts} 
\label{app:sim}
We report details of the three simulation studies that are
summarized in the manuscript (Section \ref{sec:simulation}). The discussion includes a comparison
with inference under methods that use only marginal mutation counts.

\subsubsection{Simulation 1}
\label{app:sim1}
\begin{figure}[h!]
\begin{center}
  \begin{tabular}{ccc}
    \includegraphics[scale = 0.37]{./Z_sim1_alt.pdf}
    &
    \hspace{-2mm}\includegraphics[scale = 0.275]{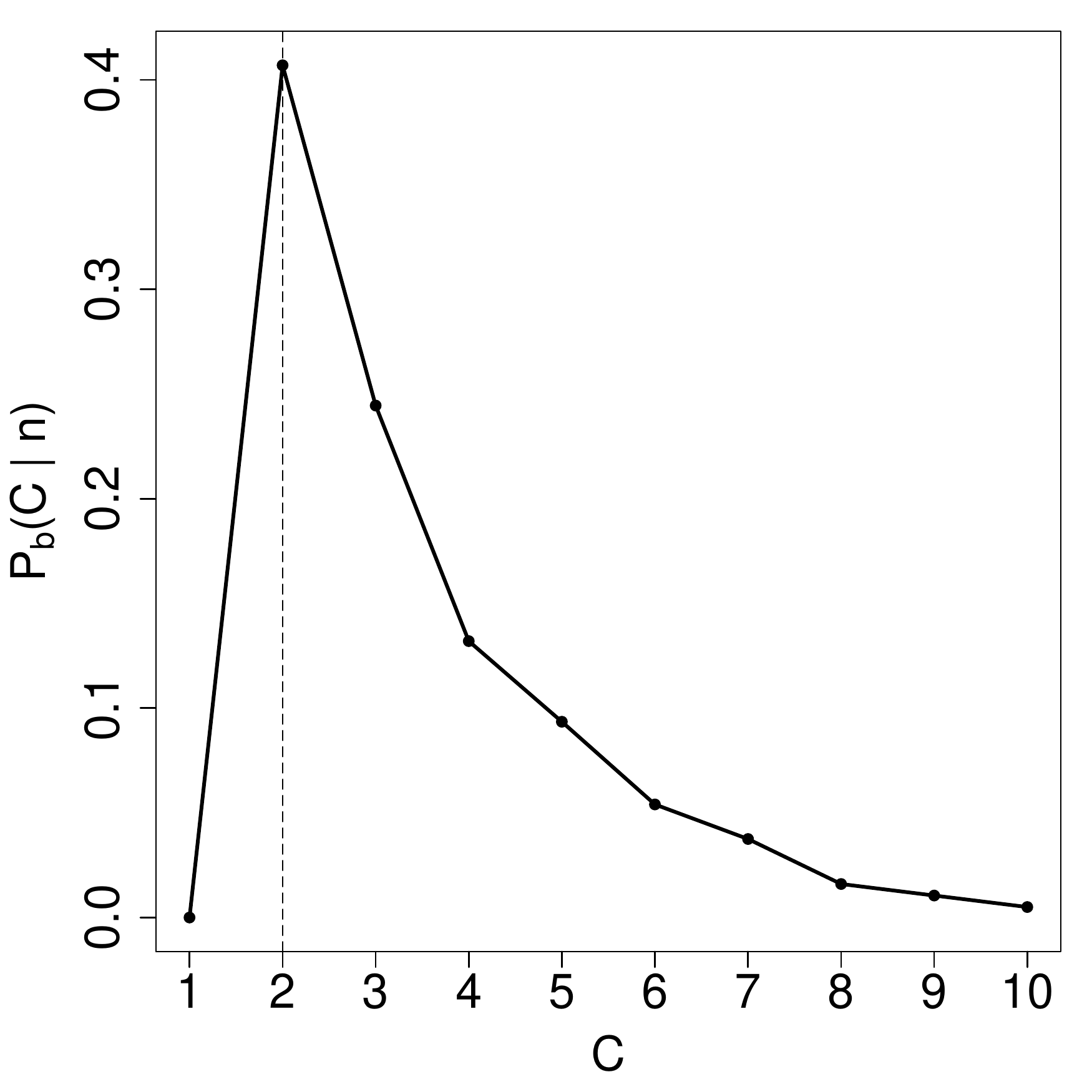}
    &
    \includegraphics[scale = 0.37]{./Zstar_sim1_alt.pdf}
    \\
    {\footnotesize{(a) $\bZ\true$}}
    &
    {\footnotesize{(b) $p_b(C \mid \bn'')$}}
    &
    {\footnotesize{(c) $\Zhat$}}
    \\
    \includegraphics[scale = 0.275]{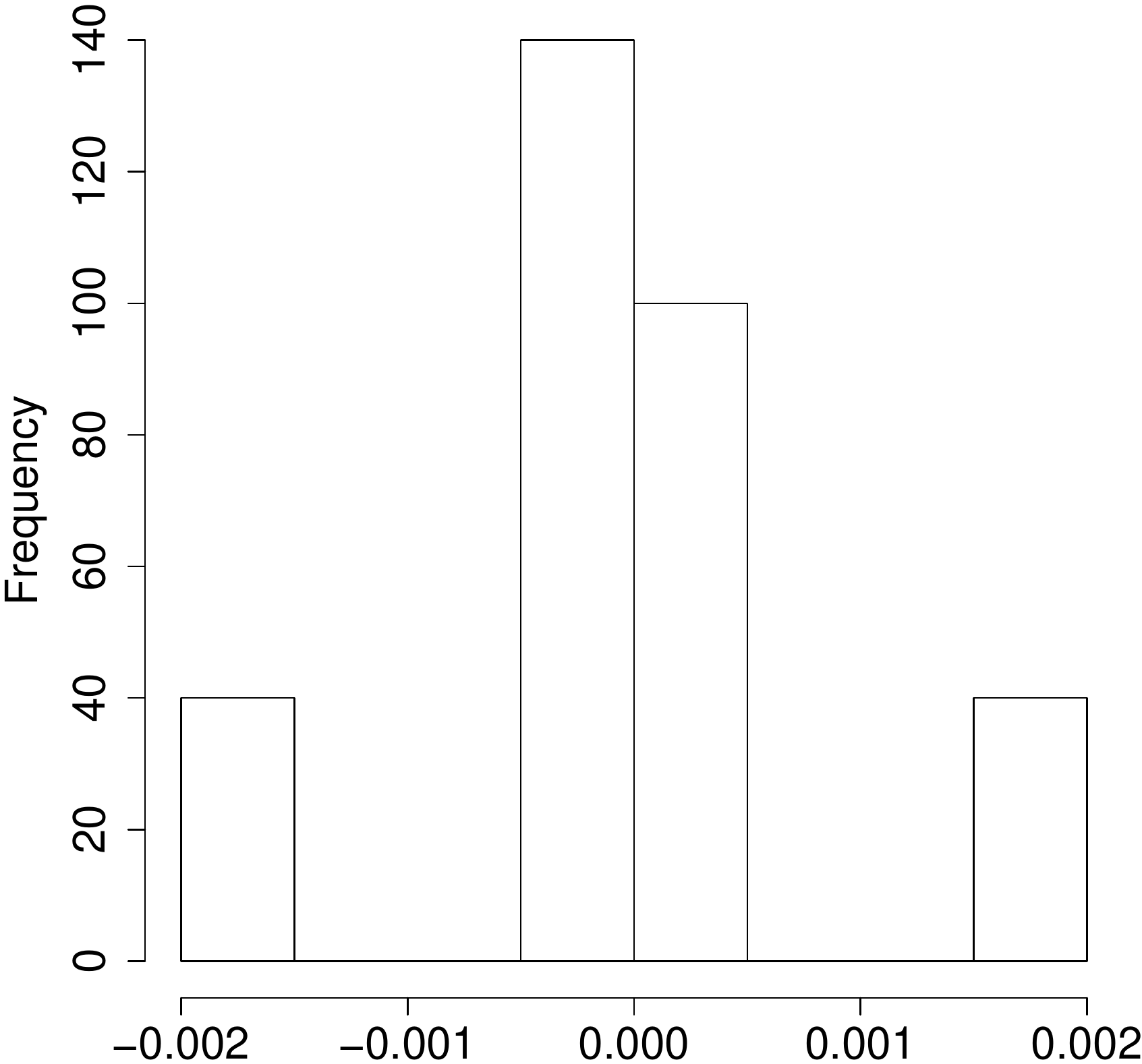}
    &
    \hspace{1mm}\includegraphics[scale = 0.37]{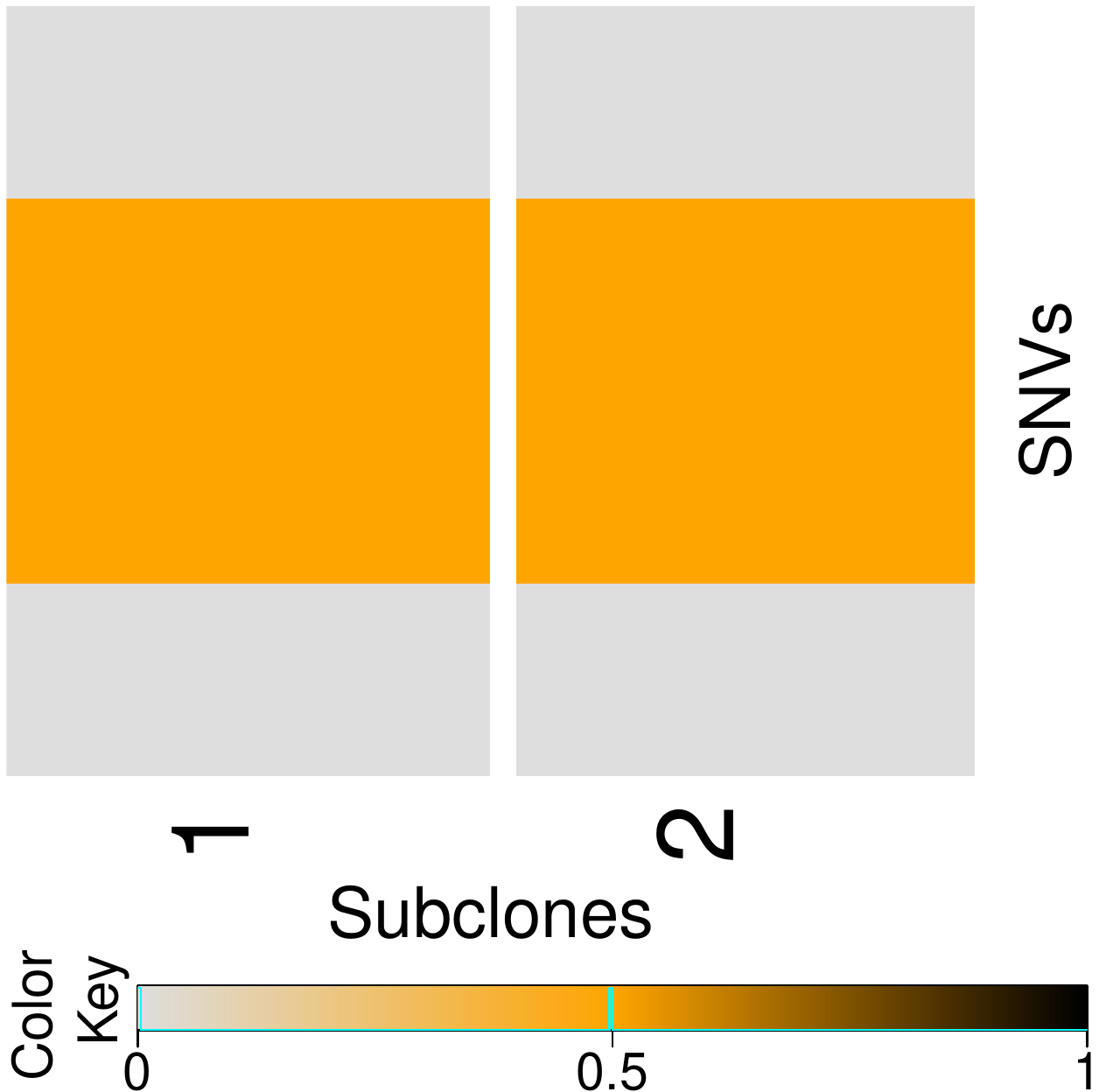}
    &
    \includegraphics[scale = 0.37]{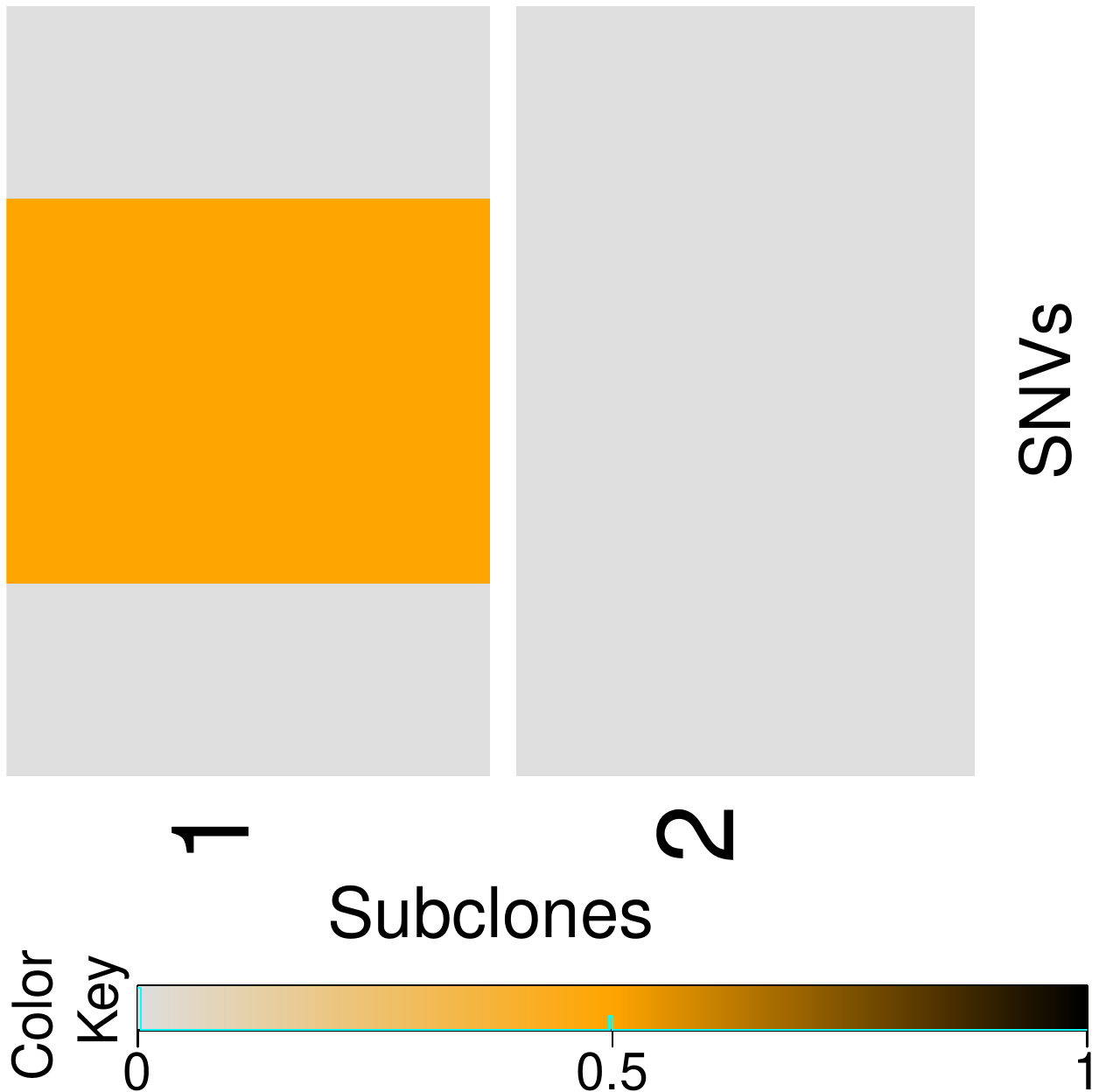}\\
    {\footnotesize{(d) Histogram of $(\ptkghat - p_{tkg}\true)$}}
    &
    {\footnotesize{(e) $\bZ_{\text{BC}}\true$}}
    &
    {\footnotesize{(f) $\Zhat_{\text{BC}}$}}
    \\
  \end{tabular}
\end{center}
\caption{Simulation 1. Simulation truth $\bZ\true$ (a, e), and posterior
  inference under PairClone (b, c, d) and under BayClone (f).}
\label{app:figsim1}
\end{figure}

\paragraph{Setup. }
In the first simulation, we illustrate the advantage of using mutation
pair data over marginal SNV counts.
We generate hypothetical short reads data for $T = 1$ sample and $K = 40$
mutation pairs.  Based on our own experiences, for a whole-exome
sequencing data set, we usually obtain dozens of mutation pairs with
decent coverage. See \cite{sengupta2015NAR} for a discussion.  We
assume there are $C\true = 2$ latent subclones, and set their
population frequencies as $\bw\true = (1.0 \times 10^{-7}, 0.8,
0.2)$, where $1.0 \times 10^{-7}$ refers to the proportion of the
hypothetical background subclone $c=0$.
The subclone matrix $\bZ\true$ is shown in
Figure~\ref{app:figsim1}(a) (as a heat map). Light grey, red and black
colors are used to represent genotypes $\bz^{(1)}$, $\bz^{(4)}$ and
$\bz^{(6)}$. For example, subclone 1 
has genotype $\bz^{(1)}$ (wild type) for mutation pairs 1--10 and 31
-- 40, and $\bz^{(4)}$ for mutation pairs 11--30.  We generate $\brho\true$ from its prior with hyperparameter $d_1 = 1$. Next we set the probabilities of observing left and right missing reads as
 $v_{tk2} = v_{tk3} = 0.3$ for all $k$ and $t$, to mimic a typical missing rate observed in the real data.  We calculate multinomial
probabilities $\{p_{tkg}\true\}$ shown in equations
\eqref{pprior1} and \eqref{eq:A} from the simulated $\bm
Z\true$, $\bm w\true$ and $\bm
\rho\true$. Total read counts $N_{tk}$ are generated as
random numbers ranging from $400$ to $600$, and finally we generate
read counts $n_{tkg}$ from the multinomial distribution given $N_{tk}$
as shown in equation \eqref{eq:multi}.

We fit the model with hyperparameters fixed as follows: $\alpha = 4$,
$\gamma_2 = \cdots = \gamma_Q = 2$, $d = 0.5$, $d_0 = 0.03$, $d_1 =
1$, and $r = 0.4$. We set $C_{\text{min}} = 1$ and $C_{\text{max}} =
10$ as the range of $C$. The fraction $b$ needs to be calibrated. We
choose $b$ such that the test sample size $(1 - b) \sum_{t = 1}^{T}
\sum_{k = 1}^{K} N_{tk}$ is approximately equal to
$160/T$. See Section \ref{app:sec:calib} for a discussion of
this choice.

We run MCMC simulation for $30,000$
iterations, discarding the first $10,000$ iterations as initial
burn-in, and keep one sample every $10$ iterations. The initial values
are randomly generated from the priors.

\paragraph{Results. }
Figure~\ref{app:figsim1}(b) shows $p_b(C \mid \bn'')$, where the
vertical dashed line marks the simulation truth. The posterior
mode $\Chat = 2$ recovers the truth. Figure~\ref{app:figsim1}(c) shows the
point estimate of $\bZ\true$, given by $\Zhat$. The true subclone
structure is perfectly recovered. The estimated subclone weights are
$\what = (2.27 \times 10^{-116}, 0.8099, 0.1901)$, which is also
very close to the truth. We use $\Zhat$ and $\what$ to calculate
estimated multinomial probabilities, denoted by $\{\ptkghat\}$.
Figure~\ref{app:figsim1}(d) shows a histogram of the differences
$(\ptkghat - p_{tkg}\true)$  as a residual plot to assess model
fitting. The histogram is centered at zero with little variation,
  indicating   a reasonably good model fit.
In summary, this simulation shows that the proposed inference
can almost perfectly recover the truth in a simple scenario with a single sample.

\paragraph{Inference with marginal read counts. }
We compare the proposed inference under PairClone versus
inference under SNV-based subclone callers ,i.e., based on marginal (un-paired) counts of point mutations, including BayClone~\citep{Bayclone2015} and 
PyClone~\citep{roth2014pyclone}. 

\begin{figure}[h!]
\begin{center}
  \begin{tabular}{ccc}
    \includegraphics[scale = 0.3]{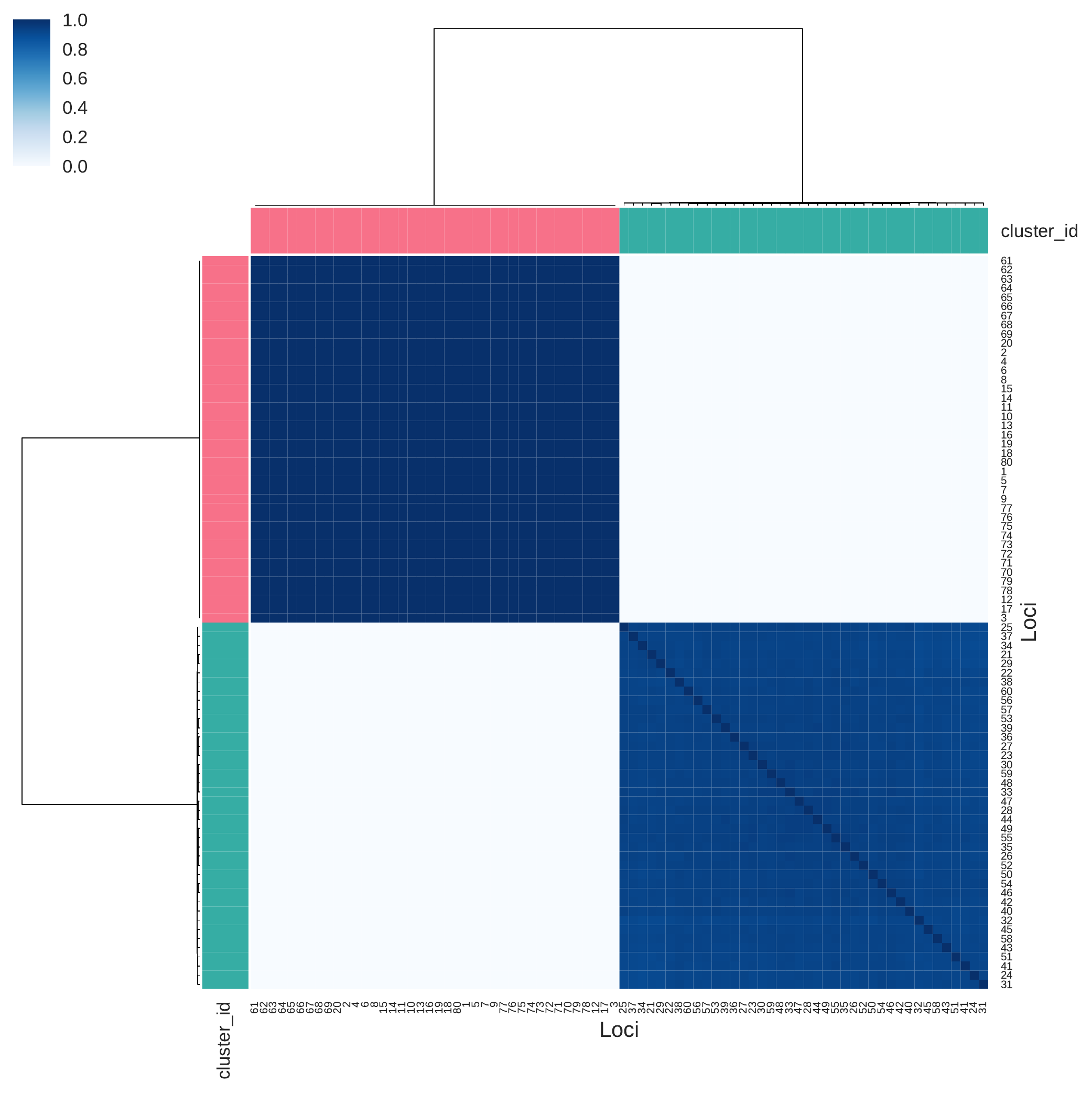}
    &
    \includegraphics[scale = 0.6]{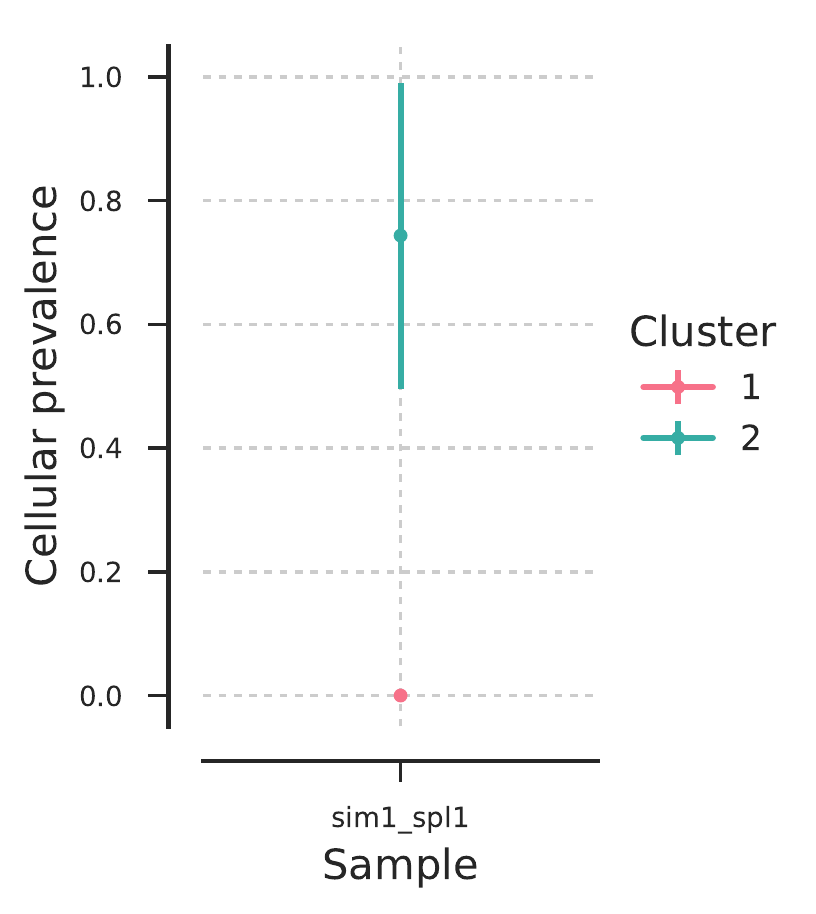}
    \\
    {\footnotesize{(a) Posterior similarity matrix}}
    &
    {\footnotesize{(b) Cellular prevalence}}
  \end{tabular}
\end{center}
\caption{Simulation 1. Posterior inference under PyClone.}
\label{app:figsim1_pyclone}
\end{figure}

\underline{BayClone} infers the subclone structure
based on marginal allele frequencies of the recorded SNVs, and chooses the
number of subclones based on log pseudo marginal likelihood (LPML)
model comparison.
Under the LPML criterion, the estimated number of subclones reported
by BayClone is $\Chat = 2$, which also recovers the truth.
Figure~\ref{app:figsim1}(e) displays
the true genotypes of the unpaired SNVs, denoted by
$\bZ_{\text{BC}}\true$, based on the true genotypes in
Figure~\ref{app:figsim1} for the mutation pairs. That is, we derive the
corresponding marginal genotype for each SNV in the mutation pair
based on the truth $\bZ\true$.  Figure~\ref{app:figsim1}(f) shows
the heat map of estimated matrix $\Zhat_{\text{BC}}$, where
$z_{sc} = 0$ (light grey), $0.5$ (orange) and $1$
(black) refer to homozygous wild-type, heterozygous variant and
homozygous variant at SNV locus $s$, respectively. The estimated
subclone proportions are $\what_{\text{BC}} = (0.008, 0.988, 
0.004)$.

\underline{PyClone}, on the other hand, clusters mutations based on
allele frequencies of the recorded SNVs using
the implied clustering under a Dirichlet process mixture model.
PyClone does not report subclonal genotypes and thus is not
directly comparable with PairClone. 
Posterior inference is summarized in 
Figure
\ref{app:figsim1_pyclone}.
Panel (a) indicates that the 80 SNV loci form two
clusters, with one cluster corresponding to loci 1--20 and
61--80, and the other cluster corresponding to loci 21--60, which
agrees with the truth. Panel (b) shows the cellular prevalence of the two
clusters across samples, where the middle point represents the
posterior mean, and the error bar indicates posterior standard
deviation. 
The cellular prevalence is defined as fraction of clonal population
harbouring a mutation.  In the PyClone MCMC samples, the estimated
cellular prevalence of cluster 2 fluctuates between 0.5 and 1 and thus
includes high posterior uncertainty, 
while the true cellular prevalence of cluster 2 is 1.

The estimates under SNV-based subclone callers do not fully recover
the simulation truth. The main reason is probably that the phasing
information of paired SNVs is 
lost in the marginal counts that are used in BayClone and PyClone,
making the subclone estimation less accurate than under PairClone.
For example, the two subclones with genotypes
$\bz^{(4)} = (00, 11)$ and $\bz^{(6)} = (01, 10)$ lead
to exactly the same allele frequency (50\%) for both loci.
BayClone can not distinguish between these two different subclones
based on the 50\% allele frequency for each locus.
Although BayClone correctly reports the number of subclones, inference
mistakenly includes a normal subclone with negligible weight, and thus
fails to recover the true population frequencies.  
On the other hand,  PyClone can not identify if cluster 2 contains homozygous (corresponding to cellular prevalence of 0.5) or heterozygous (corresponding to cellular prevalence of 1) variants. In contrast, using the phasing information, PairClone is able to infer two subclones having genotypes $(00, 11)$ and $(01, 10)$ for mutation pairs 11--30, and we know cluster 2 contains only heterozygous variants for sure.

\subsubsection{Simulation 2}
\label{app:sim2}

\begin{figure}[h!]
  \begin{center}
    \begin{tabular}{ccc}
      \includegraphics[scale = 0.37]{./Z_sim2.pdf}
      &
      \includegraphics[scale = 0.37]{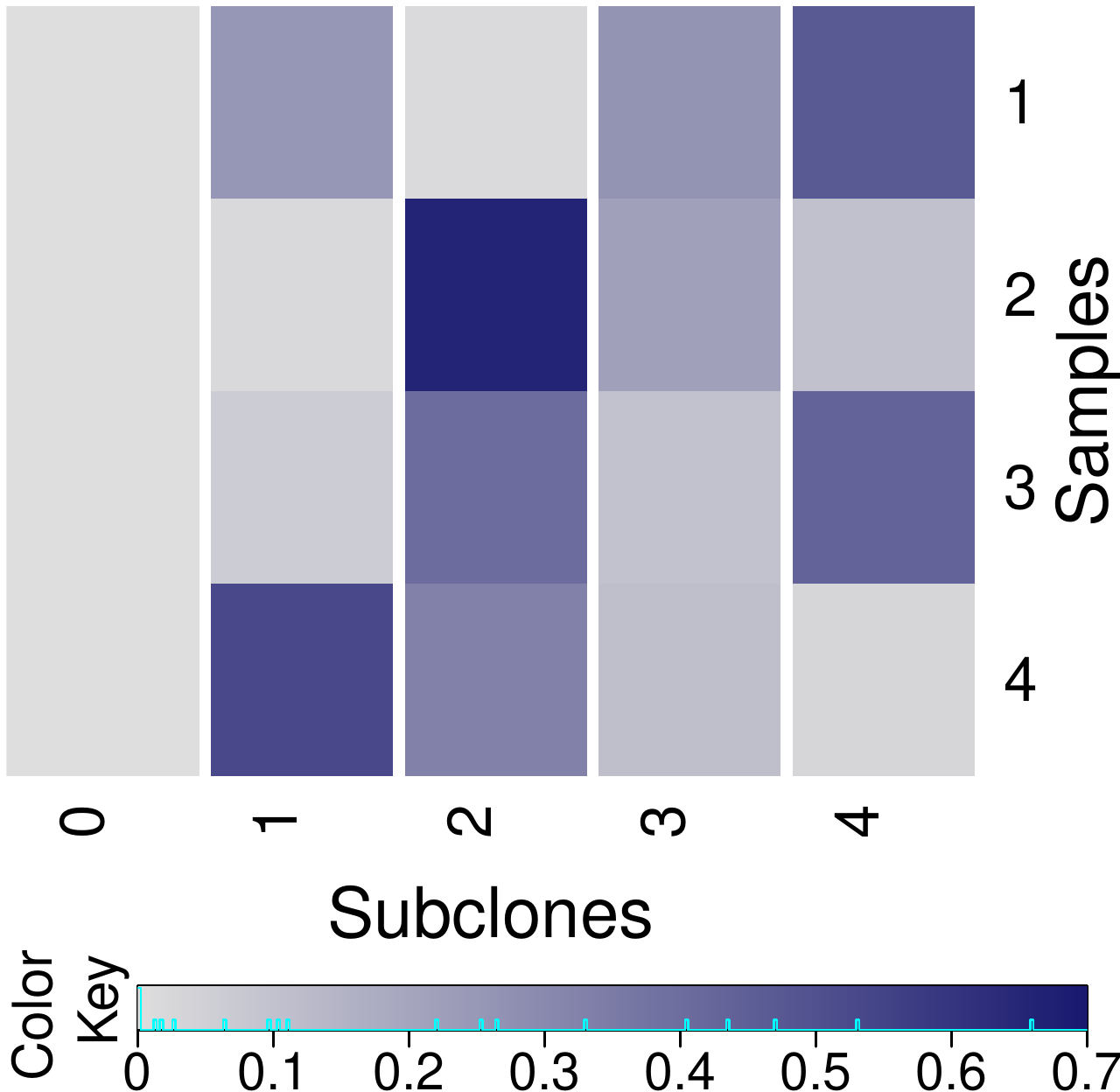}
      &
       \hspace{-1mm}\includegraphics[scale = 0.275]{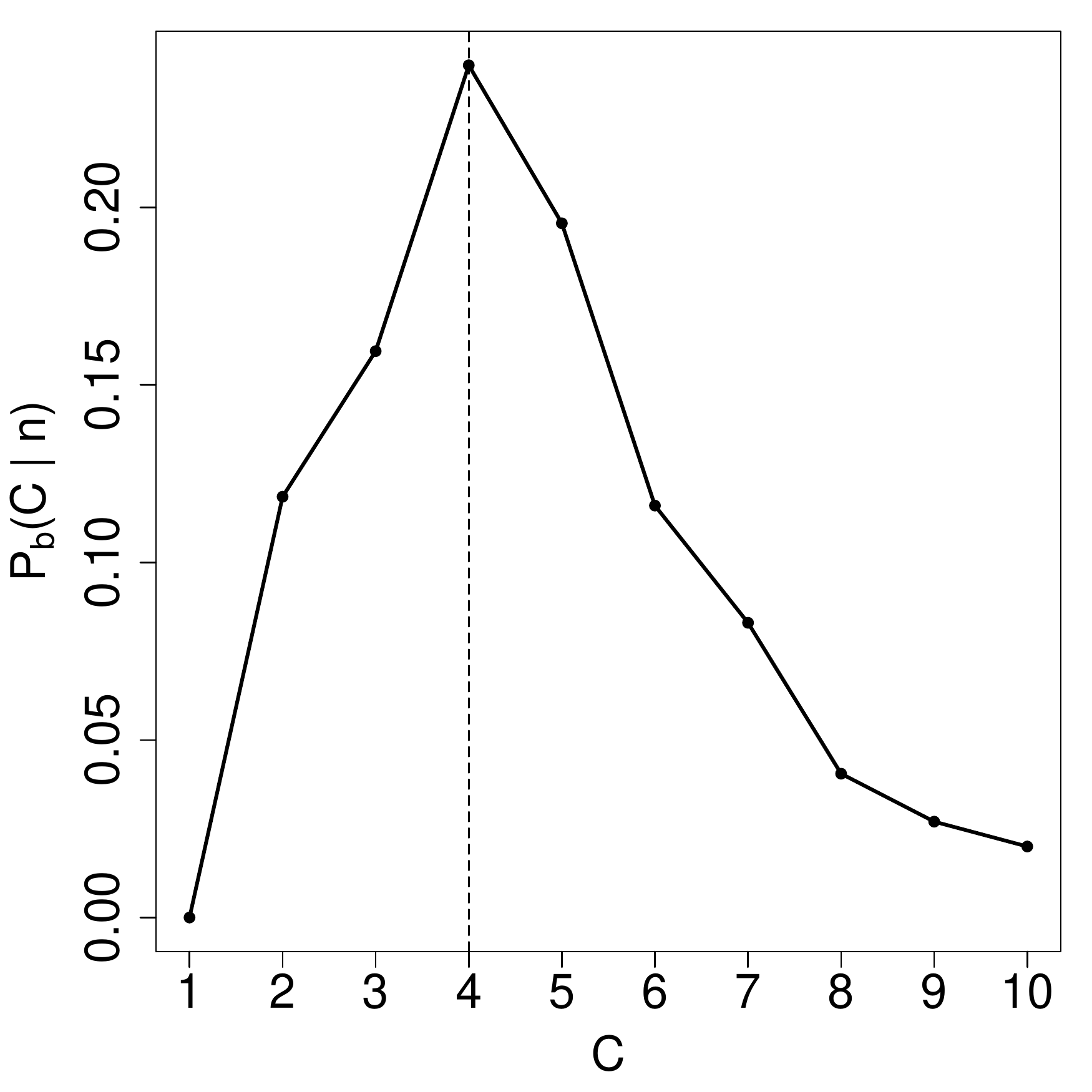}
      \\
      {\footnotesize{(a) $\bZ\true$}}
      &
      {\footnotesize{(b) $\bw\true$}}
      &
      {\footnotesize{(c) $p_b(C \mid \bn'')$}}
      \\
      \includegraphics[scale = 0.37]{./Zstar_sim2.pdf}
      &
      \includegraphics[scale = 0.37]{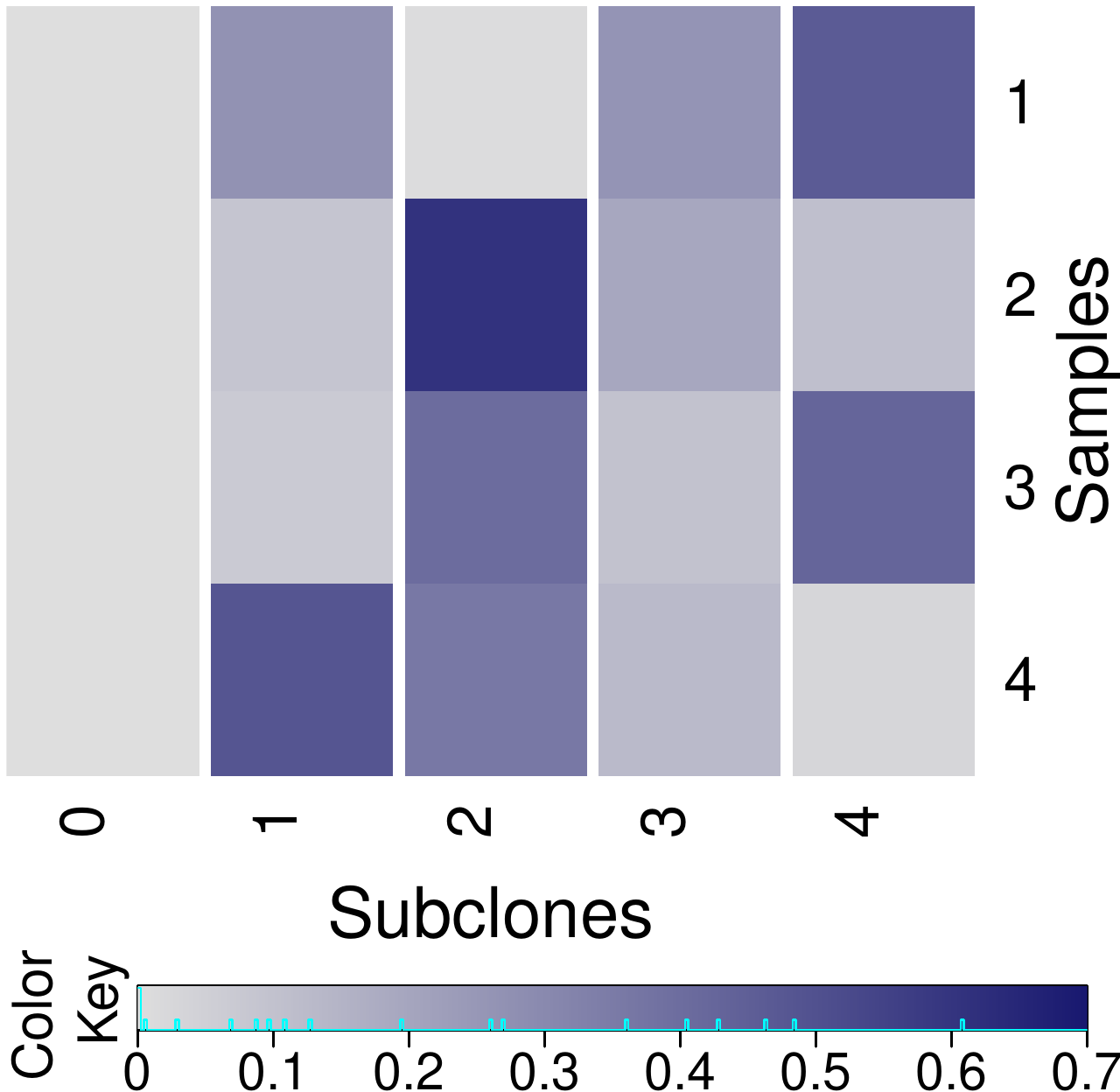}
      &
      \includegraphics[scale = 0.275]{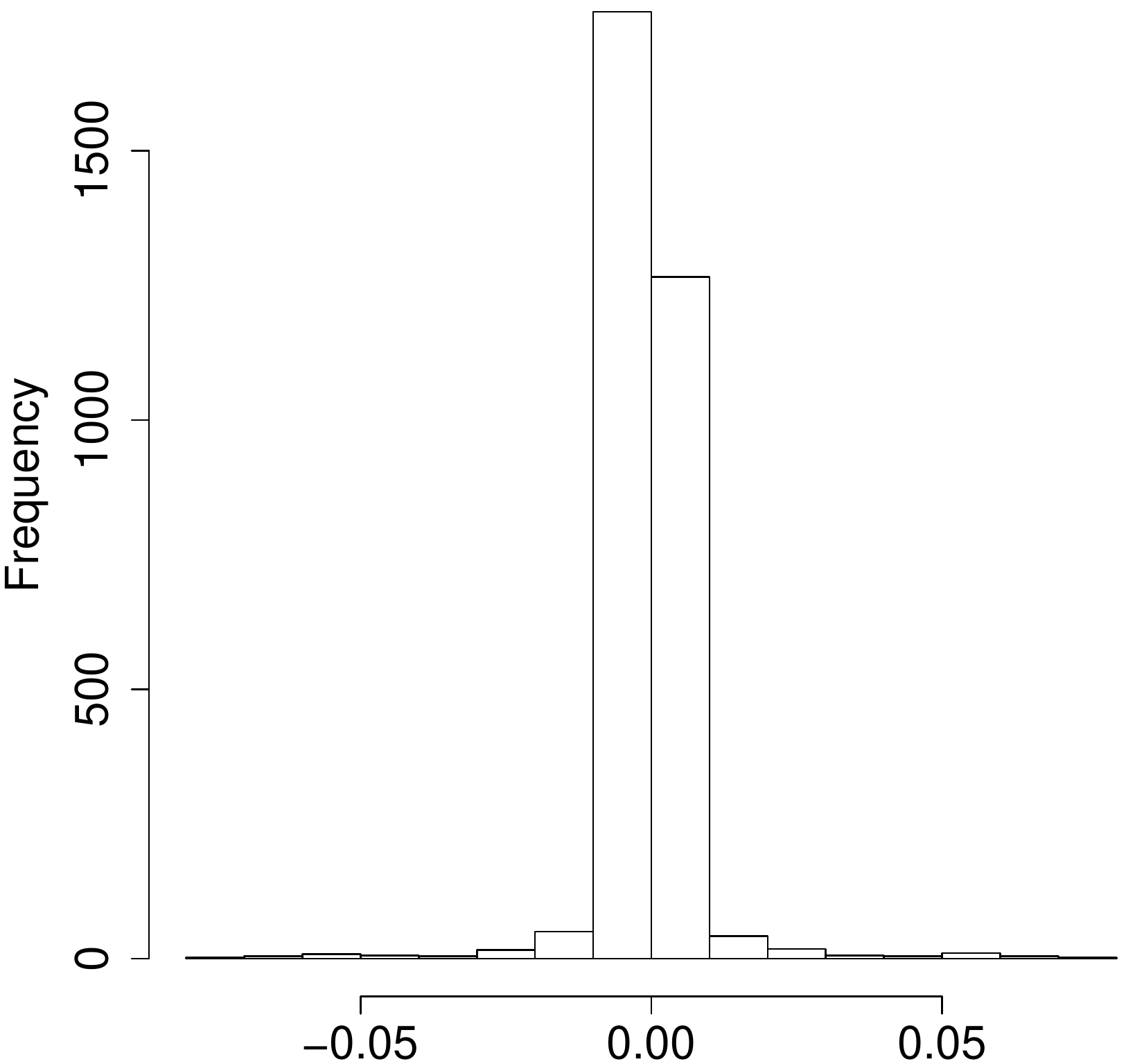}\\
      {\footnotesize{(d) $\Zhat$}}
      &
      {\footnotesize{(e) $\what$}}
      &
      {\footnotesize{(f) Histogram of $(\ptkghat - p_{tkg}\true)$}}
      \\
    \end{tabular}
  \end{center}
  \caption{Simulation 2. Simulation truth $\bZ\true$ and $\bw\true$ (a, b), and posterior
  inference under PairClone (c, d, e, f).}
  \label{app:figsim2}
\end{figure}

In the second simulation, we  consider   data
with $K = 100$ mutation pairs    and a more complicated subclonal
structure with $C\true = 4$ latent 
subclones. 
We generate hypothetical data for $T = 4$ samples. 
The subclone matrix $\bZ\true$ is shown in
Figure~\ref{app:figsim2}(a). Colors on a scale from light grey to red, to
black (see the scale in the figure) are used to represent genotype
$\bz^{(q)}$ with $q = 1, \ldots, 10$. For example, subclone 4 has
genotype $\bz^{(10)}$  for mutation pairs 1--20, $\bz^{(5)}$  for
mutation pairs 21--40, $\bz^{(8)}$  for mutation pairs 41--60,
$\bz^{(1)}$  for mutation pairs 61--80, and $\bz^{(9)}$ for mutation
pairs 81--100. 
For each sample $t$, we generate the subclone proportions from a
Dirichlet distribution, $\bw_t\true \sim \Dir(0.01, \sigma(20, 10, 5,
2))$, where $\sigma(20, 10, 5, 2)$ is a random permutation of $(20,
10, 5, 2)$.  
The subclone proportion matrix $\bw\true$ is shown in Figure
\ref{app:figsim2}(b), where darker blue color indicates higher abundance
of a subclone in a sample, and light grey color represents low
abundance.  
The parameters $\bm \rho\true$ and $N_{tk}$ are generated using the
same approach as before, and we use
$v_{tk2} = v_{tk3} = 0.3$ for $k = 1, \ldots, 50$ and all $t$, and
$v_{tk2} = v_{tk3} = 0.35$ for $k = 51, \ldots, 100$ and all $t$.
Finally, we calculate $\{p_{tkg}\true\}$ and generate read counts $n_{tkg}$ from equation (\ref{eq:multi}) similar to previous simulation.

We fit the model with the same set of hyperparameters and MCMC
parameters as in simulation 1. Figure~\ref{app:figsim2}(c) shows
$p_b(C \mid \bn'')$. Again, the posterior mode $\Chat = 4$ recovers
the truth. Figure~\ref{app:figsim2}(d) shows the estimate $\Zhat$; the
truth is nicely approximated. 
Some mismatches are
expected under this more complex subclone structure.
The estimated subclone proportions $\what$ are shown in Figure~\ref{app:figsim2}(e), again close to the
truth. Figure~\ref{app:figsim2}(f) shows the histogram of ($\ptkghat -
p_{tkg}\true$) which indicates a good model fit.

\begin{figure}[h!]
\begin{center}
\begin{tabular}{ccc}
      \includegraphics[scale = 0.37]{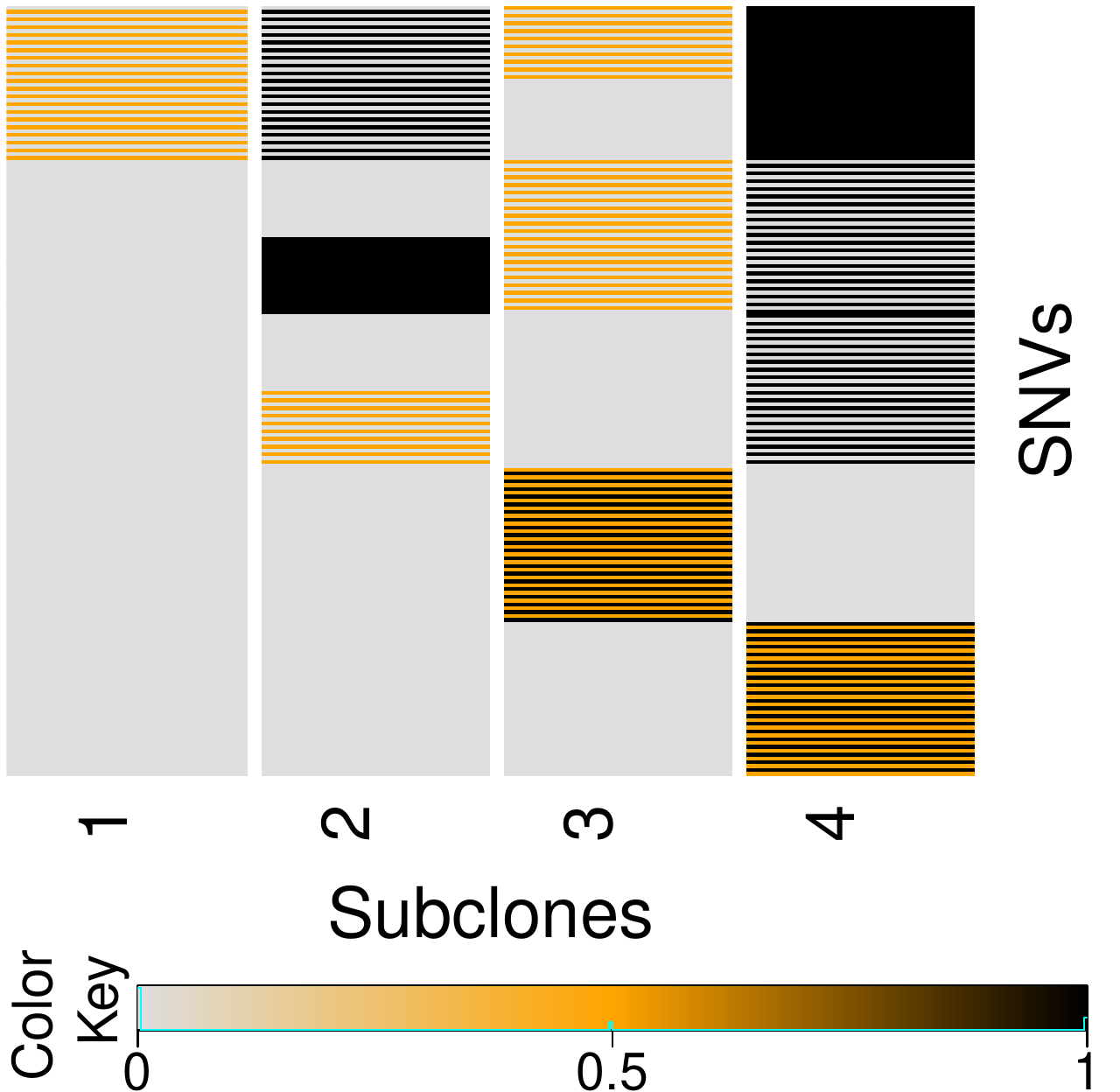}
      &
      \includegraphics[scale = 0.37]{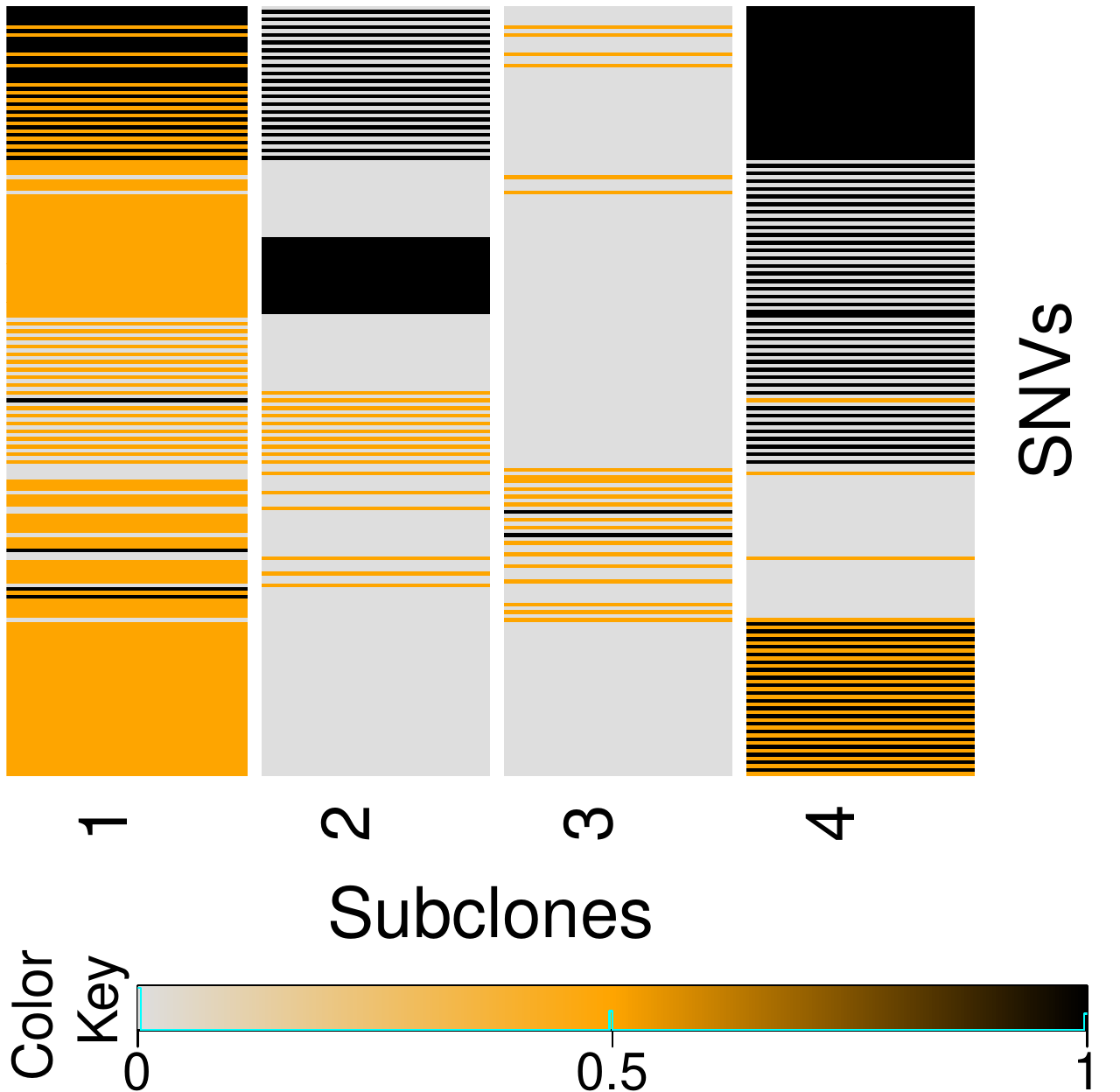}
      &
      \includegraphics[scale = 0.37]{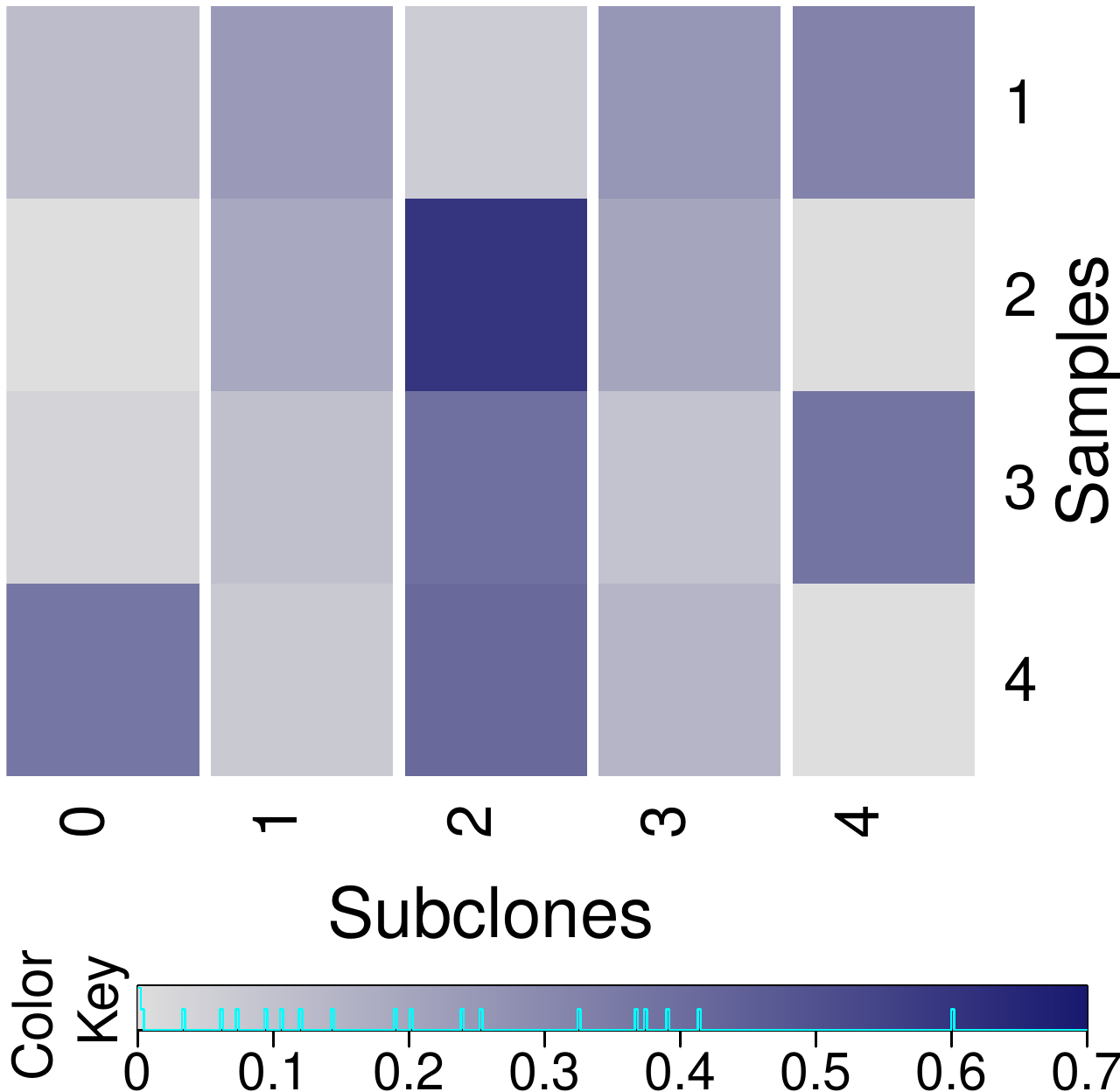}
      \\
      {\footnotesize{(a) $\bZ_{\BC}\true$}}
      &
      {\footnotesize{(b) $\Zhat_{\BC}$}}
      &
      {\footnotesize{(c) $\what_{\BC}$}}
\end{tabular}
\begin{tabular}{cc}
       \includegraphics[scale = 0.12]{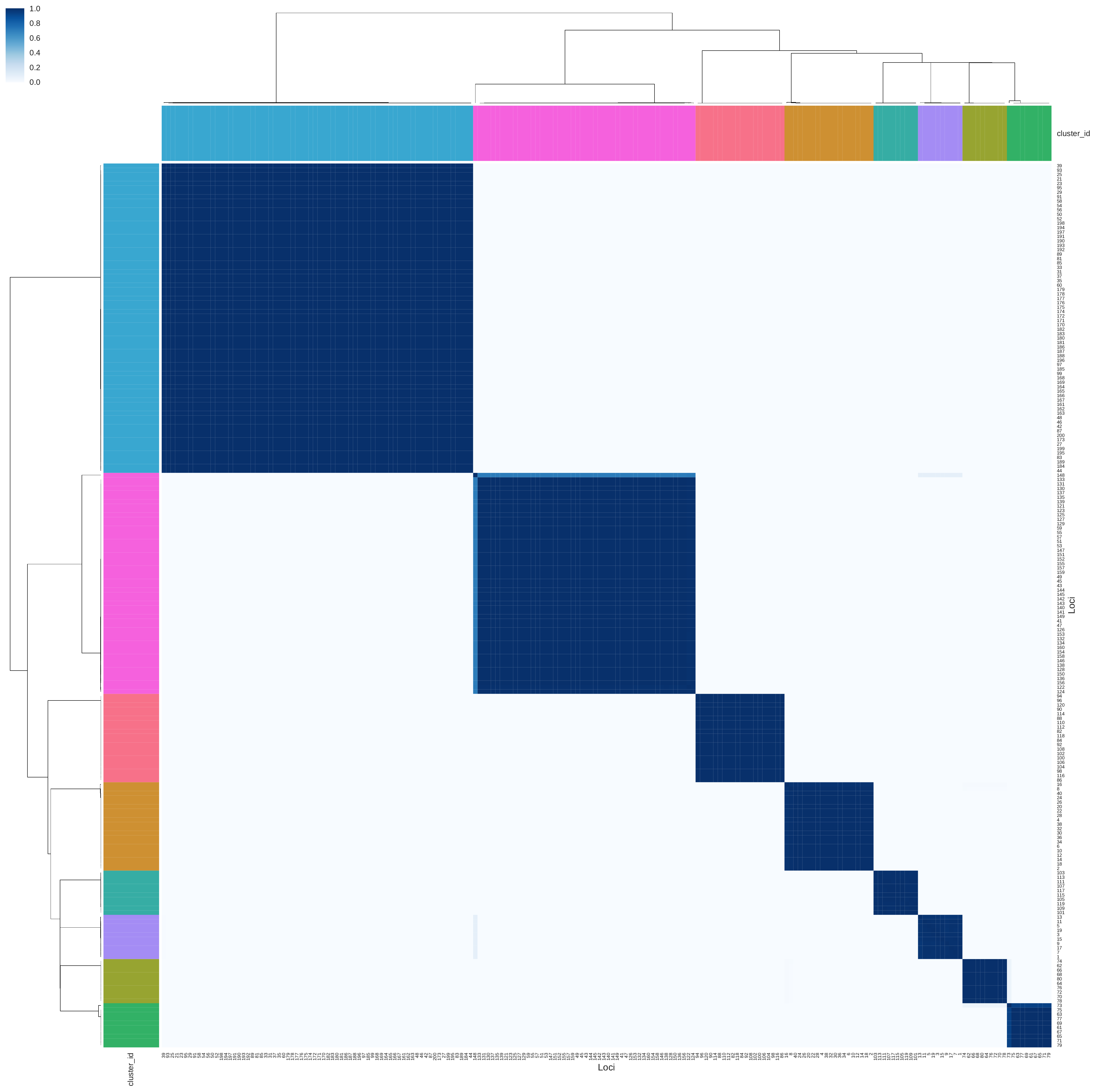}
    &
    \includegraphics[scale = 0.6]{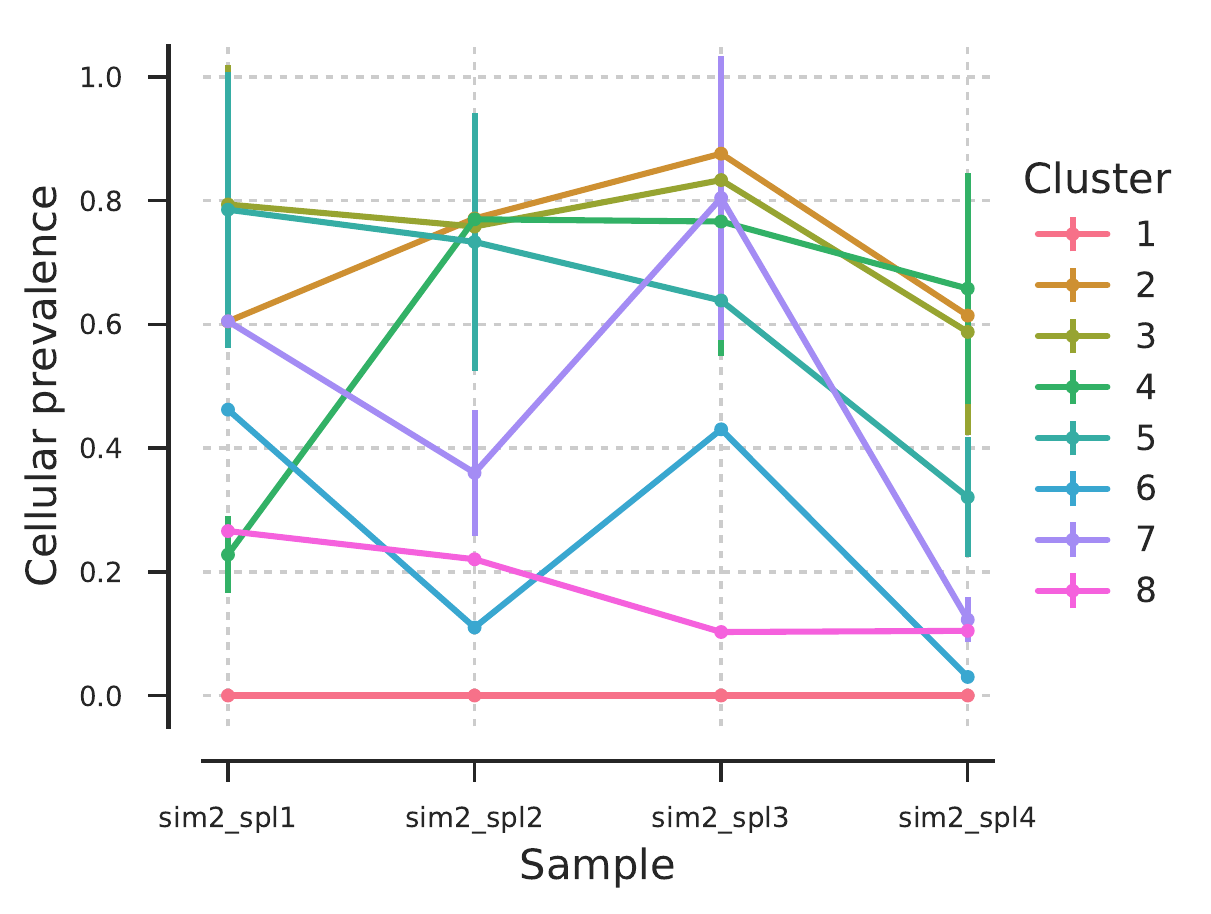}
    \\
    {\footnotesize{(d) Posterior similarity matrix}}
    &
    {\footnotesize{(e) Cellular prevalence}}
     \end{tabular}
  \end{center}
  \caption{Simulation 2. Posterior inference under BayClone (a, b, c) and PyClone (d, e).}
  \label{fig:sim2_BC}
\end{figure}

For comparison, we again fit the same simulated data with BayClone and
PyClone.  
\underline{BayClone} chooses the model with 4 subclones, which still
recovers the truth.  However, using only SNV data, BayClone can not
see the connection between adjacent SNVs, and inference fails to
recover $\bw_{\text{BC}}\true$ and therefore $\bZ_{\text{BC}}\true$,
even approximately.  
\underline{PyClone} infers 8 clusters for the 200 loci, which
reasonably recovers the truth. However, since the underlying subclone
structure is more complex, the PyClone cellular prevalence is not
directly comparable to PairClone outputs.

\subsubsection{Simulation 3}
\label{app:sim3}
\begin{figure}[h!]
  \begin{center}
    \begin{tabular}{ccc}
      \includegraphics[scale = 0.37]{./Z_sim3.pdf} &
      \includegraphics[scale = 0.37]{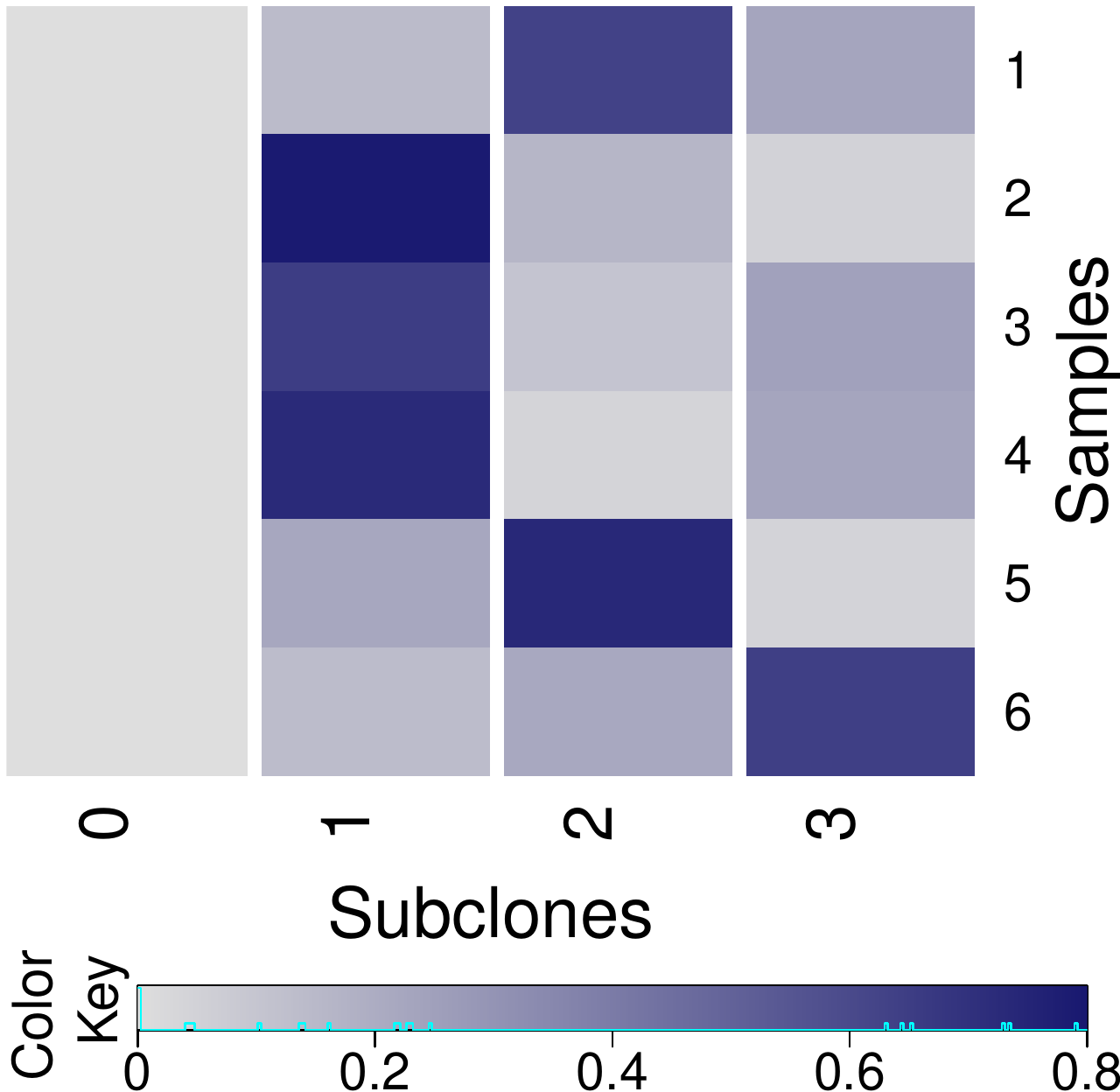} &
      \hspace{-1mm}\includegraphics[scale = 0.275]{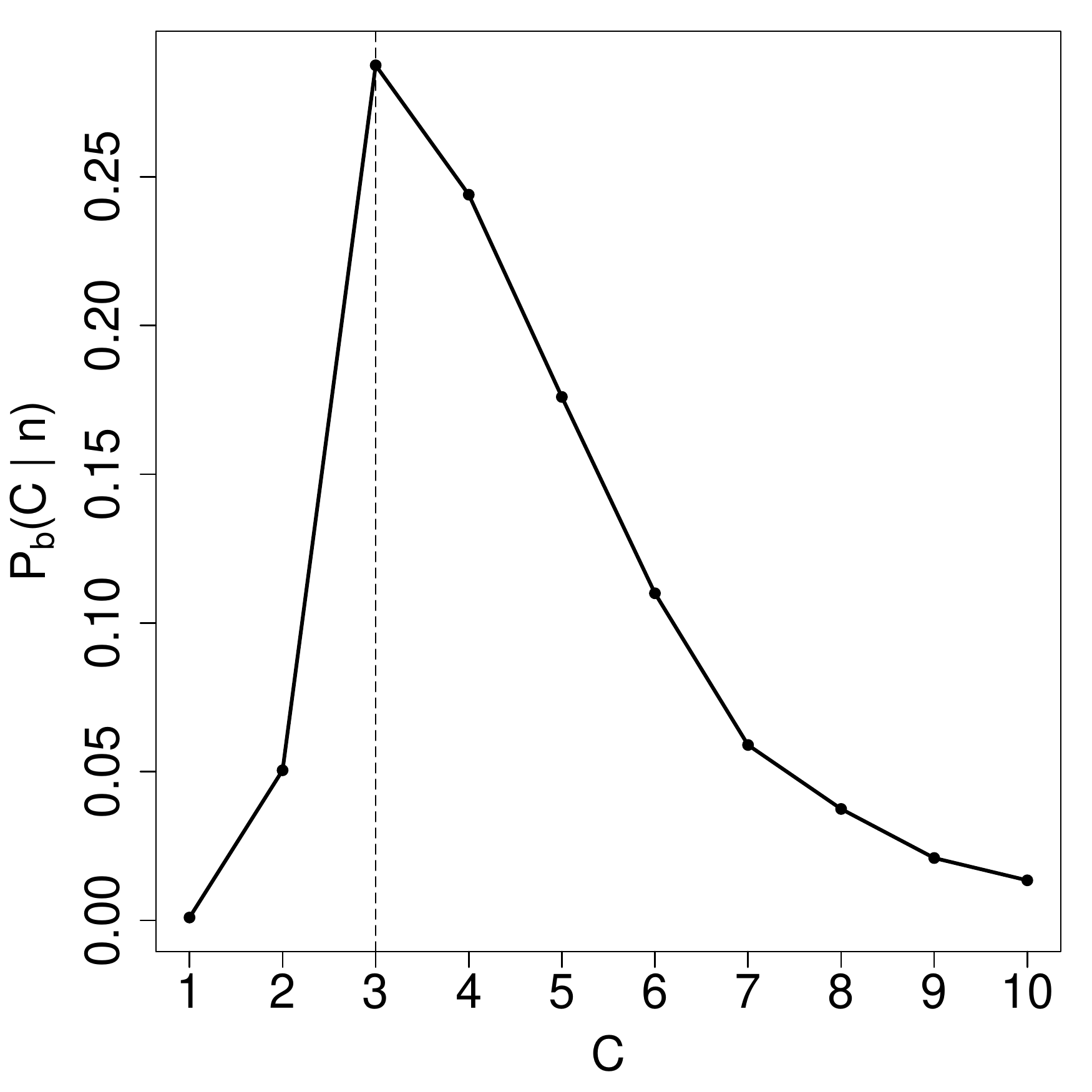} \\
      {\footnotesize{(a) $\bZ\true$}} & 
      {\footnotesize{(b) $\bw\true$}} &
      {\footnotesize{(c) $p_b(C \mid \bn'')$}} \\
      \includegraphics[scale = 0.37]{./Zstar_sim3.pdf} &
      \includegraphics[scale = 0.37]{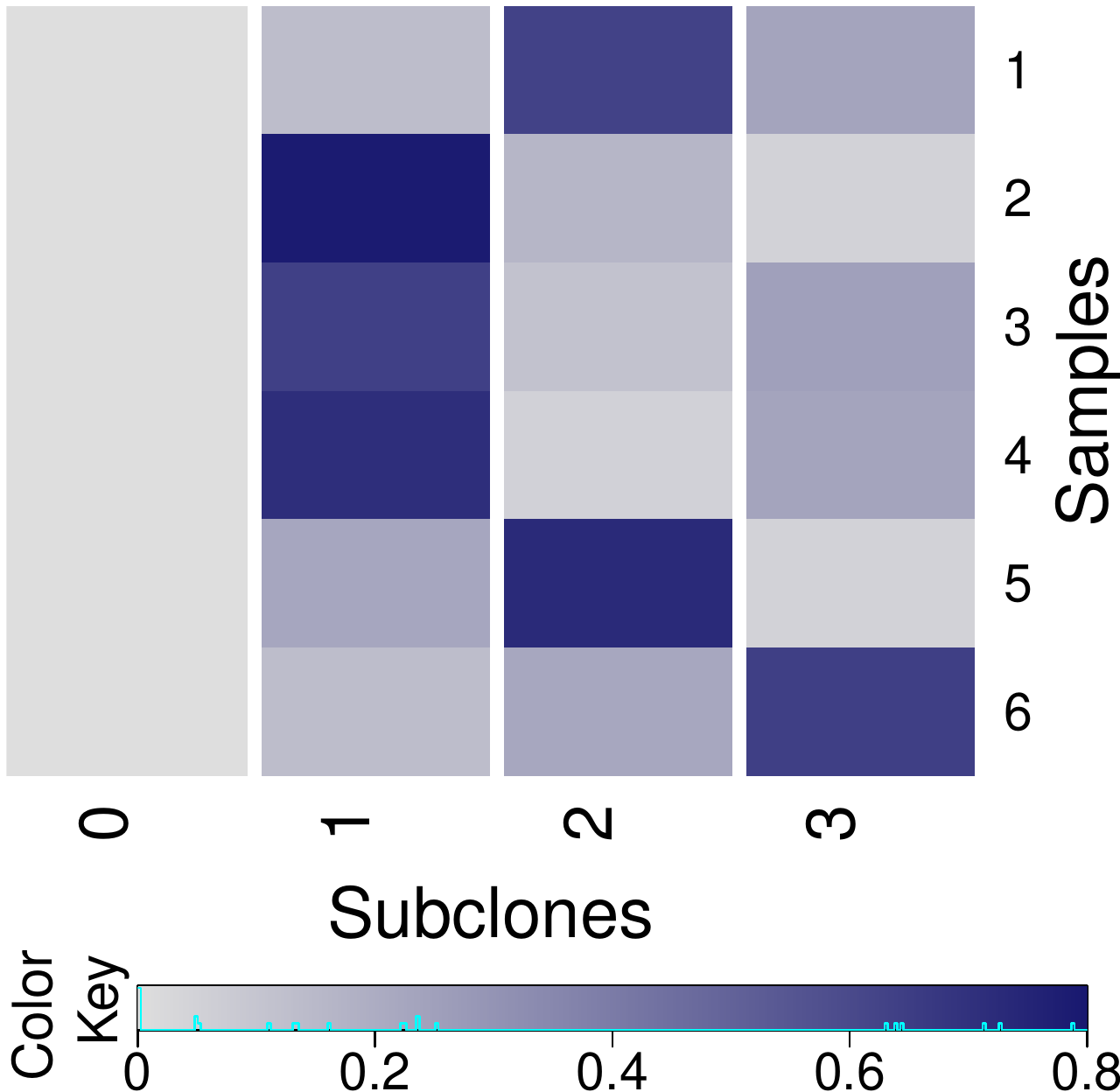} &
      \includegraphics[scale = 0.275]{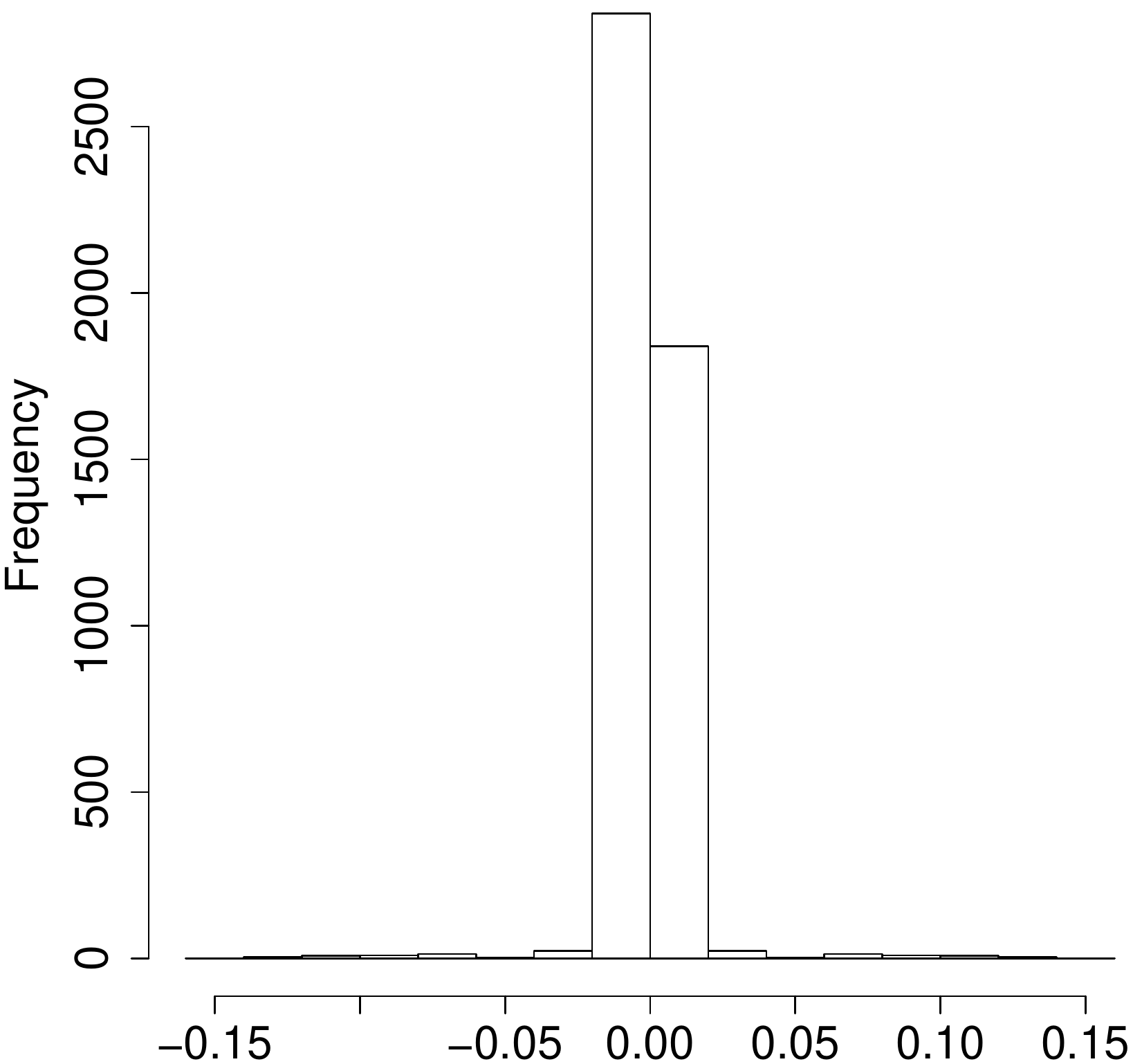}\\
      {\footnotesize{(d) $\Zhat$}} & 
      {\footnotesize{(e) $\what$}} &
      {\footnotesize{(f) Histogram of $(\ptkghat - p_{tkg}\true)$}} \\
    \end{tabular}
  \end{center}
  \caption{Simulation 3. Simulation truth $\bZ\true$ and $\bw\true$ (a, b), and posterior inference under PairClone (c, d, e, f).}
  \label{app:figsim3}
\end{figure}

In the last simulation we use  
$T = 6$ samples with $C^{\text{TRUE}} = 3$ and latent subclones.  
We still consider $K = 100$ mutation pairs.
The subclone matrix $\bZ\true$ is shown in Figure~\ref{app:figsim3}(a). For
each sample $t$, we generate the subclone proportions from
$\bw_t\true \sim \Dir(0.01, \sigma(14, 6, 3))$, where $ \sigma (14,
6, 3)$ is a random permutation of $(14, 6, 3)$. The proportions
$\bw\true$ are shown in
Figure~\ref{app:figsim3}(b).  The parameters $\brho\true$ and
$N_{tk}$ are generated using the same approach as before, and we use the same $v_{tk2}$ and $v_{tk3}$ as in Simulation 2.
Finally, we calculate $\{p_{tkg}\true\}$ and generate read counts
$n_{tkg}$ from equation \eqref{eq:multi} similar to simulation 1.

We fit the model with the same hyperparameter and the same MCMC
tuning parameters as in simulation 1.  We now use a smaller test
sample size, i.e., a smaller fraction 
$b$   in the transdimensional MCMC.  
See Section \ref{app:sec:calib} for a discussion.

Figure \ref{app:figsim3}(c) shows $p_b(C \mid \bn'')$, with the
posterior mode $\Chat = 3$ recovering the truth.
Figures~\ref{app:figsim3}(d, e) show $\Zhat$ and $\what$.
Comparing with panels (a) and (b) we can see an almost perfect
recovery of the truth. 
Figure~\ref{app:figsim3}(f) shows a histogram of
the residuals $(\ptkghat - p_{tkg}\true)$. The plot indicates a good model fit.

\begin{figure}[h!]
  \begin{center}
    \begin{tabular}{ccc}
      \includegraphics[scale = 0.37]{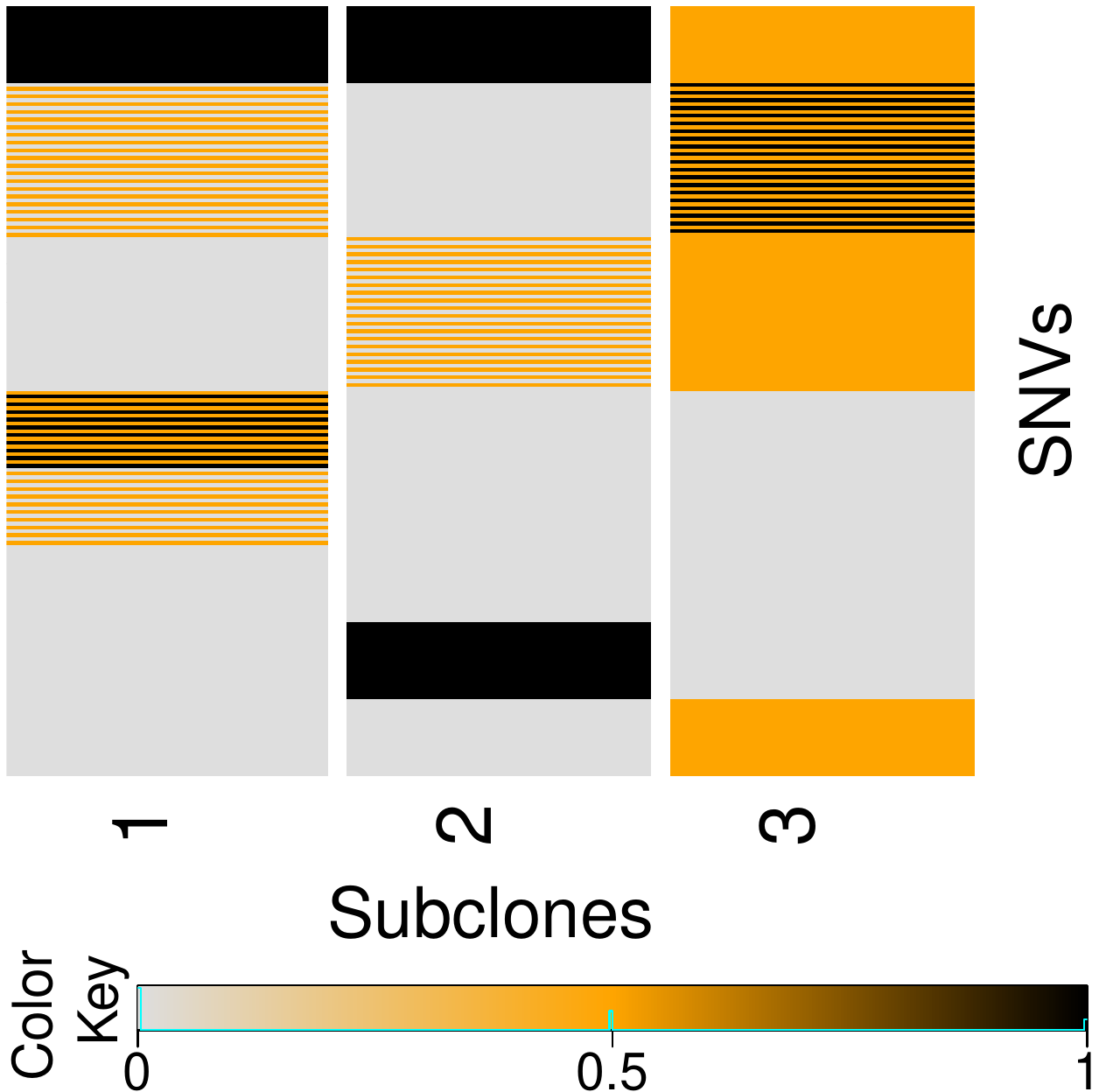}
      &
      \includegraphics[scale = 0.37]{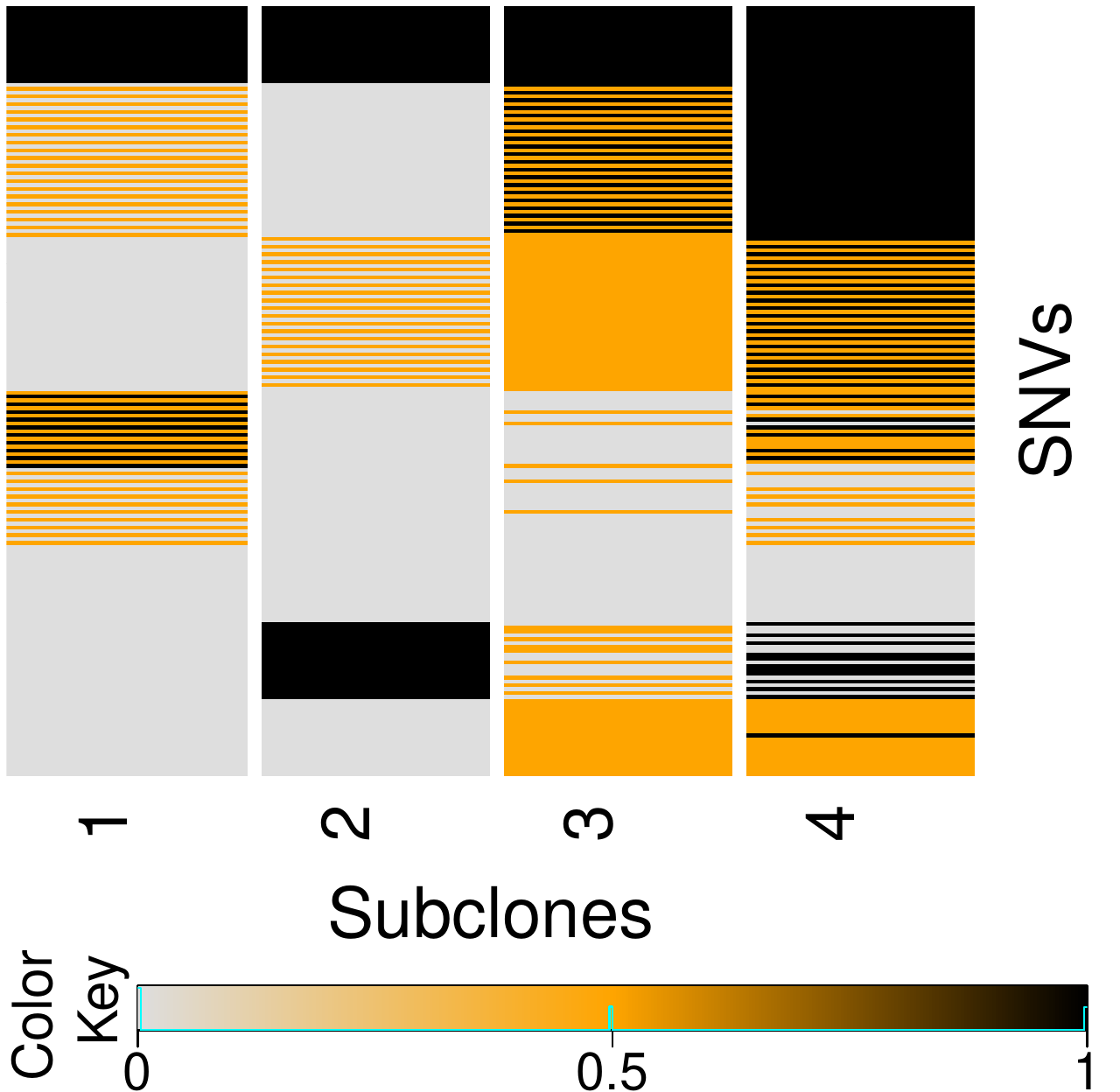}
      &
      \includegraphics[scale = 0.37]{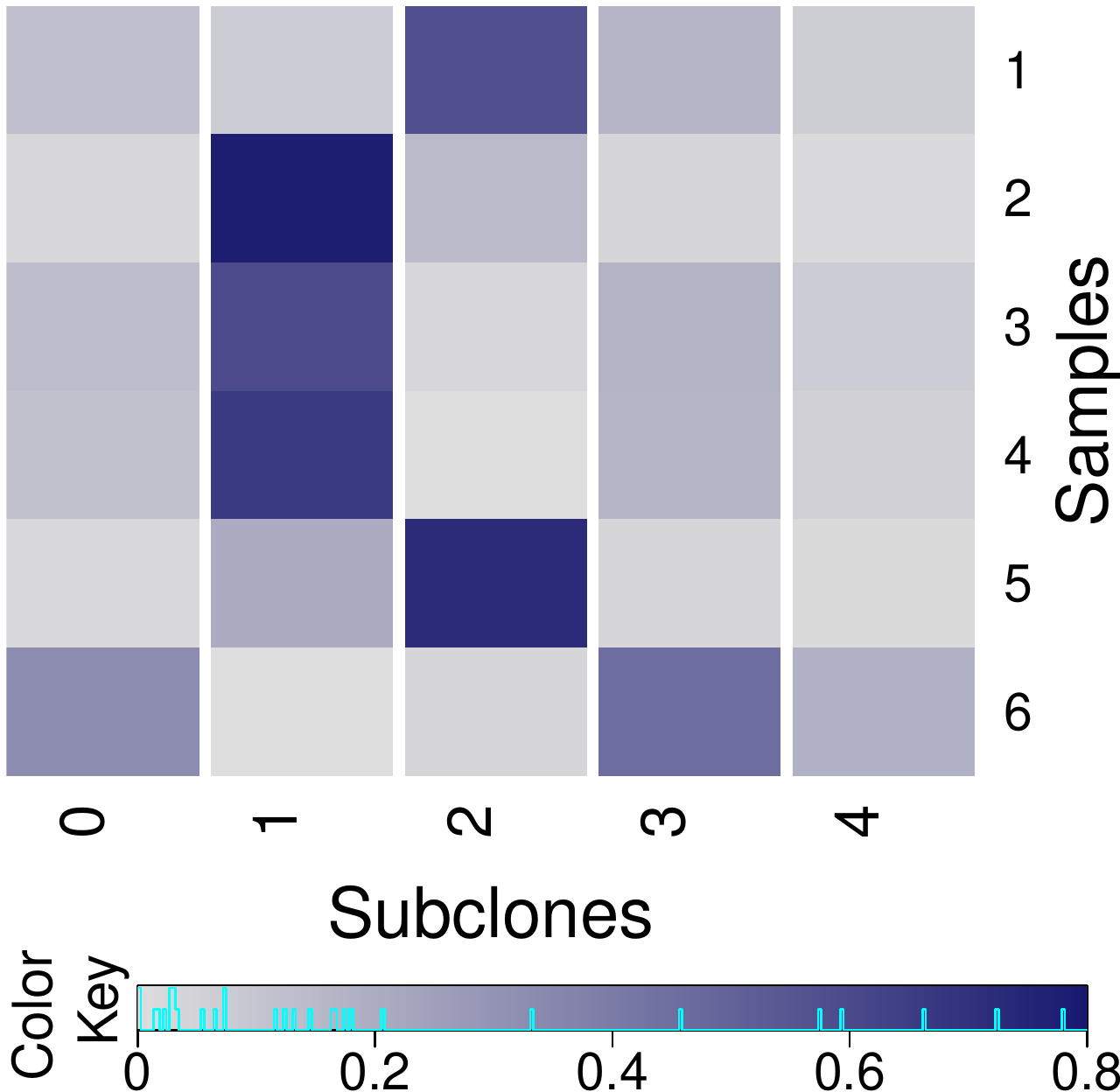}
      \\
      {\footnotesize{(a) $\bZ_{\BC}\true$}}
      &
      {\footnotesize{(b) $\Zhat_{\BC}$}}
      &
      {\footnotesize{(c) $\what_{\BC}$}}
    \end{tabular}
    \begin{tabular}{cc}
       \includegraphics[scale = 0.12]{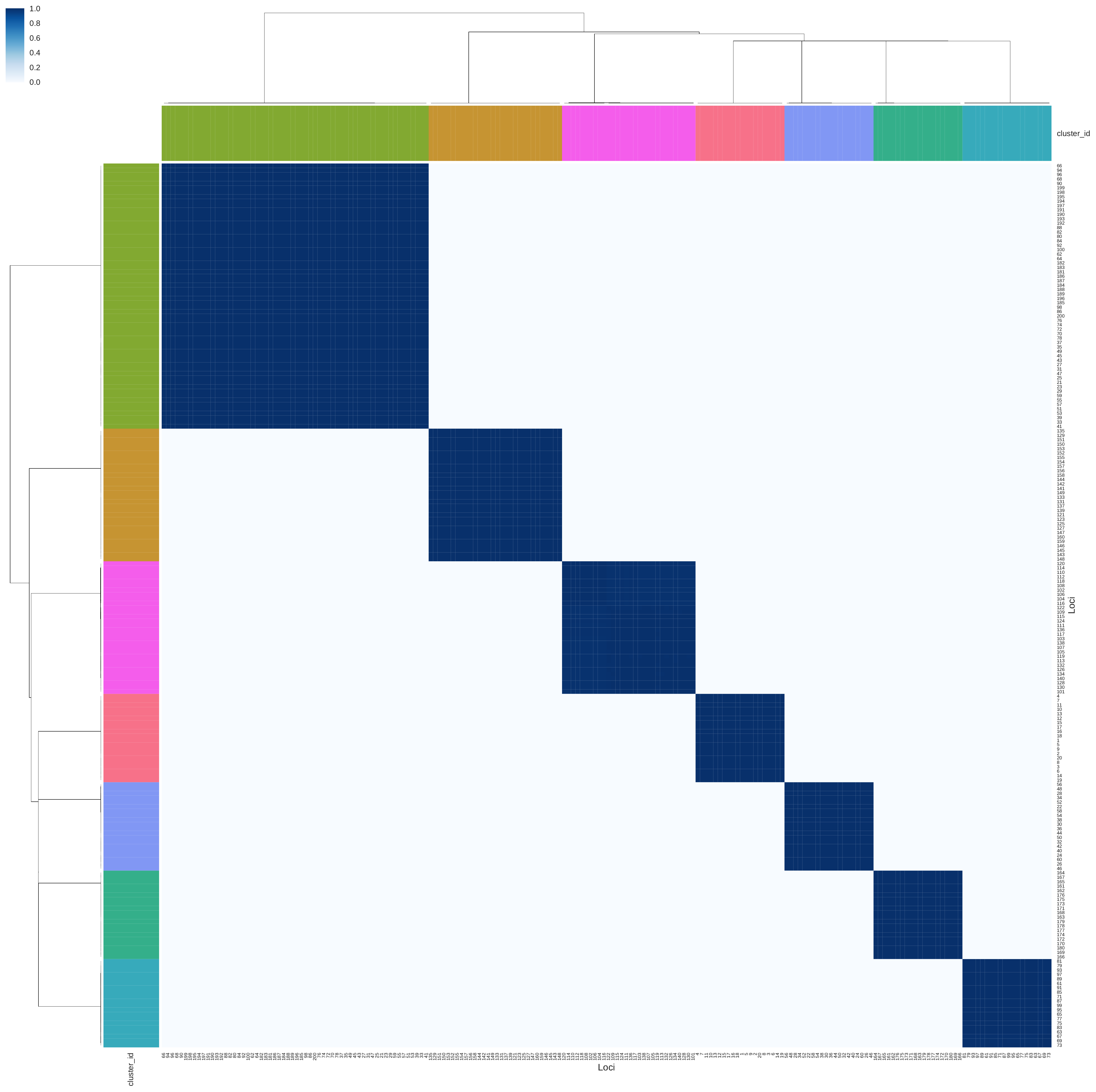}
    &
    \includegraphics[scale = 0.6]{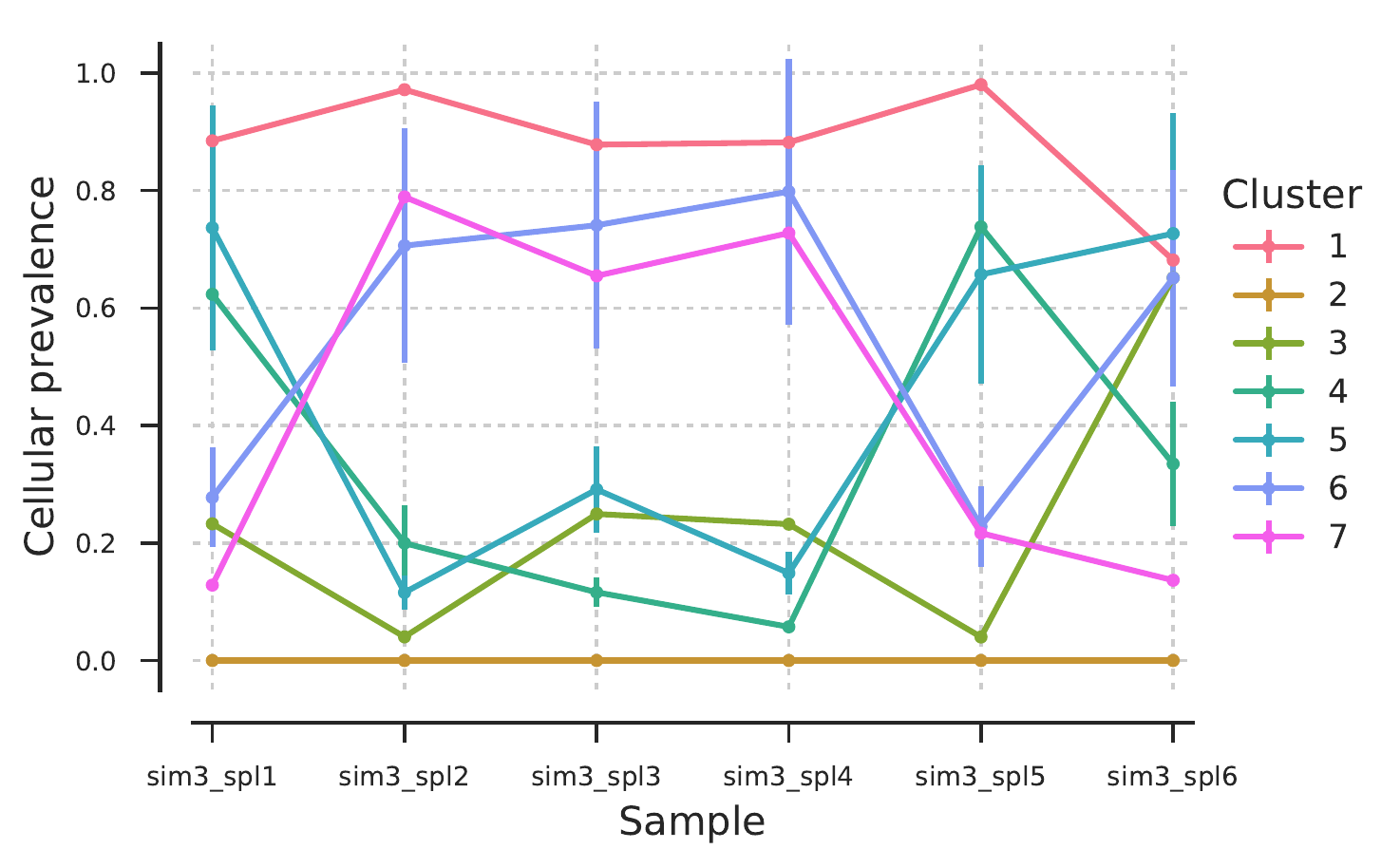}
    \\
    {\footnotesize{(d) Posterior similarity matrix}}
    &
    {\footnotesize{(e) Cellular prevalence}}
     \end{tabular}
  \end{center}
  \caption{Simulation 3. Posterior inference under BayClone (a, b, c) and PyClone (d, e).}
  \label{fig:sim3_BC}
\end{figure}

We again compare with inference under BayClone and PyClone.
In this case, \underline{BayClone} chooses the model with 4 subclones,
failing to recover the truth.
\underline{PyClone} infers 7 clusters for the 200 loci, which
reasonably recovers the truth, but the result is still not directly
comparable. 

\subsection{Simulation with tumor purity incorporated}
\label{app:sim_purity}

\begin{figure}[h!]
\begin{center}
\begin{subfigure}[t]{.3\textwidth}
\centering
\includegraphics[width=\textwidth]{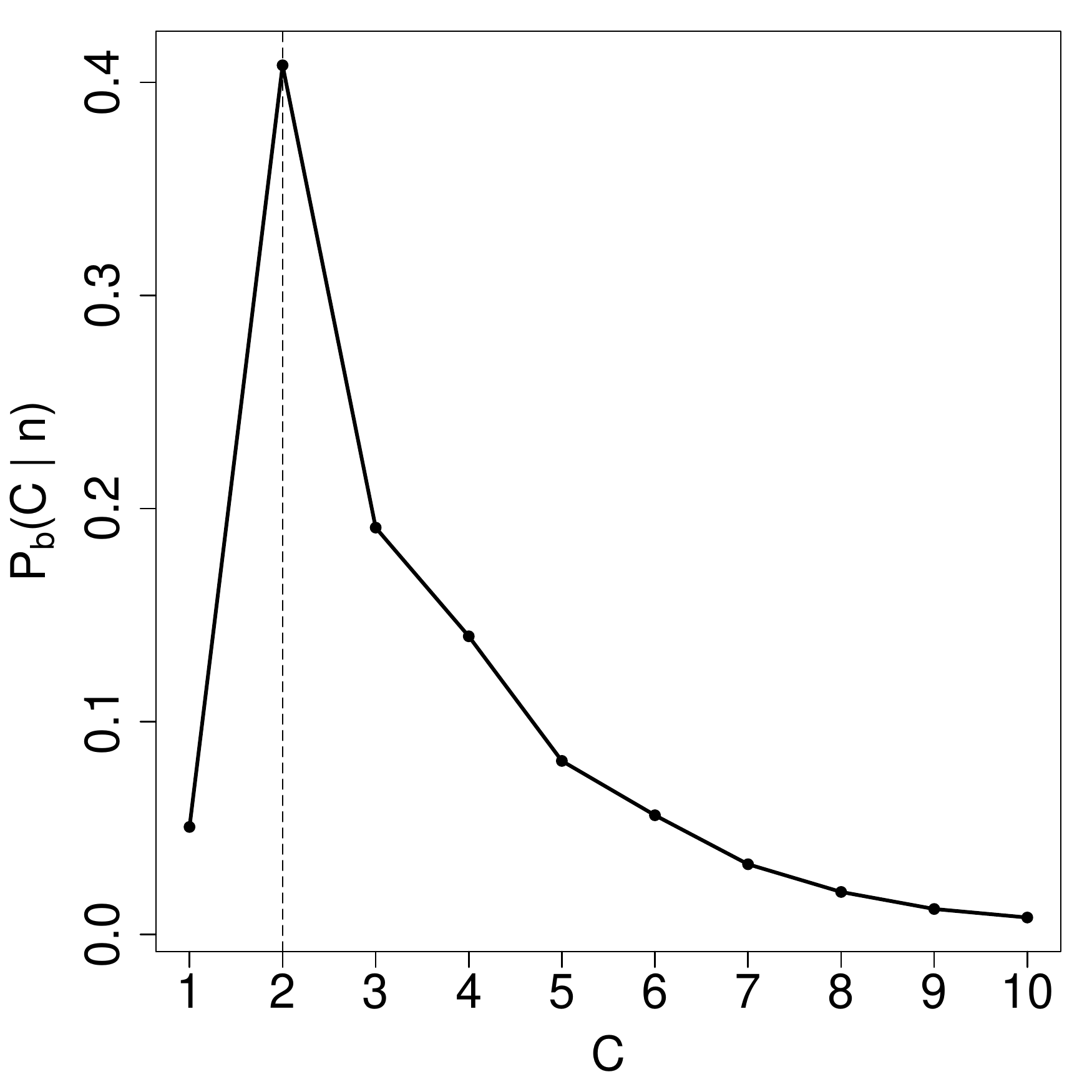}
\caption{$p_b(C \mid \bm n'')$}		
\end{subfigure}
\hspace{2mm}\begin{subfigure}[t]{.3\textwidth}
\centering
\includegraphics[width=\textwidth]{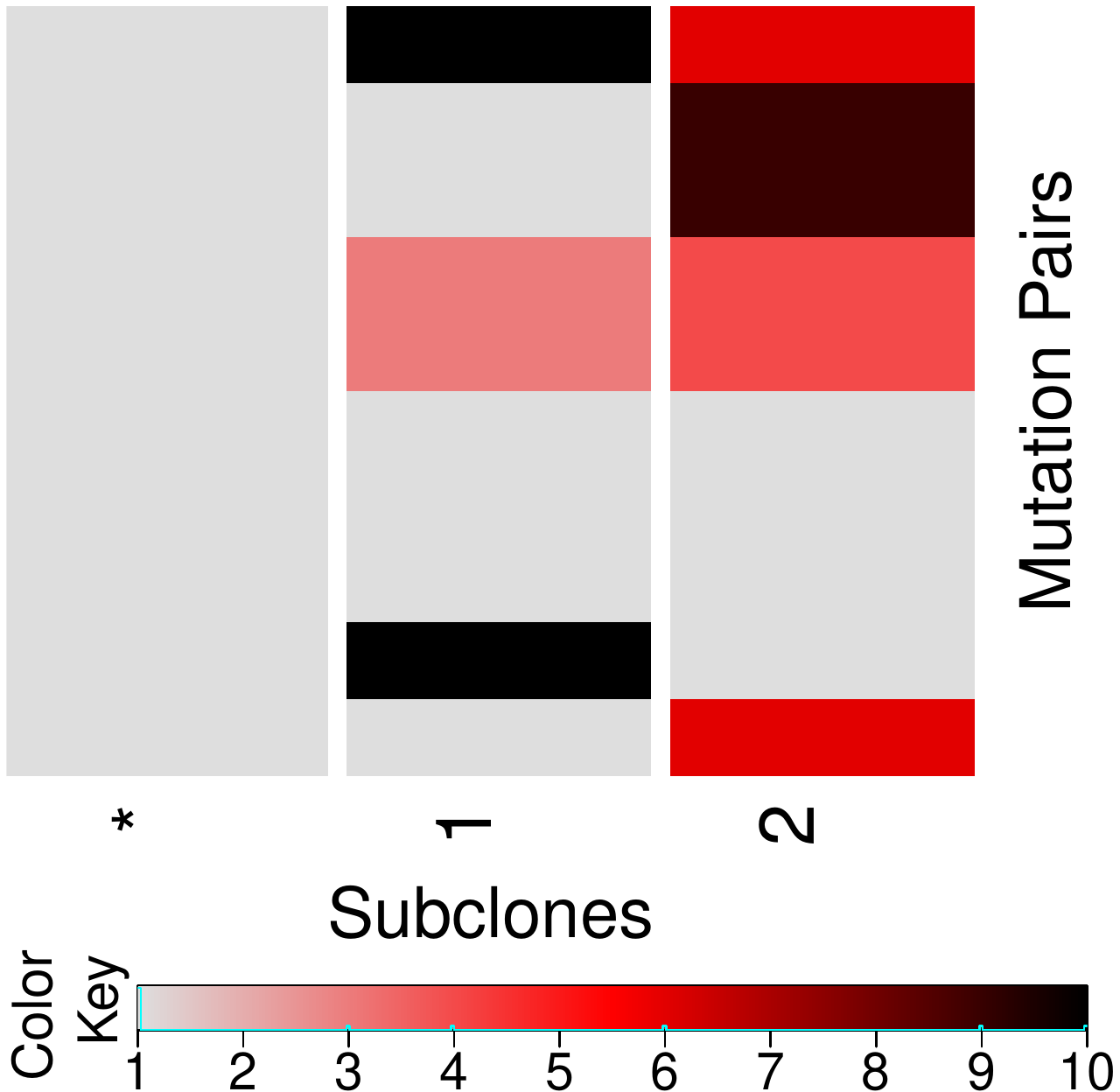}
\caption{$\Zhat$}		
\end{subfigure}
\begin{subfigure}[t]{.3\textwidth}
\centering
\includegraphics[width=\textwidth]{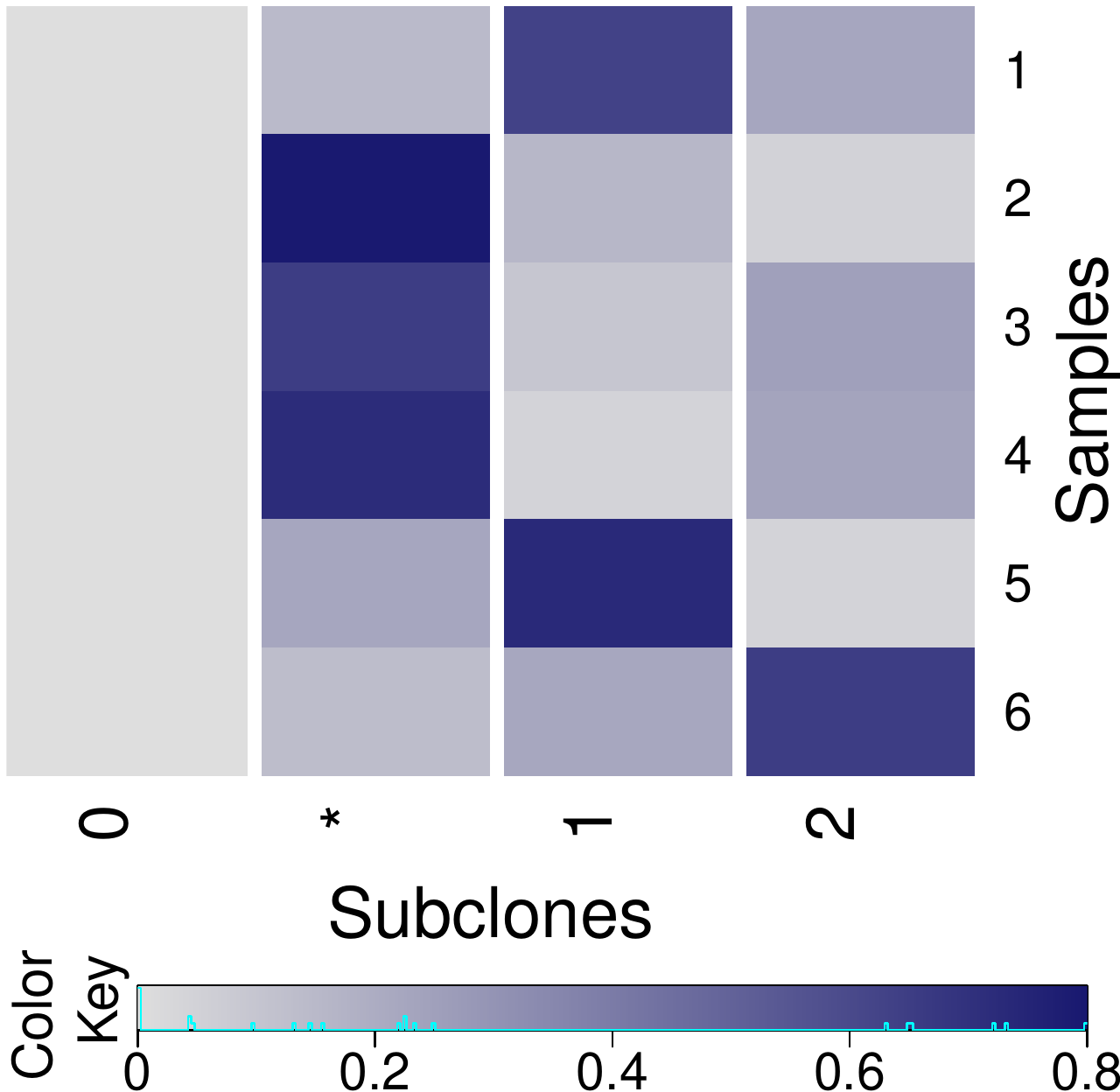}
\caption{$\what$}		
\end{subfigure}
\end{center}
\caption{Summary of simulation results with tumor purity incorporated.}
\label{app:figsimpurity}
\end{figure}

We report simulation details of the simulation study with tumor purity incorporated in the manuscript (Section \ref{sec:purity}).
The simulation setting is the same as simulation 3 in Section \ref{app:sim3},  except that we substitute the first subclone with a normal subclone. We use exactly the same hyperparameters  as those in simulations 2 and 3, and in addition we take $d_1^* = d_2^* = 1$.
Figure \ref{app:figsimpurity} summarizes inference results.
Columns in panels (b) and (c) marked with ``*'' correspond to the
normal subclone.  Panel (a) shows $p_b(C \mid \bm n'')$.
Posterior inference recovers the simulation truth, with 
posterior mode $\Chat = 2$.
Panel (b) shows $\Zhat$. Comparing with subclones 2 and 3 in
Figure \ref{app:figsim3}(a) we find a good recovery of the simulation
truth.
Panel (c) shows $\what$, which can be compared with Figure \ref{app:figsim3}(b). 

\subsection{Calibration of $b$}
\label{app:sec:calib}

The construction of an informative prior
$p_b(\bx \mid C) \equiv p(\bx \mid \bn', C)$
based on a training sample $\bn'$ is similar to the use of a training
sample in the construction of the fractional Bayes factor (FBF) of
\cite{ohagan1995}.
However, there is an important difference. In the FBF construction the
aim is to replace a noninformative prior in the evaluation of a Bayes
factor. A minimally informative prior $p_b$ with small $b$ suffices.
In contrast, here $p_b(\bx \mid C)$ is (also) used as proposal
distribution in the trans-dimensional MCMC. The aim is to construct a
good proposal that fits the data well and thus leads to good
acceptance probabilities and a well mixing Markov chain.
With the highly informative multinomial likelihood we find that we
need a large training sample, that is, large $b$.
In Appendix \ref{app:sec:updatec} we show that the effect of using
$p_b$ is that $p(C \mid \bn)$ is approximated by
$$
   p_b(C \mid \bn'') \propto
   p(C) p(\bn \mid \umle, C)^{1 - b} b^{\;   p_{C} / 2},
$$
where $\bu = \{\bm \pi, \bw, \bm \rho\}$ are the parameters other than
$\bZ$, $\umle$ is the maximum likelihood estimate of $\bu$, and $p_C$
is the number of unconstrained parameters in $\bu$.
Importantly, however, inference on other parameters, $p(\bx \mid
C,\bn)$, remains entirely unchanged.

\begin{figure}[h!]
\begin{center}
\includegraphics[scale = 0.43]{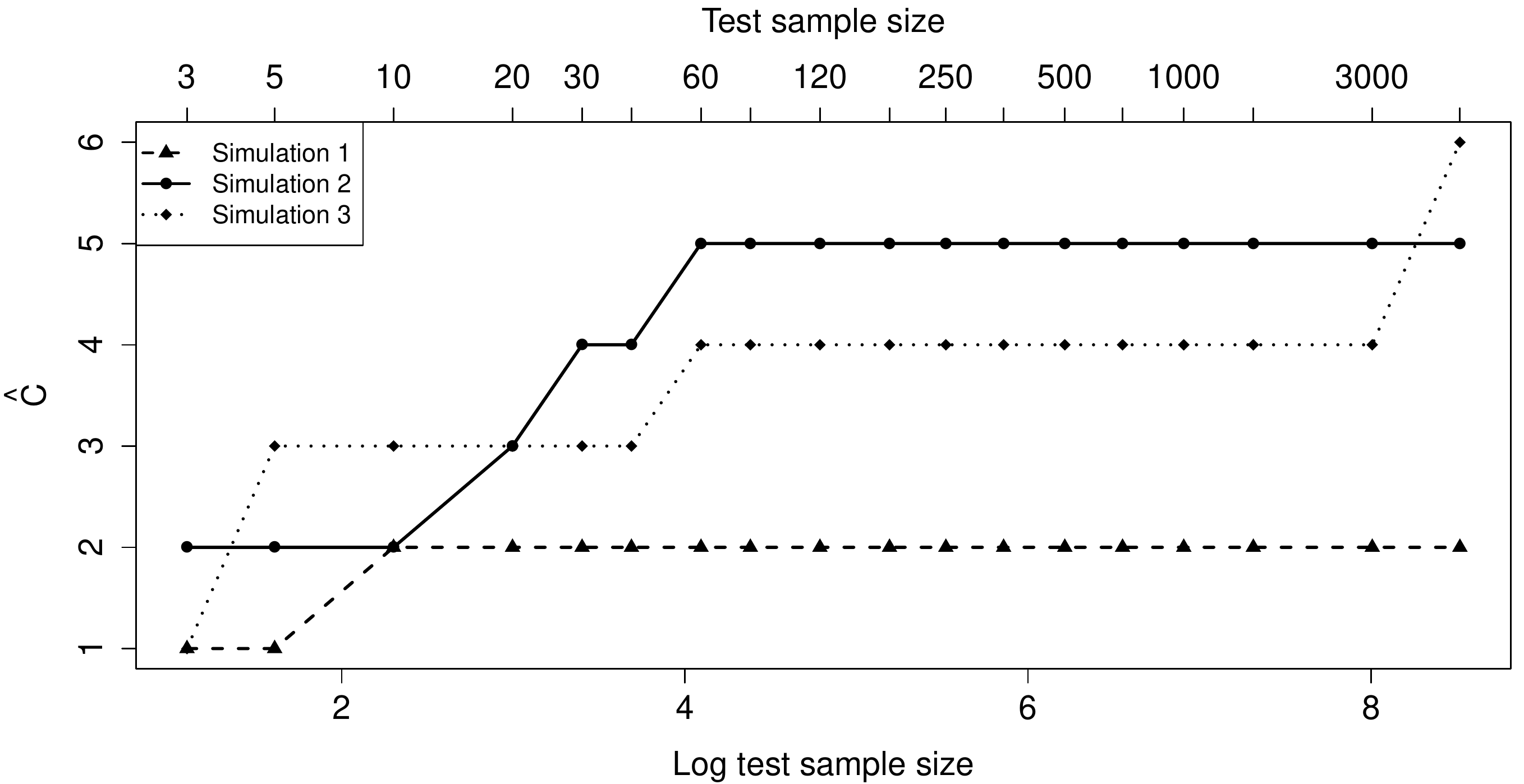}
\end{center}
\caption{Path plot of $\Chat$ with different test sample sizes for
   three simulations. The true number of subclones are 2, 4, and 3
  for simulations 1, 2, and 3, respectively. 
}
\label{fig:cpath}
\end{figure}

We therefore recommend to focus on inference for $C$ when calibrating
$b$. 
Carrying out simulation studies with single and multi-sample data, 
we find that the simulation truth for $C$ is best recovered with
a test sample size $(1-b) \sum_{t=1}^T \sum_{k=1}^K N_{tk} \approx 160 / T$, where
$N_{tk}$ is the total number of short reads mapped to mutation pair
$k$ in sample $t$.
For example, Figure \ref{fig:cpath} plots the posterior mode of $C$
against test sample sizes for simulated data in three simulations.
For multi-sample data we find (empirically, by simulation) that
the test sample size can be reduced, at a rate linear in
$T$. In summary we recommend to set $b$ to achieve a test sample
size around $160/T$. 
Following these guidelines, in our implementation in the previous
section, we used values $b = 0.992$ for simulation 1,  $b = 0.9998$
for simulation 2, and $b = 0.999911$ for simulation 3.

\subsection{Lung Cancer Data Analysis Plots}
\begin{figure}[h!]
\begin{center}
\begin{subfigure}[t]{.3\textwidth}
\centering
\includegraphics[width=\textwidth]{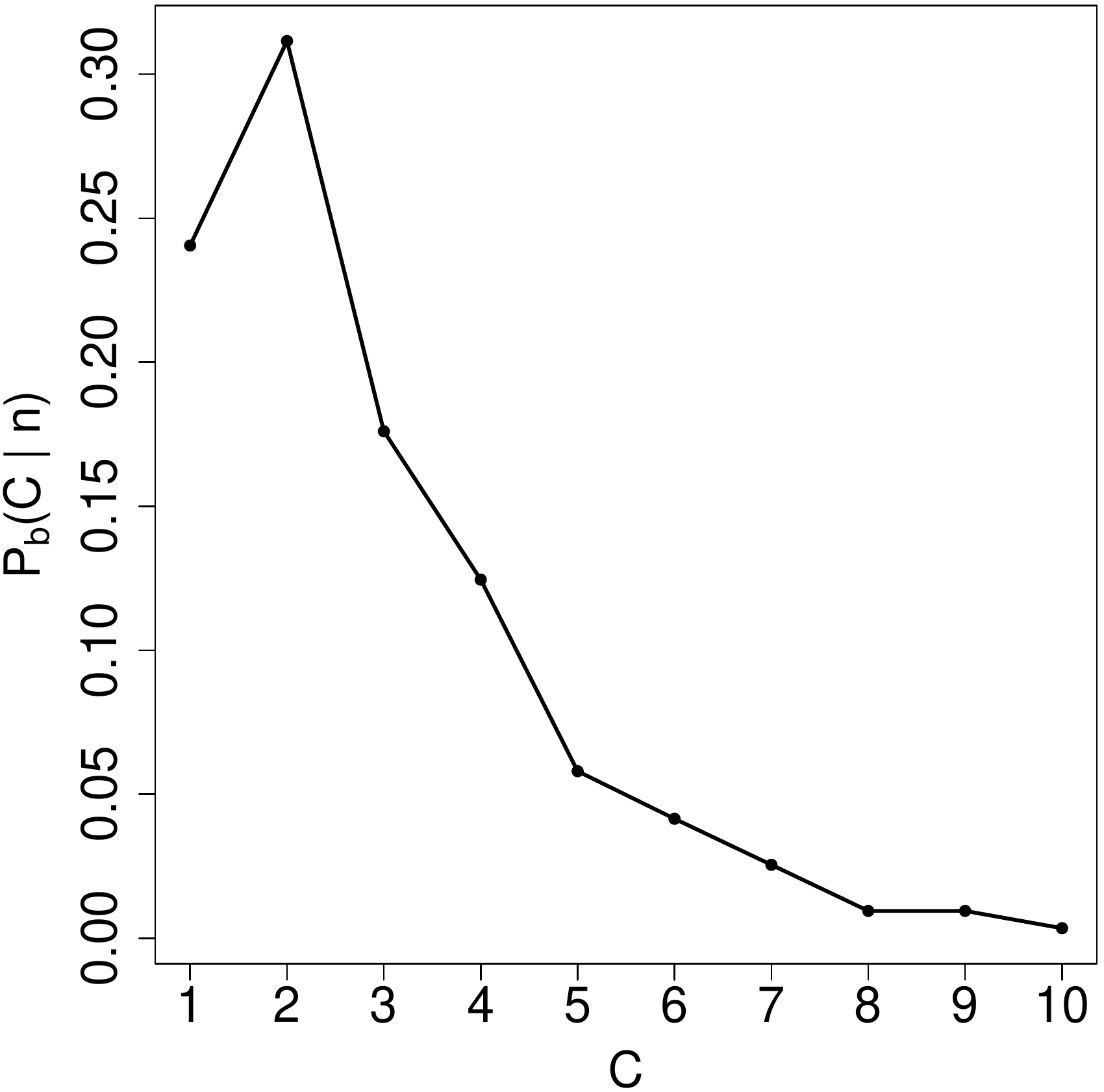}
\caption{$p_b(C \mid \bn'')$}	
\label{app:figlungC}	
\end{subfigure}
\hspace{6mm}\begin{subfigure}[t]{.3\textwidth}
\centering
\includegraphics[width=\textwidth]{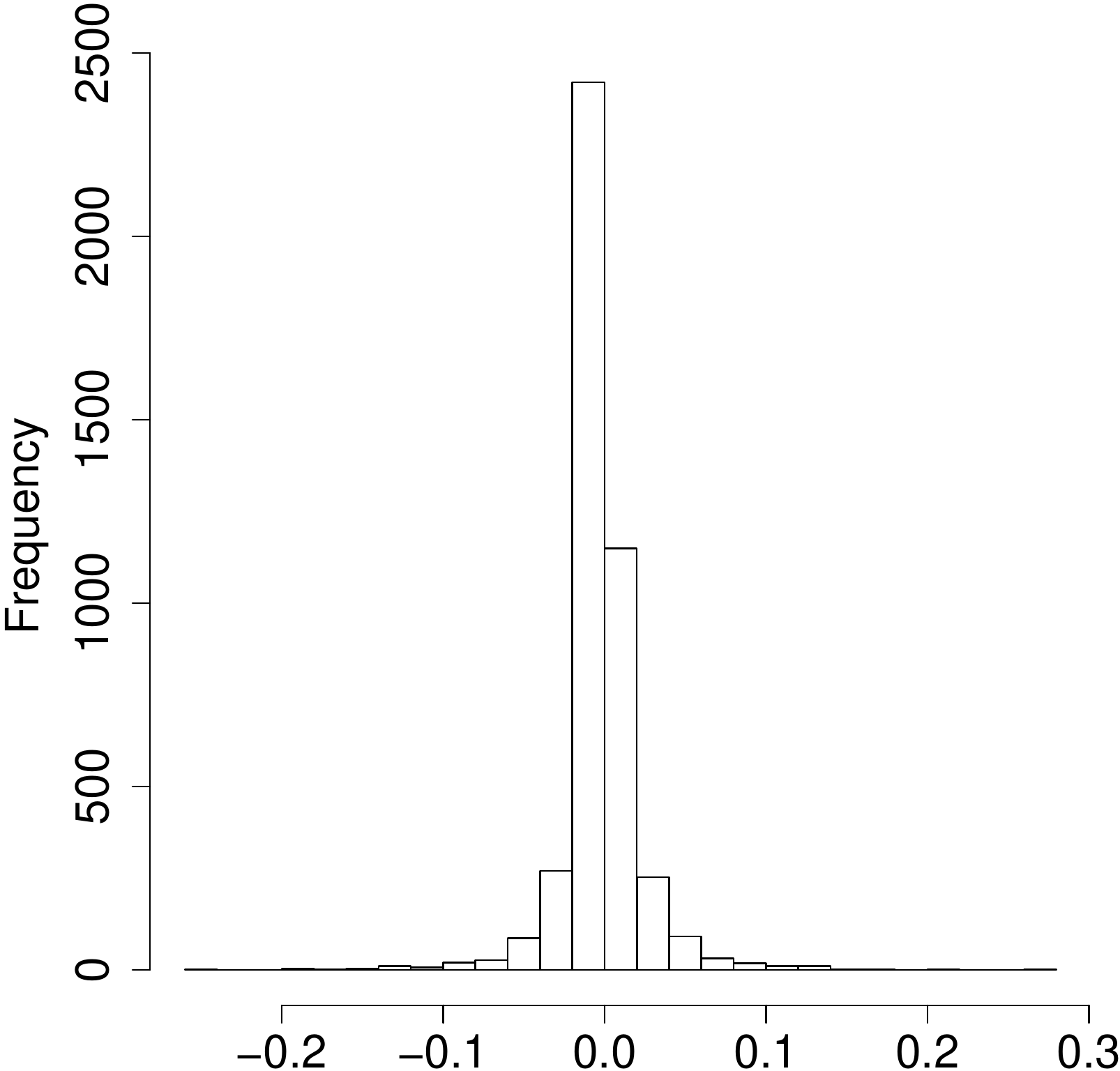}
\caption{Histogram of $(\ptkghat -\pbar_{tkg})$}		
\label{app:figlungresid}
\end{subfigure}
\end{center}
\caption{Lung cancer. Posterior inference under PairClone.}
\label{app:figlung}
\end{figure}

We present two more plots for the lung cancer data analysis (manuscript Section \ref{sec:realdata}). Figure \ref{app:figlungC} shows the posterior distribution $p_b(C \mid \bn'')$ with posterior mode $\Chat = 2$.  Figure \ref{app:figlungresid} shows the histogram of realized residuals.

\subsection{Validation of the MCMC scheme}
 \paragraph{Validation of the correctness of the sampler.}  
We first use a scheme to validate the correctness of our MCMC sampler
in the style of \cite{geweke2004getting}. The joint density of the
parameters and observed data can be written as $p(\bx, \bn) = p( \bx)
p(\bn \mid \bx)$. Let $g$ be any function $g: \mathcal{X} \times
\mathcal{N} \rightarrow \mathbb{R}$ satisfying $\text{Var}[g(\bx,
\bn)] < \infty$, where $\mathcal{X}$ and $\mathcal{N}$ represent
sample spaces of $\bx$ and $\bn$, respectively. 
Denote by $\bar{g} = E[g(\bx, \bn) ]$, which can be evaluated
by independent Monte Carlo simulation from the joint distribution, or
in some cases might be known exactly as prior mean of functions of
parameters only. 
Alternatively, the same mean can be estimated by a different Markov
chain Monte Carlo scheme for the joint distribution, 
constructed by an initial draw $\bx^{(0)} \sim p(\bx)$, followed by
$\bn^{(l)} \sim p(\bn \mid \bx^{(l-1)})$, $\bx^{(l)} \sim q(\bx \mid
\bx^{(l-1)}, \bn^{(l)})$, and $g^{(l)} = g(\bn^{(l)}, \bx^{(l)})$ ,
for $l = 1, \ldots, L$. Under certain conditions, $\{\bx^{(l)},
\bn^{(l)}\}$ is ergodic with unique invariant kernel $p(\bx, \bn)$.
If the simulator is error-free, one should have
\begin{align}
 (\bar{g}^{(L)} - \bar{g}) / \left[ L^{-1} \hat{S}_g(0) \right]^{1/2}
  \xrightarrow{d} N(0, 1),
\label{eq:geweke}
\end{align}
where $\hat{S}_g(0)$ is consistent spectral density estimate for
$\{g^{(l)}, l = 1, \ldots, L \}$.  In our application, we take $g(\bx,
\bn) = w_{tc}$ and $p_{tkg}$. We set the number of samples $T = 4$,
and the number of mutation pairs $K = 80$. Since our inference on $C$
is not a standard MCMC, we fix $C = 3$ here and only consider $\bx =
\{ \bZ, \bm \pi, \bw, \brho\}$. Table \ref{tbl:convd1} shows the
statistic \eqref{eq:geweke} for five randomly selected $w_{tc}$ and
$p_{tkg}$. 
The recorded $z$-scores show no evidence for errors in the
simulator. 
\begin{table}[h!]
\begin{center}
\begin{tabular}{ | c | c | c |}
    \hline
    Test statistic & $z$-score & $p$-value \\ \hline
    $w_{12}$ & -0.4736149 & 0.6357745 \\ \hline
    $w_{43}$ & -1.441169 & 0.149537 \\ \hline
    $p_{1,23,3}$ & 0.9413715  & 0.3465145 \\ \hline
    $p_{3,60,7}$ & 1.388424  & 0.1650079 \\ \hline
    $p_{2,13,2}$ & -0.6051894  & 0.5450532 \\ \hline
\end{tabular}
\end{center}
\caption{Geweke's statistics and the corresponding $z$-scores and $p$-values.}
\label{tbl:convd1}
\end{table}
 
\paragraph{Convergence diagnostic.}
Next, we present some convergence diagnostics of our MCMC chain, including trace plots, autocorrelation plots, and test statistics described in \cite{geweke1991evaluating}. Those convergence diagnostics are based on the posterior distribution of parameters $p(\bx \mid \bn) \propto p(\bx) p(\bn \mid \bx)$. Let $g$ be any function $g: \mathcal{X} \rightarrow \mathbb{R}$, and $g^{(l)} = g(\bx^{(l)})$ where $\{\bx^{(l)}, l = 1, \ldots, L \}$ are samples from the posterior. Let
\begin{align*}
\bar{g}^{A}_L = L_{A}^{-1} \sum_{l = 1}^{L_A} g^{(l)}, \quad \quad \bar{g}^{B}_L = L_{B}^{-1} \sum_{l = l^*}^{L} g^{(l)} \; \; \; (l^* = L - L_B + 1),
\end{align*}
and let $\hat{S}_g^A(0)$ and $\hat{S}_g^B(0)$ denote consistent spectral density estimates for $\{g^{(l)}, l = 1, \ldots, L_A \}$ and $\{g^{(l)}, l = l^*, \ldots, L \}$, respectively. If the ratios $L_A / L$ and $L_B / L$ are fixed, with $(L_A + L_B) / L < 1$, then as $L \rightarrow \infty$, 
\begin{align*}
(\bar{g}^{A}_L - \bar{g}^{B}_L) / \left[ L_A^{-1} \hat{S}_g^A(0) + L_B^{-1} \hat{S}_g^B(0) \right]^{1/2} \xrightarrow{d} N(0, 1).
\end{align*}

\begin{figure}[h!]
  \begin{center}
    \begin{tabular}{cc}
      \includegraphics[scale = 0.327]{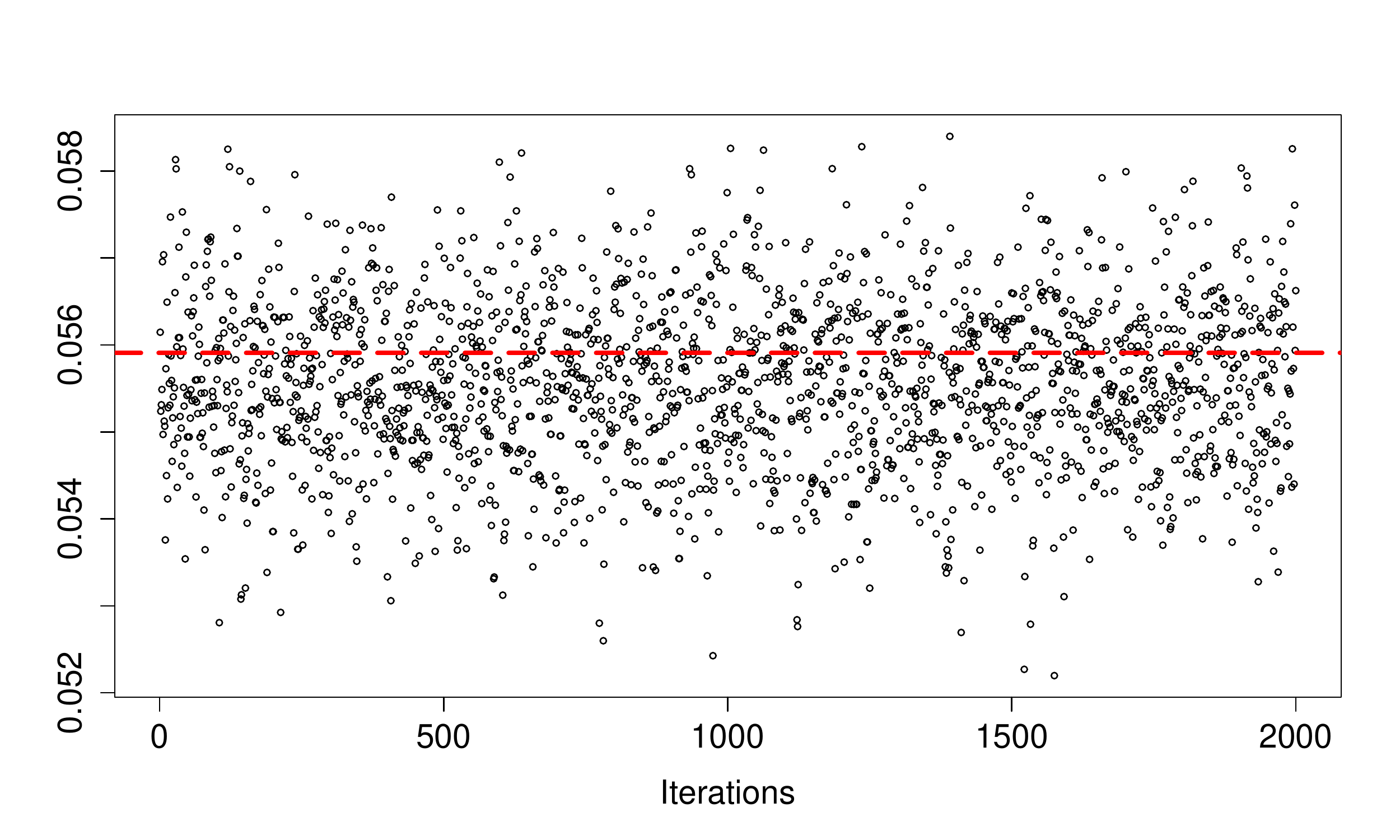} & 
      \includegraphics[scale = 0.28]{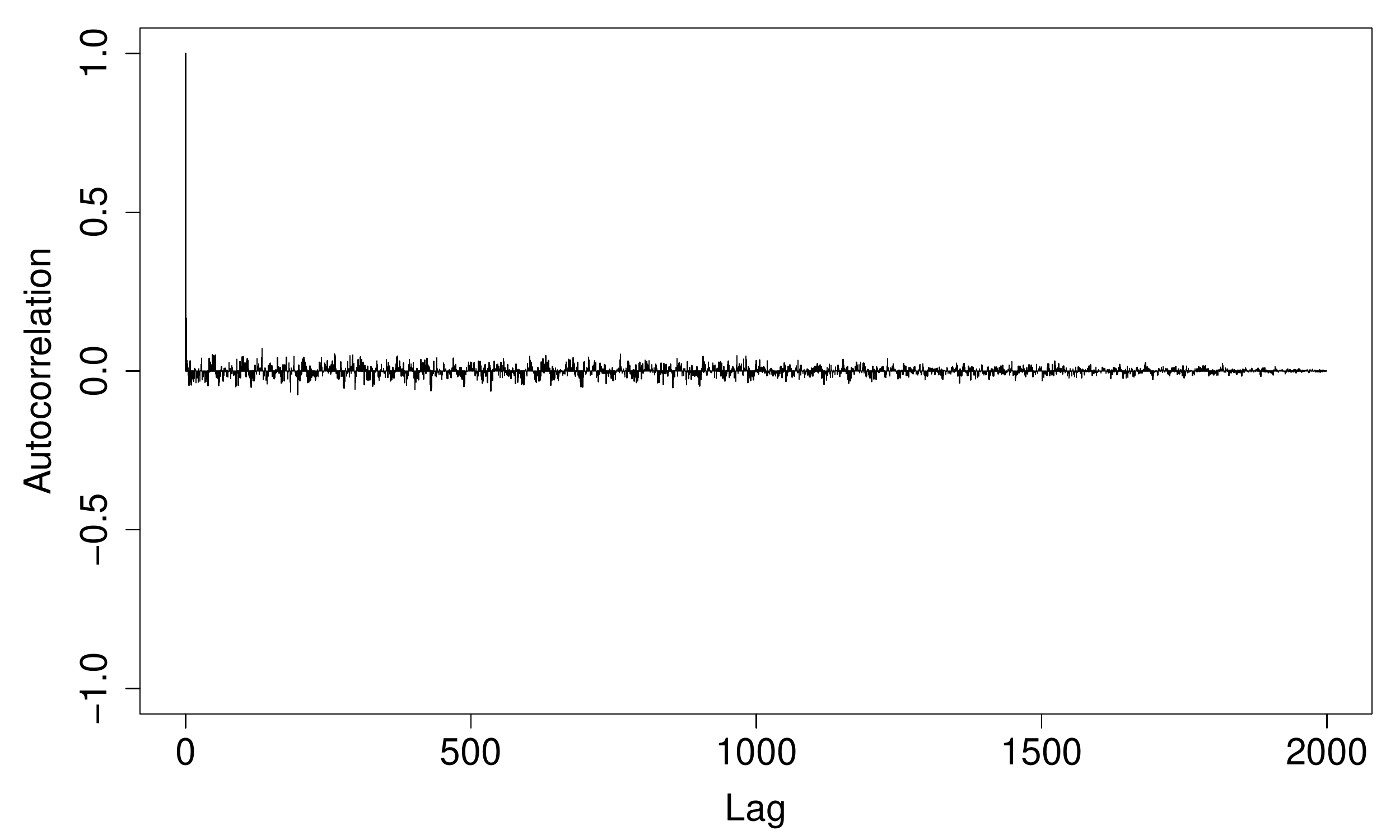}\\
      {\footnotesize{(a) Trace plot of $p_{1,5,2}$}} &
      {\footnotesize{(b) Autocorrelation plot of $p_{1,5,2}$}}\\
      \includegraphics[scale = 0.327]{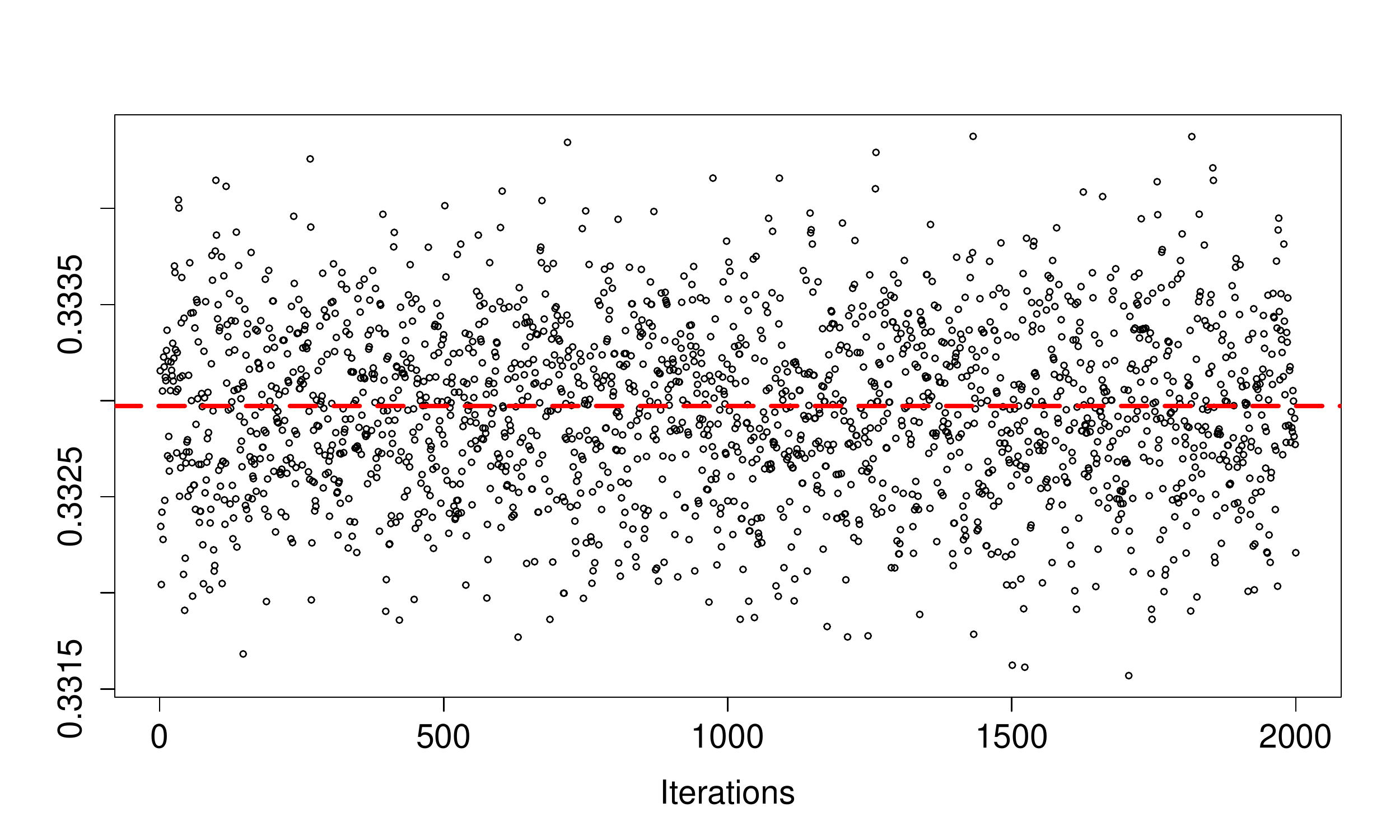} &
      \includegraphics[scale = 0.28]{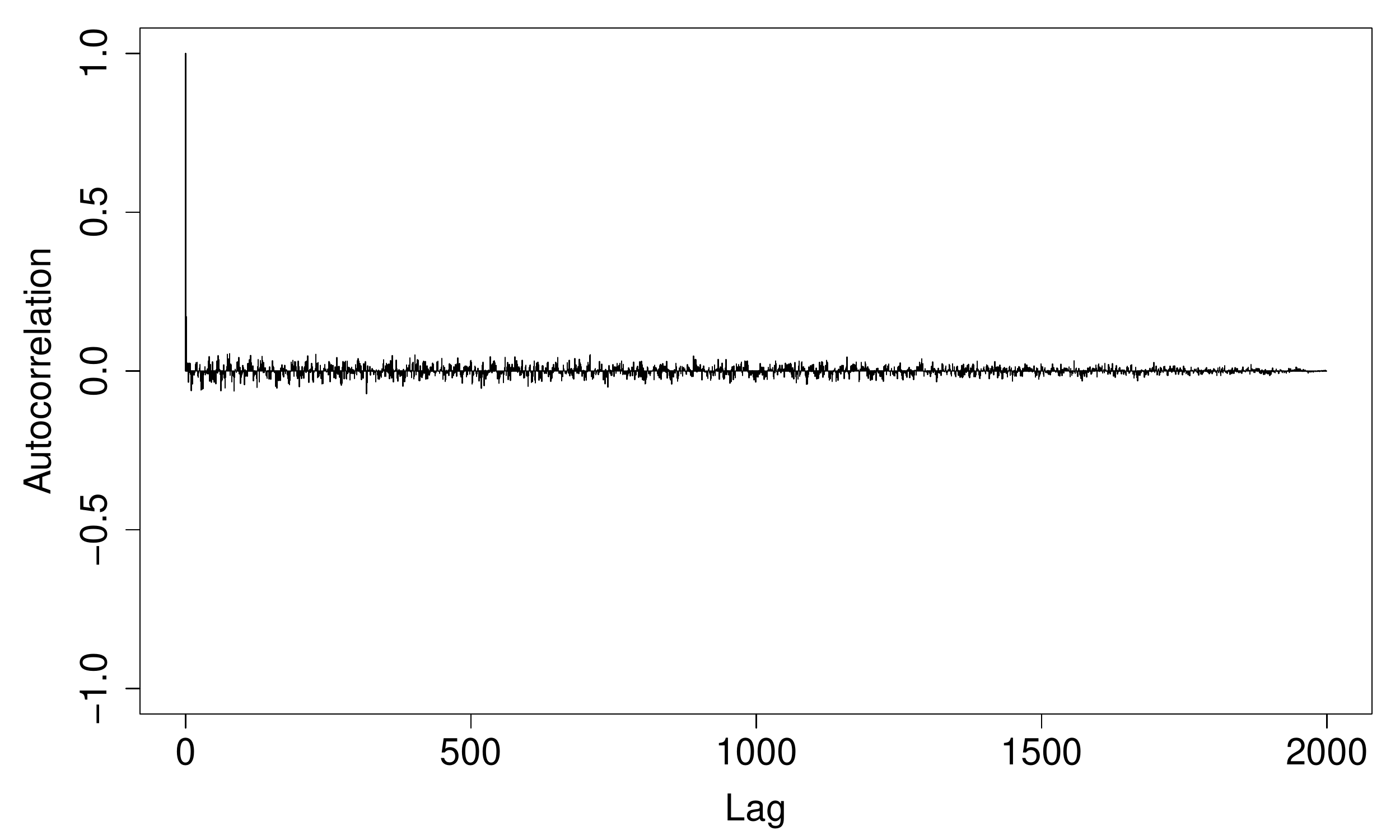}\\
      {\footnotesize{(c) Trace plot of $p_{3,68,7}$}} & 
      {\footnotesize{(d) Autocorrelation plot of $p_{3,68,7}$}}
    \end{tabular}
  \end{center}
  \caption{Convergence check for Simulation 2.}
\label{fig:convd}
\end{figure}

In our application, a reasonable choice of $g$ is $g(\bx) = p_{tkg}(\bZ, \bw, \brho)$.
We use simulation 2 as an example, and show some plots and Geweke's statistics for some randomly chosen $p_{tkg}$.
Figure \ref{fig:convd}(a, c) shows the trace plot for $p_{tkg}$, with the red dashed line denoting the true value. The posterior samples are centered around the true value and symmetrically distributed. Figure \ref{fig:convd}(b, d) shows the autocorrelation plot for $p_{tkg}$. The autocorrelations between MCMC draws are small, indicating good mixing of the chain. Table \ref{tbl:convd2} shows the Geweke's statistics for five randomly selected $p_{tkg}$. The $p$-values for them are all greater than 0.05, representing those statistics pass the Geweke's diagnostic, and there is no strong evidence that the chain does not converge.

\begin{table}[h!]
\begin{center}
\begin{tabular}{ | c | c | c |}
    \hline
    Test statistic & $z$-score & $p$-value \\ \hline
    $p_{1,5,2}$ & 0.1748906 & 0.8611656 \\ \hline
    $p_{3,68,7}$ & -0.02609703 & 0.9791799 \\ \hline
    $p_{4,25,5}$ & 0.4454738  & 0.6559774 \\ \hline
    $p_{2,96,4}$ & -1.341994  & 0.179598 \\ \hline
    $p_{1,66,1}$ & -0.2727737  & 0.7850272 \\ \hline
\end{tabular}
\end{center}
\caption{Convergence check for Simulation 2.}
\label{tbl:convd2}
\end{table}
\end{appendices}

\end{document}